\documentclass[fleqn,usenatbib,a4]{mnras}
\usepackage{newtxtext,newtxmath}
\usepackage{xspace}
 
\usepackage{hyperref}

\usepackage[T1]{fontenc}
\usepackage{ae,aecompl}
\usepackage{rviewport}
\usepackage{makecell}
\usepackage{siunitx}
\usepackage{stfloats}

\usepackage{graphicx}	
\usepackage{amsmath}	
\usepackage{amssymb}	
\usepackage{dcolumn}
\usepackage{pdflscape}

\usepackage{rotating}
\usepackage{lipsum}
\usepackage{makecell}
\usepackage{color}

\def \vareps {\varepsilon}
\def \Fermi{{\it Fermi}\xspace}
\def \Swift{{\it Swift}\xspace}
\def \TXS{TXS\,0506+056\xspace} 
\def \Sfour{S4\,0954+65\xspace}  
\def \OJ{OJ\,287\xspace}  
\def \Sfive{S5\,0716+714\xspace}  
\def \PG{PG\,1553+113\xspace}  
\def \AO{AO\,0235+164\xspace}  
\def \IES{1ES\,1959+650\xspace}
\def \IESa{1ES\,1959+650a\xspace}
\def \IESb{1ES\,1959+650b\xspace}
\def \Stwo{S2\,0109+22\xspace}
\def \mrkfour{Mrk\,421\xspace} 
\def \mrkfive{Mrk\,501\xspace} 
\def \ThreeC{3C\,66A\xspace}
\def \bllac{BL\,Lac\xspace}
\def \ie{{i.e.\xspace}}
\def \gRay{$\gamma$-ray\xspace}
\def \gRays{$\gamma$-rays\xspace}

\newcommand{\be}{\begin{equation}}
\newcommand{\ee}{\end{equation}}

\title[High-energy neutrinos from individual blazar flares]{High-energy neutrino flux from individual blazar flares}

\author[F. Oikonomou et al.]{Foteini Oikonomou$^{1}$\thanks{E-mail: foikonom@eso.org}, 
Kohta Murase$^{2,3}$, 
Paolo Padovani$^{1,4}$, 
\newauthor 
Elisa Resconi$^{5}$ 
and Peter M\'{e}sz\'{a}ros$^{2}$
\\
$^{1}$ European Southern Observatory, Karl-Schwarzschild-Str. 
2, D-85748 Garching bei M\"unchen, Germany\\
$^{2}$ Department of Physics; Department of Astronomy \& Astrophysics; Center for Particle \& Gravitational Astrophysics, \\ Institute for Gravitation and the Cosmos, Pennsylvania State University, University Park, PA 16802, USA \\
$^{3}$ Center for Gravitational Physics, Yukawa Institute for Theoretical Physics, Kyoto University, Kyoto, Kyoto 606-8502, Japan \\
$^{4}$ Associated to INAF - Osservatorio Astronomico di Roma, via Frascati 33,
I-00040 Monteporzio Catone, Italy\\
$^{5}$Technische Universit{\"a}t M{\"u}nchen, Physik-Department, 
James-Frank-Str. 1, D-85748 Garching bei M{\"u}nchen, Germany\\
}
\date{Accepted XXX. Received YYY; in original form ZZZ}
\pubyear{2019}

\begin{document}
\label{firstpage}
\pagerange{\pageref{firstpage}--\pageref{lastpage}}
\maketitle

\begin{abstract}
Motivated by the recently reported evidence of an association between a
high-energy neutrino and a \gRay flare from the
blazar TXS 0506+056, we calculate the expected high-energy neutrino signal
from past, individual flares, from twelve blazars, selected in declinations favourable for detection with IceCube. 
To keep the number of free parameters to a minimum, we mainly focus on BL~Lac objects and assume the synchrotron self-Compton 
mechanism produces the bulk of the high-energy emission. We consider a broad range of the allowed parameter space for the efficiency of proton acceleration, the proton content of BL Lac jets, and the presence of external photon fields. To model the expected neutrino
fluence we use simultaneous multi-wavelength observations. We find that in the absence of external photon fields and with jet proton luminosity normalised to match the observed production rate of ultra-high-energy cosmic rays, individual flaring sources produce a modest neutrino flux in IceCube, $ N^{\mathrm{IC,10 yr}}_{\nu_{\mu},{\mathrm{> 100~TeV}}} \lesssim 10^{-3}$~muon neutrinos with energy exceeding 100~TeV, stacking ten years of flare periods selected
in the $>800~$MeV \Fermi energy range, from each source. 
Under optimistic assumptions about the jet proton luminosity and in the presence 
of external photon fields, we find that the two most powerful sources in our sample, \AO, and \OJ, would produce, 
in total, $N^{\mathrm{IC \times 10,10 yr}}_{\nu_{\mu}, \rm all~flares, > 100~TeV} \approx 3$~muon neutrinos 
during \Fermi flaring periods, in future neutrino detectors with total instrumented volume $\sim$ ten times larger than IceCube,
or otherwise, constrain the proton luminosity of blazar jets.
  
\end{abstract}

\section{Introduction} \label{sec:intro}

The IceCube South Pole Neutrino
Observatory\footnote{http://icecube.wisc.edu} first reported the
observation of high-energy astrophysical neutrinos ~\citep{PhysRevLett.111.021103,icecubeScience, PhysRevLett.113.101101}
a few years ago. Most recently, the neutrino flux collected over 6
years with deposited energy up to about 2 PeV was reported in~\citet{Aartsen:2017mau}, 
bringing the number of high-energy starting
neutrino events detected to 82 and strengthening the significance
($>6.5\sigma$) of the observation that they are incompatible with
being of purely terrestrial origin. Many scenarios have been put
forward for the origin of the neutrinos (see
e.g.~\citealp{Ahlers:2018fkn} for a comprehensive review), but none
are statistically supported at the exclusion of all other models at
present.

The IceCube Collaboration has recently reported the observation of a
$\gtrsim 290$~TeV muon neutrino, IceCube-170922A, coincident with a $\sim6$-month-long 
$\gamma$-ray flare of the blazar TXS\,0506+056~\citep{IceCube:2018dnn}  
at redshift $z = 0.3365$~\citep{Paiano:2018qeq}. The neutrino detection prompted
electromagnetic follow-up of the event, and the blazar flare was detected
by several instruments, including MAGIC at energies exceeding > 100
GeV. The correlation of the neutrino with the flare of TXS 0506+056 is
inconsistent with arising by chance at the 3-3.5$\sigma$ level. An
archival search revealed $13\pm5$ further, high-energy neutrinos in
the direction of TXS 0506+056 during a 6-month period in 2014-2015~\citep{IceCube:2018cha}. 
These events were not accompanied by a
$\gamma$-ray flare. Such an accumulation of events is inconsistent
with arising from a background fluctuation at the 3.5$\sigma$ level.

Blazars are active galactic nuclei (AGN) hosting a strong relativistic
jet, which is oriented at a small angle with respect to the line of
sight~\citep{Urry:1995mg}. Based on their optical spectra, blazars are
divided into two main sub-classes, namely BL~Lacertae objects (BL~Lacs) and Flat Spectrum Radio Quasars (FSRQs). FSRQs display broad,
strong emission lines, while BL~Lacs exhibit at most weak emission
lines, or in many cases featureless optical spectra. According to the
value of the frequency (in the source rest-frame) at which the
synchrotron, i.e. low-energy, peak, $\nu_S$, of the spectral energy
distribution (SED) occurs, blazars are divided into Low ($\nu_S <
10^{14}$~Hz), Intermediate ($10^{14} < \nu_S < 10^{15}$~Hz) and High
energy peaked ($\nu_S < 10^{15}$~Hz), referred to in short as LSP, ISP
and HSP respectively~\citep{Padovani:1994sh,2010ApJ...716...30A}.

Blazars have long been discussed as some of the most likely sources of
high-energy neutrinos and cosmic-rays (see~\citealp{1995APh.....3..295M,1997ApJ...488..669H,Atoyan:2001ey,2003APh....18..593M,Murase:2011cy,Dermer:2012rg,2014PhRvD..90b3007M,Padovani:2014bha,Dermer:2014vaa,
  Petropoulou:2015upa,
  Padovani:2016wwn,
  Gao:2016uld,Rodrigues:2017fmu},
and references therein).

Observations of the diffuse, all-sky neutrino spectrum by IceCube have
resulted in constraints on the time-averaged emission from \gRay
bright blazars as sources of high-energy neutrinos from
analyses of the observed diffuse neutrino flux~\citep{Aartsen:2016ngq,Aartsen:2017mau}, 
and stacking analyses~\citep{Aartsen:2016lir,Aartsen:2017kru}, down to $\lesssim 10 - 30\%$ 
of the diffuse astrophysical flux observed by
IceCube at $\lesssim 100$~TeV. Thus blazars are inconsistent with being 
the dominant sources of the observed IceCube neutrino flux,
unless the faintest blazars produce a disproportionately 
large amount of the neutrinos~\citep{Palladino:2018lov}.
Constraints on blazars as dominant sources of the diffuse neutrino intensity 
observed by IceCube are additionally imposed by the lack of multiplets in the IceCube data, 
often referred to as ``clustering constraints''~\citep{2016PhRvD..94j3006M,Aartsen:2014ivk,Aartsen:2017kru,Neronov:2018wuo,Murase:2018iyl}, which limits in a complementary way the blazar contribution to the diffuse IceCube flux~\citep{Yuan:2019ucv}. 

Despite these limits the brightest neutrino sources, could still be blazars, due to, for example beaming. 
Blazar flaring periods are ideal for the detection of high-energy neutrinos both 
observationally and theoretically. Experimentally, the short, well-defined time duration of a flare means a very reduced background rate. Theoretically, in many models of neutrino emission, it is natural for the neutrino-production efficiency to be doubly enhanced during flares
because it is typically expected that the proton injection increases while
at the same time the target photon field is in an enhanced state 
(see e.g.~\citealp{2014PhRvD..90b3007M,Tavecchio:2014xha,Petropoulou:2016ujj,2016PhRvD..94j3006M,Righi:2016kio,Rodrigues:2017fmu}).
It was shown in~\citet{Murase:2018iyl} that even though blazars 
emit a small fraction of their \gRay luminosity during flares~\citep[e.g.][]{2010ApJ...722..520A,2011ApJ...743..171A},
 the distribution of flaring states of several studied
\Fermi blazars is consistent with neutrino production during flares
being dominant in canonical models of neutrino emission. 

The reported association of IceCube-170922A with, and flaring
activity by, TXS 0506+056~\citep{IceCube:2018dnn,Ahnen:2018mvi,Abeysekara:2018oub,Keivani:2018rnh,Padovani:2018acg}
prompts us to investigate here the number of
expected neutrino events from earlier blazar flares, that were also
detectable by IceCube. Our aim is to investigate what are the
conditions under which a blazar flare may produce a strong neutrino
flux and to present a strategy for searching for neutrinos from blazar flares.  
The procedure we provide is easily applicable to general stacking analyses 
on blazar flares as we illustrate in the following sections.

The detectability of individual blazar flares of neutrinos in IceCube
has been discussed in~\citet{Atoyan:2001ey,Halzen:2005pz,Dermer:2012rg,Dermer:2014vaa, Petropoulou:2016ujj,Halzen:2016uaj,Kadler:2016ygj,Gao:2016uld,Guepin:2017dfi}. 
In~\citet{Turley:2016ver,Turley:2018biv} IceCube data were analysed to
look for neutrino emission spatially and temporally correlated with TeV
emitting blazars, and \Fermi data respectively. Here, we opt for a fully self-consistent, 
``leptonic'' setup (referred to by other authors as ``leptohadronic'' or ``hybrid''), in which the
neutrinos are produced by the interactions of protons with target
radiation fields in the source environment. In this context, leptonic
refers to the fact that the majority of the observed \gRays are
produced by the interactions of leptons in the blazar jet and \gRays
from the interactions of the protons are present but not dominant. 

With respect to the time-dependent radiation modelling of e.g.~\citet{Petropoulou:2016ujj}
our semi-analytic approach has the advantage of fast computation with reasonable accuracy, which allows for parameter-space scans and application to multiple sources. 
We have benchmarked our method against the time-dependent radiation modelling of~\citet{Keivani:2018rnh} 
and find agreement within a factor of two for the expected neutrino flux between the two approaches.

In Section \ref{sec:data} we present the flare sample used
in this work. In Section \ref{sec:method} we present the adopted neutrino modelling
formalism, the blazar SED fitting method and the
neutrino flux calculation procedure. Section \ref{sec:results}
presents our results. We discuss the implications of our findings and
conclude in Section \ref{sec:discussion}. We assume a flat Universe
 with $\Omega_{M}$ = 0.3, $\Omega_{\Lambda}$ =
0.7 and $H_{0}$ = 70~km s$^{-1}$~Mpc$^{-1}$.

\section{Data Selection}
\label{sec:data}

To calculate the expected neutrino output from individual blazar
flares, an essential ingredient is the availability of
multi-wavelength, and simultaneous observations of the broadband
SED of the blazar. If hadrons exist in sufficient amounts, neutrinos are produced 
by the interactions of the hadrons with photons in the jet and external fields, when the latter are present. 

Since its launch in June 2008, {\it Fermi}-LAT surveys the entire sky
every three hours, allowing for the first time monitoring of a sizeable
number of blazars in the 30 MeV- 300 GeV energy band
\citep{2009ApJ...697.1071A}. Unfortunately, a similar monitoring
instrument does not exist in other wavelengths at present. Several
optical programs perform follow-up observations of \Fermi monitored blazars
\citep{2012ApJ...756...13B,2009arXiv0912.3621S}, and there are similar
programs in the radio band
\citep{2011ApJS..194...29R,galaxies4040047}. The \textit{Swift},
\textit{Integral}, and since more recently \textit{NuSTAR}
\citep{2013ApJ...770..103H} telescopes perform ToO observations to
better cover the multi-wavelength SEDs, generally after being
triggered due to extraordinary activity at a different
wavelength. Very-high-energy (VHE) \gRay telescopes also perform
monitoring and follow-up observations of sources seen in exceptionally
high-states by \Fermi or other instruments.

Since a complete sample of flares does not exist, except in the \Fermi
energy band, our ability to model the neutrino emission from
individual blazar flares is constrained by the availability of
simultaneous data. For this work, we collected publicly available data
for recorded flaring episodes of 12 \Fermi bright BL~Lacs, for which
simultaneous (or semi-simultaneous) observations exist, with
sufficient spectral coverage as to infer the main characteristics of
the SED with reasonable confidence (for example the peak synchrotron
frequency and peak synchrotron flux). Neutrino production is expected
to be much more efficient in FSRQs than in BL Lac objects due to the
higher-powers and existence of external photon fields
\citep{2003ApJ...586...79A}. However, these also introduce
additional free parameters. Therefore in this work we have considered
only objects classified as BL~Lacs.

We use simultaneous or quasi-simultaneous observations of the sources at all
wavelengths since we want to calculate the neutrino output during the
flaring state. All the blazars we have studied are monitored in the
optical (R-band) by the Tuorla long-term monitoring program
\citep{2018arXiv181001751N}. Additionally, most are monitored by the
Stewart Observatory \Fermi blazar monitoring program
\citep{2009arXiv0912.3621S}, and several are monitored in the optical
and near-infrared by SMARTS~\citep{2012ApJ...756...13B}. The details
of all the flares in our sample are given in Table~\ref{tab:1}. Below
we briefly present them in chronological order.

\begin{table*}
\begin{center}
\caption{The list of flares studied in this work. For each flare the
  table lists the redshift (or assumed redshift) of the source,
  declination, and a reference time that roughly corresponds to the
  onset of the flare in MJD-i.e. Modified Julian Date, $t_{\rm Ref}$. In addition, the month of
  onset of the flare, references where the flare campaigns were
  presented, IceCube configuration at the time of the flare, the
  blazar classification of each source, and selection band for each
  flare are listed. For IceCube partial configuration definitions see Table~\ref{tab:acronyms}.
\label{tab:1}}
\begin{tabular}{c c c c c c c c c}
\hline
Source    		& z 		&  decl. & $t_{\rm ref}$  [MJD] & UT		& 	Data Ref.           & Config. & Type	& Selection Band \\ 
\hline
3C 66A		 & 0.34	  &	$+43.04$	& 	 54747 &	 Oct 2008		&  1	 	& IC40	& ISP 		&	VERITAS\\
AO 0235+164 	 & 0.94	  & 	$+16.62$	&	54761 & 	Oct 2008	 	& 2	 	&  IC40	& LSP 		&	\Fermi	\\
Mrk 421  		& 0.031       &	$+38.21$	 &	55265 &	 Mar 2010		& 3 	 	& IC59	& HSP 		&	MAGIC	\\
PG 1553+113	& $\sim 0.4$ & $+11.19$ 	 & 56037 & 	Apr 2012		& 4,5	&	 IC86	& HSP		& MAGIC \\ 
1ES 1959+650a   & 0.048  & $+20.00$ 	& 56066.5	 & 	May 2012 		& 6		& IC86	& HSP 		& VERITAS  \\
Mrk 501 		& 0.033	  & 	$+39.76$ 	& 	56087 & 	Jun 2012		& 7		&       IC86	& HSP		& MAGIC	\\
S5 0716+714 	 & $<0.32$ & $+71.37$	&	 57045.5	& 	Jan 2015		&8,9		& IC86 	& ISP/LSP		&	Optical/NIR \\
S4 0954+65	 & $\geq 0.45$ & $+65.57$  	 &	57050	&	Feb 2015		& 10,11	& IC86	&HSP&	\Fermi \\
BL Lac 		 & 0.07	  & $+42.28$	& 	57192 &	 Jun 2015		& 12	 	& IC86	& ISP 		&	\Fermi \\ 
S2 0109+22	   & 0.36	 & $+22.74$		&  57228 &      Jul 2015 		& 13		& IC86 	& ISP	 	& \Fermi  \\
1ES 1959+650b   & 0.048  & $+20.00$		& 57285  & 	Sep 2015 		& 14		& IC86 	& HSP 		& \Fermi 	\\
OJ 287	 	& 0.306 	  & $+20.11$	& 	57359   &     Dec 2015 		& 15		& IC86	& LSP		& theoretical  \\
TXS 0506+056 	   & 0.3365 & $+5.70$		& 58002 & Sep 2017 	& 16 & IC86 & ISP & \Fermi \\ 
\hline
\end{tabular}
\\
\vspace{0.1cm}
{\footnotesize References: 1~-~\citet{2011ApJ...726...43A}. 
2~-~\citet{2012ApJ...751..159A}. 
3~-~\citet{Aleksic:2014rca} .  
4,5~-~\citet{Abramowski:2015ixa,2015MNRAS.450.4399A}. 
6~-~\citet{0004-637X-797-2-89}. 
7~-~\citet{2018A&A...620A.181A}.
8~-~\citet{2015ApJ...809..130C}, 9-\citet{2016Galax...4...69M}. 
10, 11~-~\citet{2016PASJ...68...51T,2018arXiv180104138M}. 
12~-~\citet{Acciari:2019vpr}.
13~-~\citet{2018MNRAS.480..879M}. 
14~-~\citet{0004-637X-846-2-158}.
15~-~\citet{2018MNRAS.473.1145K}.
16~-~\citet{IceCube:2018dnn}.}
\end{center}
\end{table*}

\subsection{3C 66A}
The blazar 3C 66A is classified as an ISP. It has recently been
possible to determine the redshift of 3C 66A by association with its
host galaxy cluster as $z = 0.34$~\citep{2018MNRAS.474.3162T}. We use
this latter value for neutrino calculations. In October 2008 \ThreeC
underwent a strong flare which triggered \Fermi, and subsequently a
multi-wavelength campaign, which we model in this work.

\subsection{AO 0235+164} 
AO 0235+164 was monitored by a multi-wavelength campaign between 2008
- February 2009. AO 0235+164 is generally classified as a BL Lac, though based
on the equivalent width of its H$\alpha$ line it would be classified
as an FSRQ~\citep{DElia:2015svu}. During the 2008-9 campaign AO
0235+164 underwent a series of high states, most prominently between
MJD 54750-54570. During this time its SED exhibited a hard X-ray
feature, consistent with arising from bulk Compton emission from cold
electrons confined in the jet~\citep{2012ApJ...751..159A}. We conservatively model the neutrino
emission during the 2008 flare of AO 0235+164 without considering this
bulk Compton component or additional external photon fields only
expected in FSRQs. \AO experienced a second, approximately year-long,
phase of flaring starting in August 2015 in the \Fermi band
\citep{2015ATel.7975....1C,2015ATel.8044....1C}. Though a
simultaneous SED with sufficient multi-wavelength coverage could not be obtained for the 2015
flare we comment on the likely neutrino production during this flare
in Section \ref{sec:results}.

\subsection{Mrk 421}

\mrkfour is an HSP BL Lac, and one of the best-studied blazars, due to
its proximity at $z=0.031$. In March 2010, it underwent a $\sim 13$-day 
flare, which triggered a very extensive multi-wavelength
observation campaign~\citep{Aleksic:2014rca}. This flare was
extensively studied numerically in~\citet{Petropoulou:2016ujj} in the
context of leptohadronic neutrino production. We include it in our
sample to facilitate a comparison between this well-studied source and other sources. 

\subsection{PG 1553+113}

PG 1553+113 is an HSP BL Lac with uncertain redshift. With declination
$+11.19^{\circ}$ it lies in a part of the sky that IceCube is most
sensitive to, and has been considered as a candidate source of
astrophysical neutrinos in several earlier studies
\citep{Ribordy:2009jj,Cerruti:2016all,Petropoulou:2015upa}. In April
2012 it underwent a major flare in the VHE $\gamma$-ray band and was
the subject of a large multi-wavelength campaign between
February-August of the same year
\citep{2015MNRAS.450.4399A}. Constraints on the redshift of PG
1553+113 come from the detection of Ly~$\alpha$ absorption in its optical
spectrum~\citep{2010ApJ...720..976D}, and the imprint of the EBL on
its VHE spectrum. Here we assume $z = 0.4$, which was deemed the most
likely value in~\citet{2015MNRAS.450.4399A}.

\subsection{\IES}

\IES is a nearby, $z=0.047$ HSP BL Lac. In June 2002 it underwent an orphan TeV  
flare~\citep{Holder:2002ru,2004ApJ...601..151K,Daniel:2005rv}, not compatible with a 
one-zone SSC interpretation (e.g.~\citealp{2005ApJ...621..176B}). A search for neutrinos 
in the AMANDA data revealed three neutrinos coincident with the flaring activity~\citep{2005ICRC....5....1A}. 
We model the 2012 flare of this source reported in~\citet{0004-637X-797-2-89}. 
A VHE flare was seen on MJD 56067, though no strong flare was seen in the
\Fermi~data. We refer to this as flare \IESa. We also make predictions for the October 2015 
outburst~\citep{0004-637X-846-2-158}, which we refer to as \IESb. The source underwent strong 
flaring in the \gRay band in 2016 (see~\citealp{Aartsen:2017kru} and references therein), 
but the SED for this time is not yet publicly available. 
A limit to the neutrino output of this flare was presented in~\citet{Aartsen:2017kru}. 

\subsection{Mrk 501}

\mrkfive, the second nearest BL Lac, at $z = 0.034$ is also an HSP. A
number of multi-wavelength campaigns have been coordinated to monitor
\mrkfive in the last ten years~\citep{Ahnen:2016hsf,2018A&A...620A.181A,Abdalla:2019krx}. The largest
observed VHE flare of this source since 1997 was analysed in
\citet{2018A&A...620A.181A} during a time when Mrk 501 was in an
extreme HSP phase and is the focus of our study of this source. The
VHE flare lasted one day (MJD 56087), while the \Swift flare lasted
until MJD 56089. The duration of this flare in the \Fermi energy range is $\sim$1~week. 

\subsection{S5 0716+714} 

The BL Lac S5 0716+714 exhibited an unprecedented optical/NIR flare on
January 2015, which triggered multi-wavelength observations, and 
detection up to VHE energies with MAGIC
\citep{2015ApJ...809..130C,Agudo:2018bhs,2016Galax...4...69M}. It is
classified as an ISP~\citep{1999A&A...351...59G}. Its redshift has not
been directly measured due to a featureless optical spectrum
\citep{0004-637X-837-2-144}, but detection of Ly$\alpha$ absorption at
the UV part of its spectrum puts it at $z<0.32$  with 95\% confidence~\citep{2013ApJ...764...57D}, 
consistent with the estimate of~\citet{2008A&A...487L..29N} 
based on marginal detection of the host galaxy.

\subsection{S4 0954+65}

During February 2015, the \Fermi flux of \Sfour increased by a factor
of $\sim 40$ (see Figure \ref{fig:Lightcurves}). The classification of
this source has been debated, but the latest high-signal-to-noise
spectroscopic observations classify it as a BL Lac with unknown redshift~\citep{1538-3881-150-6-181}.
 Even if S4 0954+65 is an FSRQ, as has been claimed previously, our SSC model results here
are valid as a minimal neutrino scenario for this source at its
assumed redshift. In the absence of a firm measurement of the redshift
of S4 0954+65, we assume the lower limit of
\citet{1538-3881-150-6-181}, who obtained $z \geq 0.45$. In this
sense, our estimate of the neutrino flux from the February 2015 flare
of S4 0954+65 is optimistic.

\subsection{BL Lac}

BL Lacertae, at the redshift of $z = 0.069$, is the prototype of the BL
Lac blazar type. The source flares very regularly in the \Fermi
band. Here we model the June 2015 flare of BL Lac, which triggered a
multi-wavelength campaign~\citep{2017arXiv170905063T,Acciari:2019vpr}.

\subsection{S2 0109+22}
The ISP S2 0109+22 was observed to be in a high state in July 2015
by the \Fermi-LAT. Multi-wavelength observations revealed a VHE flare
on July 24th (MJD 57228). The most likely redshift of S2 0109+22 has
been determined to be z = 0.36, based on association with its host cluster
\citep{2018MNRAS.480..879M}.

\subsection{OJ 287}

OJ 287 is a blazar known to exhibit a 12-year periodicity of its
optical brightness~\citep{2007Ap&SS.310...59S}, thought to host a
binary supermassive black hole~\citep{2006ApJ...643L...9V}. In December
2015, it underwent the latest of a series of quasi-periodic optical
outbursts. High variability from near-infrared to \Fermi-LAT energies
was exhibited between November 2015-May 2016, specifically MJD 57315 - 57460
\citep{2018MNRAS.473.1145K,Kushwaha:2019upf}. It further underwent
intense near-infrared and X-ray variability between September
2016-July 2017 and was at the time detected at $>100$~GeV by VERITAS~\citep{2017arXiv170802160O}.
Here we model the December 2015 flare. We were unable to fit the subsequent 
bursts of this source within our single-zone formalism. 

\subsection{\TXS}

We include in our sample the 2017 flare of \TXS to allow for
comparison of relative neutrino signal expectations in different
blazar flares. The neutrino signal from this flare, coincident with IceCube-170922A was modelled
extensively~\citep{Ahnen:2018mvi,Cerruti:2018tmc,Gao:2018mnu,Keivani:2018rnh,Murase:2018iyl,Liu:2018utd,Padovani:2018acg,Oikonomou:2019pmg}. We do not attempt to model the 2014-15 flare of \TXS here, as it would not fulfill our selection criteria based on \gRay/X-ray or optical
flaring. The 2014-15 flare of \TXS was studied in~\citet{Murase:2018iyl,Reimer:2018vvw,Rodrigues:2018tku,Wang:2018zln}.
In~\citet{Padovani:2019xcv} evidence was presented that \TXS is, despite its generally accepted classification, intrinsically an FSRQ. In this sense, the SSC model for this
source (Normalisation A) is conservative, but the comparison to other BL~Lac objects under fixed model assumptions is instructive.

\section{Model of blazar emission} \label{sec:method}

We assume that the synchrotron and synchrotron self-Compton (SSC) mechanisms produce
the bulk of the emission in the studied sources, i.e. electrons
produce synchrotron photons responsible for the low-energy bump of the
SED, and then they Compton scatter them to high energies producing the
high-energy hump. In the SSC model, the simultaneous determination of
the peak synchrotron frequency, $\nu_S$, the peak inverse-Compton
frequency, $\nu_{\rm C}$, the luminosity of the synchrotron peak,
$L_{S}$, and of the self-Compton peak $L_{C}$, allows us to uniquely
determine the physical parameters of the system.

The Doppler factor, $\delta$, characterises the relativistic effects
of the observed radiation from the blazar, and is given by $\delta =
[\Gamma ( 1 - \beta \cos \theta)]^{-1}$, where $\Gamma$ is the bulk
Lorentz factor, $\theta$ the angle of motion in the jet with respect
to the observer's line of sight, and $\beta = v/c$. In this work, we
will assume the special case where $\theta \sim 1/\Gamma$, which
implies $\delta \sim \Gamma$ (see
e.g. Appendix A and B of~\citealp{Urry:1995mg}). From this point on, we will use $\delta$ and $\Gamma$ 
interchangeably. In the SSC formalism
$\delta$ is given by~\citep{1996A&AS..120C.503G,Tavecchio:1998xw},

\begin{equation}
\delta \simeq 6.5 \left( \frac{\nu_C}{ c^{3/2} t_{\rm var,d}\nu_S^2}
\right)^{1/2} \left[ \frac{L_{S,{\rm bb}}^2}{L_{C,{\rm bb}}}
  \right]^{1/4},
\label{eq:DopplerFactor}
\end{equation}
where $t_{\rm var,d}$ the variability timescale of the system in days. 
Here, $L_{S(C),{\rm bb}}$ are the broad-band luminosities which relate to
the peak luminosities as $L_{S(C),{\rm bb}} = f(b) \cdot L_{S(C)}$,
where $L_{S(C)} = 4\pi d_L^2\nu_SF_{\nu}^{S(C)}$, with
$\nu_S(C)F_{\nu}^{S(C)}$ the peak fluxes and, $d_L$ the luminosity
distance to the source. The factor $f(b) \simeq 3 - 5$, is a constant of
integration dependent on the exact shape of each hump of the SED (see
Section \ref{subsec:SEDFitting}). Equation
\ref{eq:DopplerFactor} is valid when the inverse Compton process
proceeds in the Thomson regime. The magnetic field strength is given by, 
\begin{equation}
B \simeq 2.4 \times 10^{-11}~{\rm G}\cdot (1+z) \cdot \frac{v_S^3
  t_{\rm var,d}^{1/2}}{v_C^{3/2}} \left[ \frac{L_C/10^{45}~{\rm
      erg/s}}{(L_S/10^{45}~{\rm erg/s})^2} \right]^{1/4}. 
\label{eq:B_SSC}
\end{equation}
\noindent When the self-Compton process is in the Klein-Nishina regime, the
values of $v_{C}$ and $L_{C}$ are affected. The condition for the
Klein-Nishina regime to severely affect the luminosity at the peak
Compton frequency can be expressed in terms of a limiting Doppler
factor, $\delta_{\rm KN}$, above which revision of Equations
\ref{eq:DopplerFactor} and \ref{eq:B_SSC}, is necessary,

\begin{equation}
\delta_{\rm KN} = \left[\frac{\nu_C \, \nu_S }{(3/4) (m_ec^2/h)^2}\right]^{1/2},
\label{eq:DopplerKN}
\end{equation}
where $m_e$, $h$, and $c$ are the electron mass, Planck constant, and
speed of light in cgs units respectively. Equations 
~\ref{eq:DopplerFactor}, \ref{eq:B_SSC} and \ref{eq:DopplerKN} above are valid when 
all quantities are in cgs units, except the variability timescale 
which is in days. 

\subsection{High-energy neutrino production in the relativistic blob} 
\label{subsec:neutrino_production}

High-energy neutrinos are produced by the decay of charged pions from
the interactions of relativistic protons with photons, and matter
(gas). In blazar jets interactions of relativistic protons with
the ambient matter should be subdominant as energetics arguments constrain
gas densities to be rather low in steady jets
\citep{2003ApJ...586...79A}. The protons interact with ambient photons
in the jet, which, in the case of BL~Lacs are likely the synchrotron
photons produced by co-accelerated electrons. In FSRQs, several other
radiation fields, external to the jet, could provide an efficient
target for the protons, namely from the accretion disk, or
jet/accretion disk photons reprocessed in the Broad Line Region, or
photons from the dust torus~\citep{2014PhRvD..90b3007M,
 Dermer:2014vaa,Rodrigues:2017fmu}.

The threshold energy $E_{p,\rm th}$ for highly relativistic protons to
produce a pion in interactions with photons with energy
$\varepsilon_{\rm \gamma}$ in the laboratory frame is
\begin{equation}
E_{p,\rm th} = \frac{2m_{p}m_{\pi}+m_{\pi}^2}{4 \varepsilon_{\rm
    \gamma}} \approx 7 \times 10^{16} \left[ \frac{\varepsilon_{\rm
      \gamma}}{1~\rm eV} \right] ^{-1},
\end{equation} 
where $m_{p}$, $m_{\pi}$ are the the proton and pion mass
respectively, and we have given the expression in the case of a
head-on collision. On average, in a $p\gamma$ collision, $\sim 20\%$
of the proton energy goes to the pion, which decays to four leptons,
which share the energy of the pion almost equally. Thus the
characteristic neutrino energy with respect to the energy of the
parent proton is given by
\begin{equation}
E_{\nu} \approx \frac{1}{20} E_{p}. 
\end{equation}
\noindent For the $\sim$10 TeV-10 PeV neutrinos observed by IceCube 
the parent protons will have
had characteristic energies $E_p \sim 10^{14} - 10^{17}$~eV. 

We convert the observed photon spectral energy flux per decade of frequency
of the source, $\nu_{\gamma}
F_{\nu_{\gamma}}$, to the cosmic rest-frame luminosity per decade of photon energy,
$\vareps_{\gamma} L_{\vareps_{\gamma}}$, using
\begin{equation}
\nu_{\gamma} F_{\nu_{\gamma}} (\nu_{\gamma})
=\frac{\vareps_{\gamma} L_{\vareps_{\gamma}}}{4\pi d_L^2},
\end{equation}
\noindent where, $\vareps_{{\gamma}}$, is the energy of the photon in the
cosmic rest frame, and $d_L$ the luminosity distance. 
The comoving luminosity of the blob is calculated using $\Gamma^4
L_{\varepsilon_{\gamma}}^{\prime}\varepsilon_{\gamma}^{\prime} =
\vareps_{\gamma} L_{\vareps_{\gamma}}$. Here, and in what
follows we denote comoving frame quantities as primed, observer-frame
quantities are unscripted, and quantities in the cosmic rest frame are
scripted.

We assume that the neutrino production region is a relativistic blob
in the blazar jet, spherical in the comoving frame. The comoving
target photon density $n'_{\varepsilon_{\gamma}}$ in the jet zone,
where neutrino production is expected to be taking place is given by

\begin{equation}
n^{\prime}_{\varepsilon_{\gamma}} = \frac{3
  L'_{\varepsilon_{\gamma}}}{4 \pi r^{\prime 2}_{b} c
  \varepsilon'_{\gamma}}.
\label{eq:comoving_photon_density}
\end{equation}

\noindent Here $r_b^{\prime}$ is the comoving radius of the neutrino
emission region, which we infer from the (comoving) variability
timescale $t^{\prime}_{\rm var}$. We calculate $t^{\prime}_{\rm var}$
as $t^{\prime}_{\rm var} = \Gamma \cdot t_{\rm Var}/(1+z)$, with
$t_{\rm Var}$ the observed variability timescale, and
$r^{\prime}_{b}$,
\begin{equation}
r^{\prime}_{b} \sim c \cdot t^{\prime}_{\rm Var}.
\end{equation}
\noindent We use the shortest available variability timescale for each of the
studied flares. If the true variability of the system is smaller than
can be inferred from observations due to low statistics, and
$r^{\prime}_{b}$ is smaller than in our calculations, our predicted
neutrino fluxes in the SSC model are conservative, as the number
density of target photons, and therefore the optical depth to
photo-meson production is underestimated, as shown by Equation
\ref{eq:comoving_photon_density} (the situation is however different
in the external-Compton model in our setup, where typically the
external-Compton neutrino component is the dominant one. In the latter
case, a larger $r^{\prime}_{b}$ implies a larger maximum proton energy
and this affects the resulting neutrino spectrum).

\noindent The maximum proton energy for each flare is determined by
comparing the acceleration timescale to the timescales of dominant
cooling processes $t_{\rm cool}$,
\begin{equation} 
t_{\rm acc}^{'-1} > t_{\rm cool}^{-1} \equiv t_{\rm dyn}^{'-1} + t_{p,
  {\rm syn}}^{'-1} + t_{p \gamma}^{'-1}+t_{\rm BH}^{'-1},
\end{equation}
where the synchrotron cooling time of species $i$ is given by
$t'_{i,{\rm syn}} = 6 \pi m_i^4c^3/(m_e^2 \sigma_T B'^2
\varepsilon_i)$, and $t'_{\rm dyn} = r'_{b}/c$ is the dynamical
timescale of the system, which sets the timescale for adiabatic
cooling of the emitting region. 

Finally, $t'_{\rm acc} = \eta \varepsilon'_p/(ceB')$, 
is the proton acceleration timescale, for
protons assumed to undergo second order Fermi acceleration. The
acceleration efficiency is parametrised by $\eta \geq 1$, with $\eta =
1$ corresponding to the fastest possible acceleration time, when
diffusion proceeds in the Bohm limit. For a typical BL Lac emitting
region in our model with $B' = 1$~G, $r'_b \sim 10^{17}$~cm, and
$\Gamma = 10$, this leads to proton maximum energy, $\varepsilon'_{p} \approx 1$~EeV, when $\eta = 1$, subject to details of the photon density of the source. 

We also consider a higher value of $\eta$, namely $\eta = 10^{4}$. A large value of $\eta$ in the blazar emitting region was inferred
 by~\citet{Inoue:1996vv,Inoue:2016fwn} and a natural interpretation of these results was given by~\citet{Dermer:2014vaa} in the context of stochastic acceleration.
 Physically, a large value of $\eta$ corresponds to a low level of turbulence or otherwise inefficient scattering of the cosmic rays in the acceleration zone,
 leading to inefficient acceleration. The case in which $\eta$ is large is further motivated by the results of leptonic models, 
 where the maximum proton energy is shown to be $\lesssim 100$~PeV to provide a good fit to the source SED
\citep{Dimitrakoudis:2013tpa}\footnote{Referred to as {\it leptohadronic} therein.}, and by
the observation that for the 2017
flare of TXS 0506+056 it is impossible to have UHECR acceleration and
efficient sub-PeV neutrino production simultaneously. Namely, a maximum proton
energy higher than $\varepsilon^{\prime}_p \sim$~EeV would produce a
neutrino spectrum that peaks beyond 50~PeV and the sub-PeV neutrino
flux would be highly suppressed~\citep{Keivani:2018rnh,Gao:2018mnu}.

The photomeson energy-loss timescale is estimated as,
\begin{equation} 
t_{p\gamma}^{'-1} = \frac{c}{2 \gamma_p^2}
\int^{\infty}_{\bar{\varepsilon_{\rm th}}} {\rm d}
\bar{\varepsilon'_{\gamma}} \sigma_{p \gamma} \kappa_{p \gamma}
\bar{\varepsilon'_{\gamma}}
\int^{\infty}_{\bar{\varepsilon'_{\gamma}}/(2\gamma_p)} {\rm d}
\varepsilon'_{t} \varepsilon^{'-2}_{t} 
  n'_{\varepsilon_{t}}.
\label{eq:pgammaRate}
\end{equation}

\noindent Here, $\bar{\varepsilon}_{\rm th} \sim 145$~MeV is the threshold
energy for pion production in the rest frame of the proton, and
$\sigma_{p \gamma}$ and $\kappa_{p \gamma}$ are the cross-section and
inelasticity of photon-meson interactions, respectively. We adopt the
parametrisations used by~\citet{Murase:2005hy} for $\sigma_{p \gamma}$ and
$\kappa_{p \gamma}$. The quantity
$n'_{\varepsilon_{t}}$ is the target photon density
given in Equation~\ref{eq:comoving_photon_density}, differential in
energy. We calculate the fraction of energy converted to pions, as
$f_{p \gamma} \equiv t'_{\rm cool} / t'_{\rm p\gamma}$.

Our treatment does not include Bethe-Heitler photo-pair production. 
Due to the low value of the effective inelasticity $\kappa_{\rm BH}\sim m_e/m_p$ in comparison to that 
of photo-pion interactions with $\kappa_{p\gamma}\gtrsim 0.2$, the relative energy-loss rate for protons interacting 
with photons at the peak of the $\nu F_{\nu}$ spectrum is, 
\be
\frac{\kappa_{p\gamma}\sigma_{p\gamma}}{\kappa_{\rm BH}\sigma_{\rm BH}} \sim 100,
\ee
where $\sigma_{\rm BH}$ is the Bethe-Heitler photo-pair cross-section. 
The relative important between Bethe-Heitler and photo-pion interactions depend on photon spectra~\cite{Murase:2018iyl}. 
The Bethe-Heitler photo-pair production is subdominant as a proton cooling process and can be neglected when the target 
photon spectrum is sufficiently hard~(see e.g. Fig. 2 of~\citealp{petro_2015} and Equation~10 of \citealp{Murase:2018iyl}). 
On the other hand, it can produce a radiative signature on the blazar SED, for example, when the proton content of the jet becomes extremely large as shown in \citet{petro_2015,Keivani:2018rnh}. 
We do consider this effect in later sections.

We assume a power-law proton spectrum, ${\rm d} N'_p / {\rm d}{\vareps}'_p
\propto {\vareps}_p^{\prime -2}$, with an exponential cut-off beyond the
maximum attainable proton energy, determined by comparing the
acceleration timescale to the cooling time at each proton energy. We
consider the existence of protons with Lorentz factor as low as
$\gamma_{p,\rm min} = 1$.

We normalise the total proton luminosity of the source,
$\mathcal{L}^{\prime}_{\rm p}$, to the total (isotropic equivalent) comoving
photon luminosity $\mathcal{L}^{\prime}_{\rm \gamma}$, assuming
$\mathcal{L}^{\prime}_{\rm p} = \xi_{\rm cr} \mathcal{L}^{\prime}_{\rm \gamma}$, with $\xi_{\rm cr}$ the
baryon loading factor.

We consider two possible scenarios for the physical parameters
in the sources, detailed below:

\begin{itemize}
\item {\bf Normalisation A (UHECR / conservative):} Here, the baryon loading
  is relatively low, $\xi_{\rm cr}$ = 10, and the maximum proton energy is high
  (the acceleration efficiency is high, with $\eta$ = 1). This roughly
  corresponds to the average baryon loading needed for blazars to
  power the locally observed UHECR production rate of $E_{\rm UHE}
  Q_{E_{\rm UHE}} \sim 10^{44}~{\rm erg}~{\rm s}^{-1}~{\rm
    yr}^{-1}$ at $ \gtrsim 5 \times 10^{18}$~eV~\citep{Waxman:1995dg,
    Berezinsky:2002nc, Murase:2008sa, Katz:2013ooa, Aab:2016zth}, as calculated
  in~\citet{2014PhRvD..90b3007M}. The neutrino spectra obtained with
  this model typically peak beyond 10~PeV, which is typical in
  canonical blazar neutrino models. In general, in models that link
  the neutrino flux to the UHECR flux it is typically expected that
  $\xi_{\rm cr} \lesssim100$, sensitive to the power-law index of the injected
  proton spectrum~\citep{2014PhRvD..90b3007M}.

\item {\bf Normalisation B (\TXS~/ optimistic):} This is inspired by the most
  optimistic model of~\citet{Keivani:2018rnh}, LMPL2b, for the
  September 2017 flare of TXS 0506+056 (see also their Table
  7) if IceCube-170922A was indeed produced by \TXS. 
  Assuming this model,~\citet{Keivani:2018rnh} estimated the
  number of neutrinos expected from \TXS during the $\sim$180-day flare in
  2017, below 10~PeV, $N_{\nu_{\mu} <
    10~\mathrm{PeV}}/\mathrm{180~days} = 10^{-2}$ in the EHE alert
  channel. The baryon loading factor is high, $\xi_{\rm cr} = 1540$. The
  maximum proton energy is low, with $\eta = 10^{4}$, and as a result the
  neutrino spectrum typically peaks at lower, $\sim$~sub-PeV, energies
  than with Normalisation A.

In general, the baryon loading of the jet cannot much exceed the value 
considered in Normalisation B as the interactions of the protons would produce
not only neutrinos but also secondary \gRays and electron-positron pairs from the
interactions of the protons inside the source. 
These high-energy secondary leptons when produced in the 
interactions of protons further interact redistributing power from high to lower
energies, typically in the keV and MeV energy range. Too 
large a proton content then means that these electromagnetic cascade products would
exceed the simultaneous SED in the keV and MeV energy range, as shown in
e.g.~\citet{petro_2015,Murase:2018iyl,Reimer:2018vvw,Rodrigues:2018tku}. 

For some sources in our sample, Normalisation B contradicts the SED, in the
sense that such a high proton luminosity in the source would produce 
electromagnetic cascade emission from the electrons, positrons (and photons) 
produced by the Bethe-Heitler photopair (photopion) 
interaction of the protons, that would 
exceed the observed SED in the keV energy range in single-zone, standard emission scenarios. 
In what follows we have chosen to fix the
baryon loading factor and maximum proton energy across all sources to
facilitate an intuitive comparison between the sources. Where the cascade
emission overshoots the SED within the model assumptions of Normalisation B, we lower the baryon loading to 
an appropriate level to not exceed the cascade bounds and note the different treatment in the results that follow. 
\end{itemize}

\begin{figure}
\includegraphics[width=9cm,clip,rviewport=0 0 1 1]{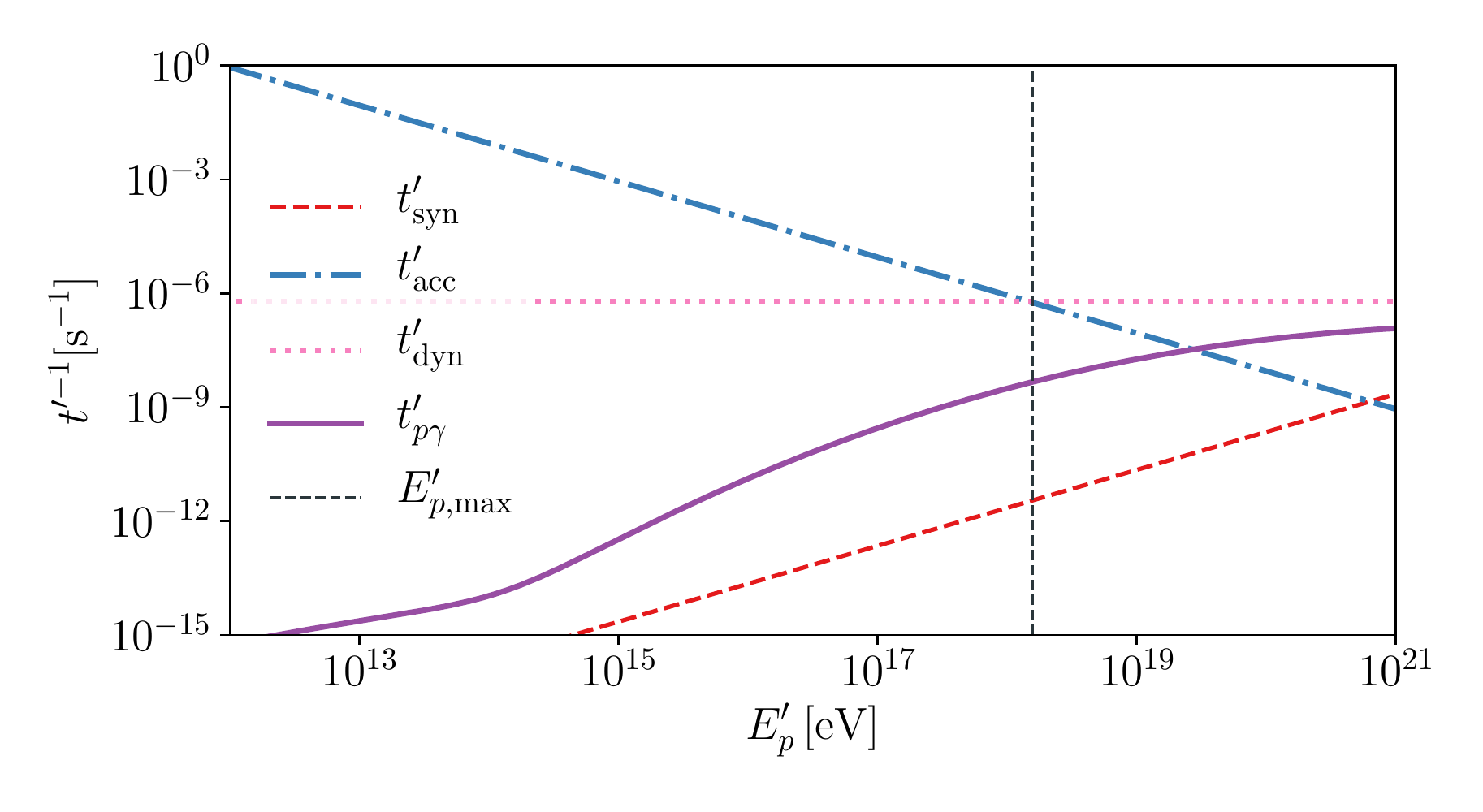}
\caption{Cooling timescales relevant for neutrino production during
  the January 2015 flare of \Sfive in the SSC only scenario (Normalisation
  A). The definition of the cooling timescales is given in Section~\ref{subsec:neutrino_production}. \label{fig:timescales}}
\end{figure}

A summary of the models considered, and relevant model parameters
is given in Table \ref{tab:models}. Figure \ref{fig:timescales} shows
the relevant cooling timescales for the January 2015 flare of \Sfive
in the SSC only scenario (Normalisation A). The maximum
proton energy, in this case, is set by the escape timescale,
$t^{\prime}_{\rm dyn}$, and other cooling processes, including
synchrotron cooling of protons are subdominant. This is typically the
case in the blazar blob in the leptonic models investigated
here. Using Normalisation B, the main differences are that the $p\gamma$
interaction timescale, $t^{\prime}_{\rm p \gamma}$ is significantly
enhanced, due to the presence of an additional target photon field and
the maximum proton energy is low because the acceleration efficiency
is lower and therefore $t^{\prime}_{\rm acc}$ is much larger.

\begin{table}
\caption{Table of parameters for the fiducial models tested in this
  work, namely the baryon loading factor $\xi_{\rm cr}$, the efficiency of
  particle acceleration $\eta$, the minimum proton Lorentz factor,
  $\gamma_{p, \rm min}$, and the energy density of the external
  radiation field in the comoving frame $u'_{\rm ext}$.
\label{tab:models}}
\begin{center}
\begin{tabular} {c c c c} 
  Normalisation & A (UHECR) & B (LMPL2b) \\
  \hline $\xi_{\rm cr}$ & 10 & 1540 \\
  $\eta$ & 1 & $10^{4}$ \\
  $\gamma_{p, \rm min}$ & 1 & 1 \\
$u'_{\rm ext}$ & N/A & \begin{tabular}{c} varies by source \\ see Table~\ref{tab:uext} \end{tabular} \\
\end{tabular}
\end{center}
\end{table}

The muon neutrino spectrum produced by pion decay following $p\gamma$ interactions is calculated as,
\begin{equation}
\varepsilon_{\nu_{\mu}}^{\prime 2} \frac{{\rm d}N'_{\nu_{\mu}}}{{\rm d}\varepsilon^{\prime}_{\nu_{\mu}}} \approx \frac{1}{8} \cdot f'_{p\gamma} \cdot f'_{\pi, {\rm cool}}  \cdot\varepsilon_p^{\prime 2} \frac{{\rm d}N'_p}{{\rm d}\varepsilon'_p},
\label{eq:NuMu}
\end{equation}
for the muon neutrinos produced by the decay of the pion, and, 
\begin{equation}
\varepsilon_{\nu_{\mu}}^{\prime 2} \frac{{\rm d}N'_{\nu_{\mu}}}{{\rm d}\varepsilon^{\prime}_{\nu_{\mu}}} \approx \frac{1}{8} \cdot f'_{p\gamma} \cdot f'_{\pi, {\rm cool}} \cdot f'_{\mu, {\rm cool}} \cdot\varepsilon_p^{\prime 2} \frac{{\rm d}N'_p}{{\rm d}\varepsilon'_p},
\label{eq:NuMu2}
\end{equation}
and 
\begin{equation}
\varepsilon_{\nu_{e}}^{\prime 2} \frac{{\rm d}N'_{\nu_{e}}}{{\rm d}\varepsilon^{\prime}_{\nu_{e}}} \approx \frac{1}{8} \cdot f'_{p\gamma} \cdot f'_{\pi, {\rm cool}} \cdot f'_{\mu, {\rm cool}} \cdot \varepsilon_p^{\prime 2} \frac{{\rm d}N'_p}{{\rm d}\varepsilon'_p},
\end{equation}
for muon and electron neutrinos produced by the subsequent decay of the muon respectively. In the above, $f_{\pi, {\rm cool}} = 1 - \exp{(-t_{\pi,{\rm cool}}/t_{\pi,{\rm dec}})}$, and equivalently $f_{\mu, {\rm cool}}$, are suppression factors due to the cooling of pions and muons respectively. The decay time of the pions is, $t_{\pi,{\rm dec}} \approx \gamma_{\pi}\tau_{\pi}$, with $\gamma_{\pi} = \varepsilon_{\pi}/(m_{\pi} c^2)$, and $\tau_{\pi} = 2.6 \times 10^{-8}$~s the mean rest frame lifetime of the pions. The cooling time for pions is obtained from $t^{-1}_{\pi,{\rm cool}} = t_{\pi,syn}^{-1} + t^{-1}_{\rm dyn}$. In the same way we obtain the cooling of the muons, with $\tau_{\mu} \approx 2.2 \times 10^{-6}$~s. 

Neutrino oscillation over large distances modifies the initial flavour
distribution of neutrino fluxes. The measurable muon neutrino flux on
Earth is given by,
\begin{equation}
\varepsilon_{\nu_{\mu},{\rm Earth}}^{\prime 2} \frac{{\rm
    d}N'_{\nu_{\mu},{\rm Earth}}}{{\rm
    d}\varepsilon^{\prime}_{\nu_{\mu},{\rm Earth}}} = P_{\nu_{\mu}
  \rightarrow \nu_{\mu}} \cdot \varepsilon_{\nu_{\mu}}^{\prime 2}
\frac{{\rm d}N'_{\nu_{\mu}}}{{\rm d}\varepsilon^{\prime}_{\nu_{\mu}}}
+ P_{\nu_{e} \rightarrow \nu_{\mu}} \cdot
\varepsilon_{\nu_{e}}^{\prime 2} \frac{{\rm d}N'_{\nu_{e}}}{{\rm
    d}\varepsilon^{\prime}_{\nu_{e}}},
\end{equation}

\noindent where $P_{\nu_{i} \rightarrow \nu_{j}}$ are mixing
probabilities, and $P_{\nu_{i} \rightarrow \nu_{i}}$ unitarity
relations, the relevant ones for our calculation being $P_{\nu_e
  \rightarrow \nu_{\mu}} \approx 0.24$, $P_{\nu_{\mu} \rightarrow
  \nu_{\mu}} \approx 0.37$~\citep{Pakvasa:2007dc}.
  
As explained in
Section \ref{subsec:NeutrinoCounts}, we consider only ``up-going'' muon
neutrinos, that is muon neutrinos from the Northern sky entering the
detector from below, in this work, primarily due to the higher
exposure than e.g. HESE events.

\subsection{Neutrino production in interactions with external fields}

From the point of view of neutrino production, the SSC model is
conservative in the sense that the hadrons in the source can only
interact with co-accelerated photons in the relativistic blob. In the
comoving frame this radiation field is not boosted, and therefore the
energy density, $u'_{\rm syn}$, relevant for neutrino production, is
relatively low. In addition to neutrino production in the relativistic blob, we study
an optimistic scenario for the production of neutrinos, by considering
the existence of external photon fields.  

In BL~Lac objects, the broad-line region is either 
non-existent or too weak to be observed.  
But some of the sources we studied, even though 
classified as BL~Lacs are likely FSRQs in reality, and thus possess
 a strong broad-line region. In fact, for several of the sources in
 our sample, the broad-line region luminosity
has been measured (e.g.~\citealp{Sbarrato:2011ps,
Sbarrato:2014uxa,Padovani:2019xcv}). On the other hand, even when a powerful broad line region does exist,
it is not guaranteed that it will contribute to neutrino production 
because the emitting region of the jet may be at larger jet radii. 
This is for example, on average, the case in the study of~\citet{Costamante:2018anp}. 

Though true BL Lac objects are not observed to possess a broad-line region, 
the existence of a spine-layer, or spine-sheath model, has
been proposed as an explanation for the observation of strongly
superluminal motion in the TeV \gRay emitting region, and slower
motion on parsec scales measured with radio observations
\citep{Ghisellini:2004ec}. The sheath can also act as an efficient
target for protons to interact with and produce neutrinos
\citep{Tavecchio:2014xha,Tavecchio:2014eia,Ahnen:2018mvi}.
Additionally, BL Lac objects possess radiatively inefficient accretion disks, 
thought to produce a complex, relatively broadband photon spectrum which 
can also act as a target field for neutrino production~\citep{Righi:2018xjr}.

We consider a generic external photon field
field, inspired by the spine-sheath model, with a power-law spectrum, $n^{\prime}
_{\vareps_{t}} \propto \vareps_{t}^{\prime -\alpha}$, extending from 
$\vareps_{t_{\rm min}}^{\prime} = 50$~eV to $\vareps_{t_{\rm max}}^{\prime} =
5000$~eV in energy, and $\alpha = 2$ as in the LMPL2b model of~\citet{Keivani:2018rnh}. 
Though the exact shape of the emerging neutrino spectrum depends 
on the details of the target photon spectrum, 
the total neutrino production is proportional to the integral 
of the energy density of the external field. In this sense, the scenario we
explore here is analogous to a physical set-up where the external target field is
a portion of the broad-line region as may be the case for \TXS~\citep{Padovani:2019xcv}.

We only consider scenarios where the SSC mechanism is the dominant 
mechanism responsible for the observed SED, in
other words, $u'_{\rm syn} > u'_{\rm ext}$. The luminosity of the external field
 in the observer frame is calculated as,

\be
L_{\rm ext} =  4 \pi R_{\rm ext}^2 c u'_{\rm ext} 
/ \Gamma^2,
\ee

\noindent with $R_{\rm ext}$ the size of the system in its proper
frame. For definiteness we fix $R_{\rm ext} = 3 \times 10^{19}$~cm,
which would encompass the large scale jet, noting that the setup is 
equivalent to one where the external radiation field is
denser, but traversed over a shorter scale, as is the case if the
emitting region is situated inside the broad-line region from the
point of view of neutrino production and associated cascade/opacity
constraints. As a starting estimate we use $u'_{\rm ext} = u'_{\rm
  syn} / 2$. For all SEDs, we ensure, that $L_{\rm ext}$ is lower than
the superluminal blob emission at all energies, if necessary by
lowering $u'_{\rm ext}$ with respect to the starting estimate. The
final values of $u'_{\rm ext}$ are given in Table~\ref{tab:uext}. For \AO, 
\Stwo and \OJ the opacity for 1~TeV \gRays is greater than one. 
This is consistent with the SEDs of these sources whose \gRay spectra show a cutoff 
at lower energies.  An additional constraint on the energy density of any external radiation 
field in BL Lac objects comes from the relative strength and peak frequency of the two bumps 
of the SED~\citep{Tavecchio:2019nvg}. In the case of HSP objects the upper limit on $u'_{\rm ext}$ is particularly tight, 
and often stronger than the constraint imposed by requiring transparency to TeV photons. 
We have checked that our assumed $u'_{\rm ext}$ is consistent with this additional requirement. 

The presence of the stationary or quasi-stationary, meaning
slow-moving, external field enhances the neutrino production
efficiency, but at the same time the external photon field provides a
target for $\gamma \gamma$ interactions, enhancing the opacity of the
source to \gRays. The optical depth for \gRays with energy,
$\vareps_{\gamma}$, via the $\gamma + \gamma \rightarrow e^{+} +
e^{-}$ process, in interactions with the external photon field, is given approximately by,

\be \tau_{\gamma \gamma} (\vareps_{\gamma}) \simeq \eta_{\gamma
  \gamma}(\alpha) \sigma_T (R_{\rm ext}/\Gamma) u'_{\vareps'_t, \rm
  ext}|_{\vareps'_{\rm t}= m_e^2
  c^4/\vareps^{\prime}_{\gamma}} \ee

\noindent (see e.g. \citealp{1967MNRAS.135..345R,1992MNRAS.258P..41R,Murase:2015xka}). 
Here, $\eta_{\gamma \gamma} (\alpha)$ is an integration
constant. For a power-law spectrum with $1 < \alpha < 7$ to a very
good approximation $\eta_{\gamma \gamma} (\alpha) \approx
7/[6\alpha^{5/3}(1+\alpha)]$~\citep{Baring_2006}. We check that
the opacity to \gRays in our assumed external field is not too
high, and therefore consistent with the \gRay observations of
all the flare SEDs. The optical depth for \gRays with energy 100
GeV, and 1 TeV is also given in Table~\ref{tab:uext}.

\begin{table}
\caption{The energy density of the external radiation field in the
  comoving frame $u'_{\rm ext}$ and the photon field comoving with the
  blob $u'_{\rm syn}$, both in units of erg~cm$^{-3}$, and optical
  depth for \gRays with energy 100 GeV, and 1 TeV. 
\label{tab:uext}}
\begin{tabular}{c c c c c}
&  $u'_{\rm syn}$ 
& $u'_{\rm ext}$ 
& \begin{tabular}{c} $\tau_{\gamma \gamma}$ \\ (100 GeV) \end{tabular} 
& \begin{tabular}{c} $\tau_{\gamma \gamma}$\\  (1 TeV) \end{tabular} \\
\hline
\ThreeC & $10^{-4}$ &  $5 \cdot 10^{-5}$  &  $< 0.1$  &  $< 0.1$ \\
\AO  &  $6 \cdot 10^{-3}$ & $3 \cdot 10^{-3}$ &  0.3  &  3.0  \\
\mrkfour  &  $2 \cdot 10^{-4}$ &  $9 \cdot 10^{-5}$  &  $< 0.1$  &  0.1  \\
\PG  &  $6 \cdot 10^{-4}$  &  $3 \cdot 10^{-3}$  &  $< 0.1$  &  0.1  \\
1ES\,1959+650a &  $5 \cdot 10^{-5}$  &  $2 \cdot 10^{-5}$  &  $< 0.1$  &  $< 0.1$ \\
\mrkfive  &  $10^{-3}$ &  $3 \cdot 10^{-4}$  &  $< 0.1$  &  0.5  \\
\Sfive  &  $3 \cdot 10^{-3}$  &  $2 \cdot 10^{-3}$  &  0.1  &  0.8  \\
\Sfour  &  $2 \cdot 10^{-3}$  &  $9 \cdot 10^{-4}$  &  $< 0.1$  &  0.3  \\
\bllac  &  $2 \cdot 10^{-2}$ &  $4 \cdot 10^{-4}$  &  $< 0.1$  &  0.1  \\
\Stwo  &  $5 \cdot 10^{-3}$  &  $5 \cdot 10^{-3}$  &  0.2  &  1.7  \\
1ES\,1959+650b &  $3 \cdot 10^{-5}$  &  $10^{-5}$  &  $< 0.1$  &  $< 0.1$  \\
\OJ &  $6 \cdot 10^{-3}$  &  $2 \cdot 10^{-3}$  &  0.4  &  3.6  \\
\TXS &  $7 \cdot 10^{-4}$  &  $7 \cdot 10^{-4}$  &  $< 0.1$  &  0.2  \\
\hline
\end{tabular}
\end{table}

\subsection{Blazar SED fitting method}
\label{subsec:SEDFitting}
We fit the blazar spectral energy distributions with two log-parabolic functions of the form
\begin{equation}
\nu_{\gamma} F_{\nu_{\gamma}} = \nu_{\gamma, \rm pk} F_{, \nu_{\gamma,\rm pk}} \cdot 10^{b \cdot \left(\log{(\nu_{\gamma} / \nu_{\gamma, \rm pk})}\right)^2},
\label{eq:logParab}
\end{equation}

\noindent where $b$ is the shape parameter and
$\nu_{\gamma, \rm pk}$ is the peak frequency of the log-parabola. It was shown
in~\citet{2004A&A...413..489M} that such a log-parabolic function is
one of the simplest ways to represent curved spectra, while fitting
the blazar spectral energy distributions in different luminosity
states very well, and providing good estimates of the energy and flux
of the SED. The observed log-parabolic shape of the SED can be
explained as deriving from stochastic acceleration processes
\citep{2004A&A...413..489M,2007A&A...466..521T,2008ApJ...681.1725S,Dermer:2014vaa}
which produce a curved lepton spectrum from injection. 

\noindent We perform two independent fits for the two humps of the
SED. For the second hump of the SED we determine the best fit
parameters of the expression $\nu_{\gamma} F_{\nu_{\gamma}} \cdot \exp(-\tau_{\gamma \gamma}
(\varepsilon, z))$, where $\tau_{\gamma \gamma} (\varepsilon, z)$ is
the optical depth of the extragalactic background light for photons of
energy $\varepsilon$ originating at a source at redshift $z$. We use
the $\tau_{\gamma \gamma} (\varepsilon, z)$ calculated by
\citet{Franceschini:2008tp}. 
We perform a numerical $\chi^2$ minimisation~\citep{James:1994vla} to determine 
the parameters $b$ and $\nu_{\gamma, \rm pk}$ that provide the best fit for each SED.

\noindent The broad-band flux can be obtained by analytically integrating
Equation \ref{eq:logParab} over the entire frequency range, which
gives~\citep{2004A&A...413..489M},
\begin{equation}
F_{\rm bb} = \sqrt{\pi \ln 10} \frac{ \nu_{\gamma, \rm pk} F_{\nu_{\gamma, \rm pk}}}{\sqrt{b}}.
\label{eq:bb_corr}
\end{equation}

\noindent The simple log-parabola model does not fit the data well for
energy distributions that do not decrease symmetrically with respect to
the peak frequency. In this case, we fit the SEDs with a log-parabola
with an exponential cutoff of the form

\begin{equation}
\nu_{\gamma} F_{\nu_{\gamma}} = \nu_{\gamma, \rm pk} F_{, \nu_{\gamma,\rm pk}} \cdot 10^{\beta \cdot \left(\log{(\nu_{\gamma} / \nu_{\gamma, \rm pk})}\right)^{2}} \cdot e^{-\nu_{\gamma}/\nu_{\gamma, \rm cut}}.
\label{eq:logparabexp}
\end{equation}
We allow the optimal value of $\nu_{\rm cut}$, the extra-free
parameter, to be determined independently for the two peaks through
the minimisation procedure described above. The cutoffs in the two
peaks need not necessarily have the same origin, even in the SSC
model, which justifies our choice to fit them separately. The required
cutoff in the synchrotron peak is likely due to cooling of the
electrons, particularly for the conditions likely in HSPs, where the
synchrotron emission extends to higher energies. On the other hand, a
cutoff in the inverse Compton peak could be due to internal absorption
in the various photon fields in the source. This can occur inside the
emitting region, but also by interactions with external photon fields.

We can calculate the Doppler factor, $\delta$, and the magnetic-field strength, 
$B$, using Equations~\ref{eq:DopplerFactor} and~\ref{eq:B_SSC}, and
the broad-band luminosity for each hump of the SED as given by
Equation~\ref{eq:bb_corr}. However, when the cutoff is very close to the peak 
we cannot accurately determine $\delta$ and $B$ from the SED with this method. 
This is the case for most of the HSP blazars in our sample. 
We have therefore opted to use literature values for $\delta$ and $B$, 
obtained from the modelling of the specific flares we study, throughout most of this work. 
The assumed Doppler factor $\delta$, and magnetic field strength for
each of the studied sources are given in Table~\ref{tab:deltaB}.
In Section~\ref{subsec:systematics}, where we discuss the effect of systematic uncertainties 
on $\delta$ and $B$ on neutrino flux expectations, we also compare the Doppler factors 
obtained with our log-parabolic model using Equations~\ref{eq:DopplerFactor} and~\ref{eq:B_SSC} with those 
from Table~\ref{tab:deltaB}. 

We also state in Table~\ref{tab:deltaB} whether a simple SSC model can fit the SED of
each of the studied flares. In the cases where an external Compton
component is required to fit the SED (denoted SSC+EC) our Normalisation A is
conservative as interactions with additional external photon fields in the
source environment are disregarded.

\begin{table}
\caption{Table of the parameters for each of the flares studied in
  this work. The Doppler factor, $\delta$, and magnetic field strength
  $B$ in Gauss are taken from the literature, as stated in the
  references column. In addition, it is indicated whether the
  parameters were derived assuming only synchrotron-self Compton (SSC)
  emission or additional external-Compton emission (EC).\label{tab:deltaB}}
\begin{center}
\begin{tabular}{ccccccc}
\hline
Source  &$\delta$ & $B$ & Model & Ref.  \\  
\hline
3C 66A \rule{0pt}{3ex} & 40 & $0.02$ & SSC & 1                                        \\
AO 0235+164 	\rule{0pt}{3ex} & 20& $0.22$ & SSC+EC & 2                        \\
Mrk 421  	\rule{0pt}{3ex} 	&  $21$ & $0.04$  & SSC & 3                          \\
PG 1553+113\rule{0pt}{3ex} & $40$ & $0.045$ & SSC & 4                                  \\
1ES 1959+650a \rule{0pt}{3ex} & $25$ & $0.01$ & SSC & 5                                \\
Mrk 501 	\rule{0pt}{3ex}  & $10$ & $0.02$ & SSC & 6 \\
S5 0716+714 \rule{0pt}{4ex} & $25$ & $0.10$ & SSC & 7                  \\
S4 0954+65 \rule{0pt}{3ex} & $30$ & $0.6$  & SSC+EC  & 8,9                              \\
BL Lac  \rule{0pt}{3ex}  &  $25$ & $0.14$  & SSC+EC & 10                               \\
S2 0109+22   \rule{0pt}{3ex} & $22$ & $0.05$ & SSC & 11        \\
1ES 1959+650b \rule{0pt}{3ex}& $25$ & $0.01$ &SSC & 5                                       \\
OJ 287a \rule{0pt}{3ex}& $14$ & $0.9$ & SSC+EC & 12                                      \\
TXS 0506+056 \rule{0pt}{3ex} & $24$ & $0.4$ & SSC+EC & 13                                  \\
\hline
\end{tabular}
\vspace{0.2cm}
\\
{\footnotesize References: 
1~-~\citet{2011ApJ...726...43A}. 
2~-~\citet{2012ApJ...751..159A}. 
3~-~\citet{Aleksic:2014rca}.  
4~-~\citet{2015MNRAS.450.4399A}.
5~-~\citet{0004-637X-797-2-89}. 
6~-~\citet{2018A&A...620A.181A}.
7~-~\citet{2016Galax...4...69M}. 
8,9~-~\citet{2016PASJ...68...51T,2018arXiv180104138M}. 
10~-~\citet{Acciari:2019vpr}.
11~-~\citet{2018MNRAS.480..879M}.
12~-~\citet{2018MNRAS.479.1672K}.
13~-~\citet{Keivani:2018rnh}.}
\end{center}
\end{table}

\subsection{Expected neutrino event counts}
\label{subsec:NeutrinoCounts}
In this section, we outline the method we use to calculate neutrino
counts in IceCube and future neutrino telescopes for a given source
neutrino spectrum. We consider here only through-going muon neutrinos from
the Northern sky. The better arrival direction reconstruction, higher
statistics, and lower energy threshold make them more suitable for the
search for faint signals from individual blazar flares than other event selections. 
Approximately 100000 neutrinos from the Northern
sky are detected by IceCube each year~\citep{Aartsen:2018ywr}. Most of
them are atmospheric neutrinos that form an irreducible
background. However, the atmospheric neutrino spectrum is softer than
the spectrum expected by astrophysical sources, including
blazars. Hence, by constraining neutrino point-source searches to the highest-energies and
the time-bin to the time of interest (in our case the blazar flare
duration), the background rate is reduced significantly.

 \begin{figure}
 \includegraphics[width=9cm,clip,rviewport=0 0 1 1]{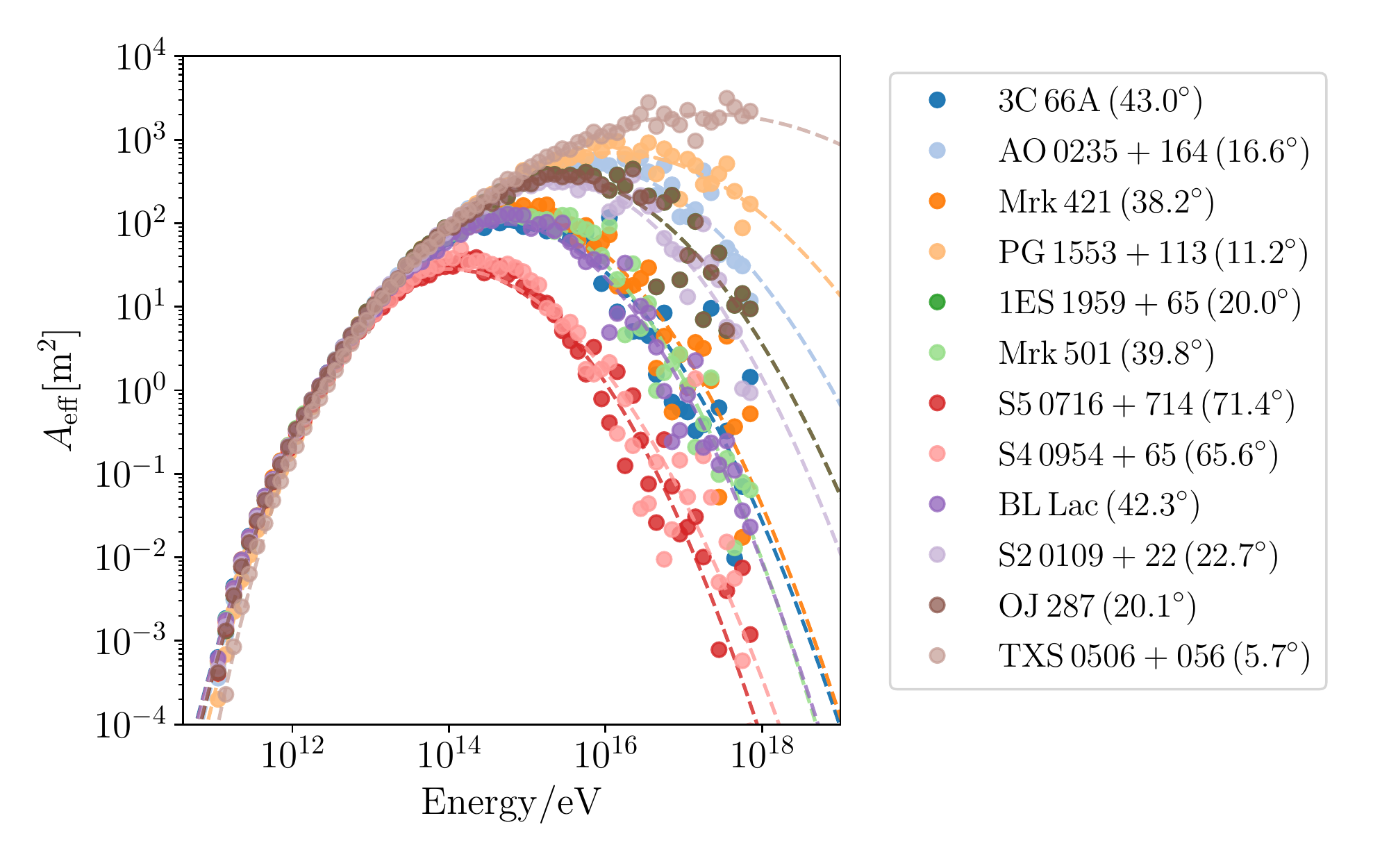} 
 \caption{Effective area as a function of energy in IceCube
   IC86 configuration at the declination of each of the sources in our
   study. The filled circles show the IceCube Monte Carlo points from
  ~\citet{Aartsen:2018ywr}, and the dashed lines the parametrisation
   used in this work. \label{fig:Aeff}}
\end{figure}

The instantaneous number of signal-only, {\it through-going}
muon-neutrino events, $\mathrm{d}N_{\nu,t_0}/\mathrm{d}t$, expected to
be detected by IceCube during a small time-interval $\mathrm{d}t$
during a neutrino flare is,
\begin{equation}
  \frac{\mathrm{d}N_{\nu_{\mu},t_0}}{\mathrm{d}t} =
  \int_{E_{\nu,{\rm min}}}^{E_{\nu,{\rm max}}} {\rm d} E_{\nu}  \frac{1}{3} A_{\rm eff}(E_{\nu},\theta) F_{E_{\nu}} ,
\label{eq:Nnu}
\end{equation}
where $E_{\nu,{\rm min}}$ and $E_{\nu,{\rm max}}$ are the minimum and
maximum energy considered respectively, and $F_{E_{\nu}}$, the all-flavour neutrino flux, differential in energy.

\noindent The effective area of the neutrino detector for a neutrino
arriving with zenith angle $\theta$ and energy $E_{\nu}$ is $A_{\rm
  eff}(E_{\nu},\theta)$. We use the full, zenith- and energy-
dependent simulated effective area of the IceCube point-source
analysis, $A_{\rm eff}(E_{\nu},\theta)$, given in
\citet{Aartsen:2018ywr}.\footnote{Available online at
  \url{https://icecube.wisc.edu/science/data}} For each period when
the IceCube detector was in incomplete configuration we use the
corresponding simulated $A_{\rm eff}(E_{\nu},\theta)$ for that
period. For years beyond 2011, we assume the IceCube configuration of
2011, IC86. We are interested in the effective area of IceCube out to,
and beyond $10^{18}$~eV. For this reason, we fit the IceCube effective
area beyond $10^{18}$~eV, which is not publicly available, with a log
parabolic function.

The left panel of Figure \ref{fig:Aeff} shows the effective area of
IceCube-IC86 as a function of energy at the declination of each of the
sources in our sample. Filled circles show the IceCube Monte Carlo
points from~\citet{Aartsen:2018ywr}, and the dashed lines the fitted
function used in this work.

We estimate the total number of neutrinos expected during the
flare by integrating over the time duration of each flare, $\Delta T$, as defined in Table \ref{tab:timescales} (based on the
analysis of the FAVA lightcurves shown in Figure \ref{fig:Lightcurves}
as described in Appendix~\ref{sec:FAVA}). The estimate of the expected
number of neutrinos takes into account the IceCube exposure during
each flare given the source's declination and time of occurrence. To
integrate over the duration of the flare we use the instantaneous SED
obtained at the specified time of multi-wavelength observations,
typically averaged to the nearest day, and use the relation,

\begin{table}
\caption{Time duration of each of the flares investigated, $\Delta T$, in days,
  based on the analysis of the high-energy FAVA data as detailed in
  Appendix~\ref{sec:FAVA}, or, in the absence of significant flux
  enhancement in the FAVA data, the flare timescale determined in a
  different waveband as stated.  The Table also gives the timescale of
  smallest detected time-variability in days, $t^{\rm obs}_{\rm var}$,
  which we use to determine the size of the emitting region and
  waveband at which it was detected. \label{tab:timescales}}
\begin{tabular}{ccccc}
\hline

Source  & $t^{\rm obs}_{\rm var}$ & Band & $\Delta T$ &  Band \\
\hline
3C 66A  &  1 & VERITAS & 14 & FAVA \\
AO 0235+164  & 3 &\begin{tabular}{c} \Fermi \\ and others \end{tabular} & 84 & FAVA \\
Mrk 421    & 1 & MAGIC & 13 & MAGIC \\
PG 1553+113  & 1 & \begin{tabular}{c} \Swift / \\ MAGIC \end{tabular}  & 30 &\begin{tabular}{c} \Swift \\ (XRT) \end{tabular}  \\
1ES 1959+650a  &   1 & VERITAS & 46 & Optical \\
Mrk 501   & 0.2  & MAGIC & 21 & FAVA \\
S5 0716+714  &1 &  MAGIC & 14 & FAVA \\
S4 0954+65  & 1& \begin{tabular}{c} \Fermi \\ and others \end{tabular}  & 28 & FAVA \\
BL Lac\rule{0pt}{0pt} & 0.1 & MAGIC & 7 & FAVA  \\
S2 0109+22\rule{0pt}{0pt}   & 1 & \begin{tabular}{c} \Fermi \\ and others \end{tabular}  & 21 & FAVA \\
1ES 1959+650b\rule{0pt}{0pt}&  1 & \begin{tabular}{c} \Fermi \\ and others \end{tabular}  & 84 & FAVA \\
OJ 287   \rule{0pt}{0pt} & 1 & \begin{tabular}{c} near-IR/UV \end{tabular}  & 7 & FAVA \\
TXS 0506+056  \rule{0pt}{0pt}  & 1 & \begin{tabular}{c} \Fermi \\ and others \end{tabular}  & 175 & FAVA \\
\hline
\end{tabular}
\end{table}
\begin{equation}
\frac{\mathrm{d}N_{\nu_{\mu},t}}{\mathrm{d}t}  = \frac{\mathrm{d}N_{\nu_{\mu},t_0}}{\mathrm{d}t} \cdot \frac{A_{\rm eff,t}}{A_{\rm eff,t_0}} \left(\frac{F_{\rm HE}}{F_{{\rm HE},t_0}}\right)^{\alpha},
\label{eq:NnuT}
\end{equation}
\noindent to estimate the instantaneous expected number of muon neutrinos, 
at all other times $t$. Here, $F_{\rm HE}$ is the FAVA flux in the high-energy bin at time
$t$, and $F_{{\rm HE},t_0}$ the high-energy-bin FAVA flux at the time when
the multi-wavelength SED was obtained. For $\alpha$ we assume $\alpha
= 2.0$, which is typically expected in BL Lac objects in both in the
SSC model, and in the presence of a sheath field, because in canonical
models the proton luminosity increases simultaneously with the density
of the target field during high-states
\citep{Tavecchio:2014xha,Petropoulou:2016ujj,2016PhRvD..94j3006M}. The
ratio $A_{\rm eff,t}/A_{\rm eff,0}$ gives the relative exposure at a
given time $t$, with respect to that at time $t_0$, to allow for the
possibility that the IceCube configuration changed during the time of
the flare.

We further calculate the total expected number of neutrinos during the
ten years that IceCube and \Fermi have been in operation, integrating
Equation~\ref{eq:NnuT} over the entire time-duration of the
\Fermi lightcurve.

In addition to IceCube, we consider the sensitivity of planned and
proposed future neutrino telescopes to the expected signal from blazar
neutrino flares. We consider the following
future facilities (see Table~\ref{tab:acronyms} for abbreviations) which could 
form a future neutrino monitoring network~\citep{resconi19}:
\begin{itemize}
\item IceCube Gen2~\citep{Aartsen:2014njl}. We assume that the
  extended IceCube detector will have an instrumented volume six times larger
  than IceCube in IC86 configuration.
\item KM3NeT~\citep{Bagley:2009wwa}. We assume that in its final
 configuration it will have an instrumented volume identical to that of IC86 but at latitude, $l = 32.27^{\circ}$N.
\item Baikal-GVD~\citep{2018arXiv180810353B}. We assume that in its
  final configuration it will have an instrumented volume identical to that of
  IC86, but at latitude, $l = 53.56^{\circ}$N.
\item ONC~\citep{Bedard:2018zml}. We assume that in its final configuration, it will have have an instrumented volume identical to that of IC86, at latitude, $l =
  48.43^{\circ}$N. 
\end{itemize}
\noindent For all four facilities, we further assume that the effective area as
a function of energy and zenith angle is identical to that of IC86. 

In order to calculate the number of neutrinos expected in the
different detectors from identical blazar flares using Equation~\ref{eq:Nnu}, 
we calculate the instantaneous zenith angle, $\theta$,
of a neutrino point source at declination $\delta$, at time $t$, in a
detector with acceptance uniform in right ascension, located at
latitude $l$, using the expression,
\begin{equation}
\cos \theta = \sin \delta \cdot \sin{l} + \cos \delta \cdot \cos{l} \cdot \sin{\left(2\pi t \right)},
\end{equation}
and obtain the time-averaged effective area as \be A_{\rm
  eff}(E_{\nu}, \delta,l) = \frac{1}{N} \sum_i^{N} A_{\rm eff}
(E_{\nu},\theta_i, l), \ee where $\theta_i$ is the zenith angle of a
source at declination $\delta$ as seen by a detector at latitude, $l$,
at time-interval $i$, and the average, $A_{\rm eff}(E_{\nu},
\delta,l)$ is obtained by summing over a large number, $N$, of
time-bins.

\section{Results}
\label{sec:results}
\begin{figure*}
\includegraphics[width=5.8cm,clip]{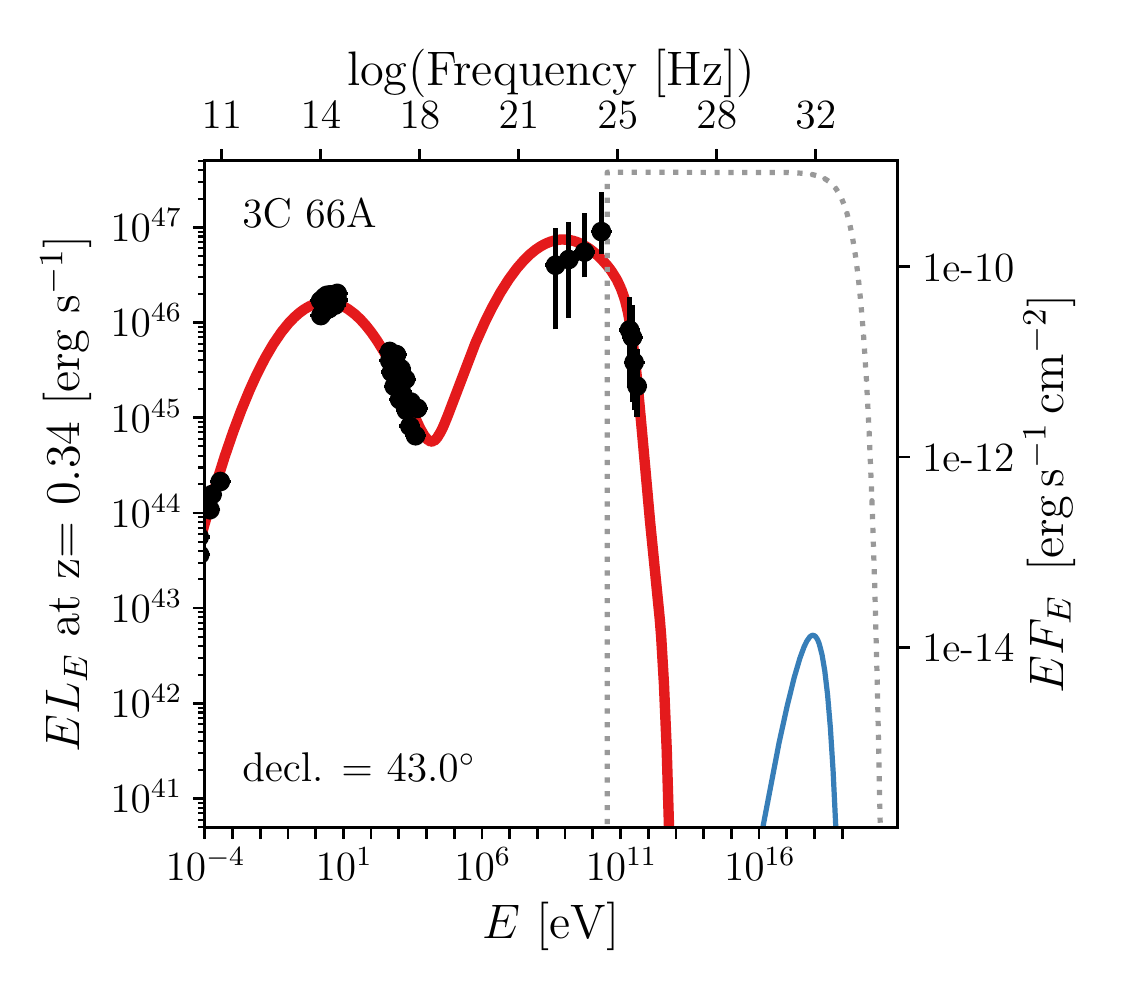}
\includegraphics[width=5.8cm,clip]{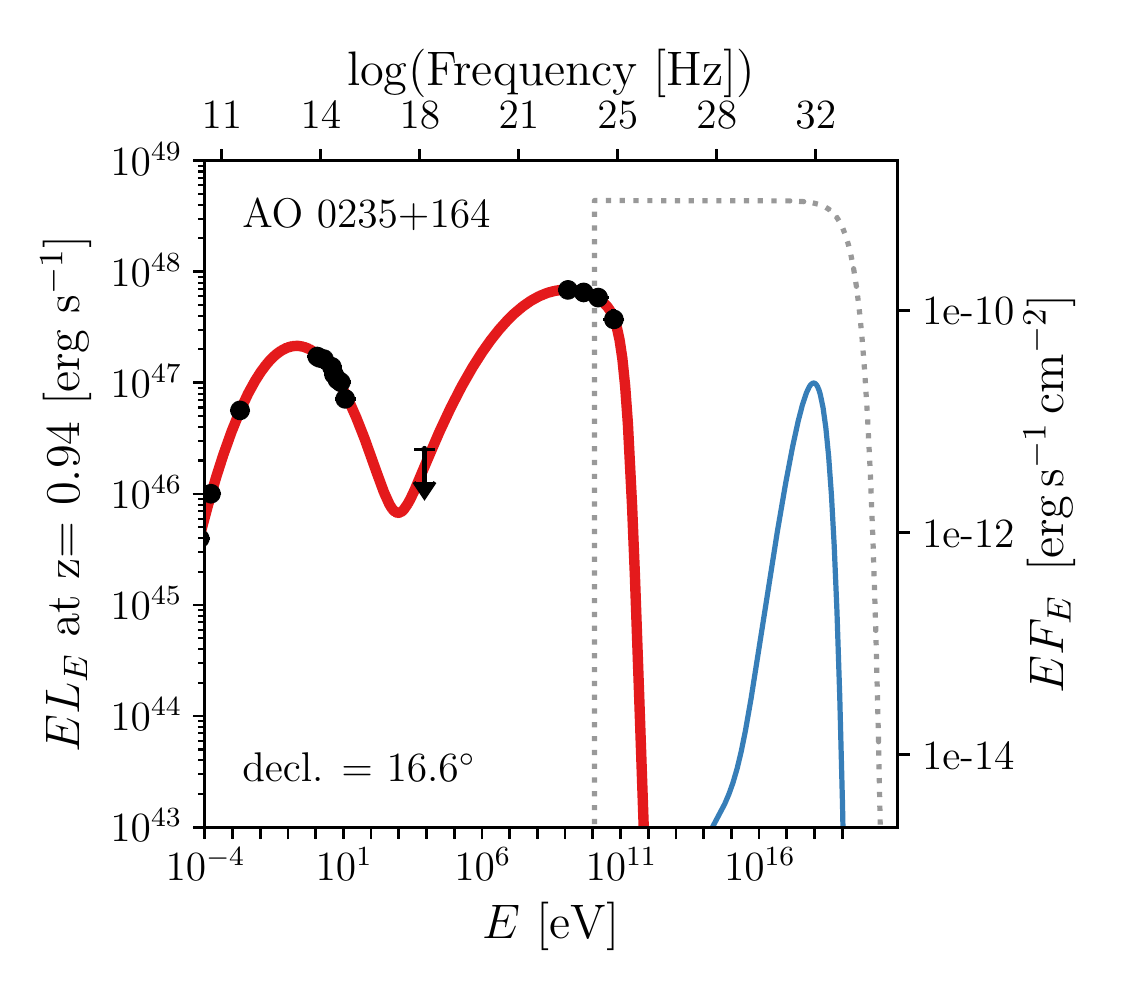}
\includegraphics[width=5.8cm,clip]{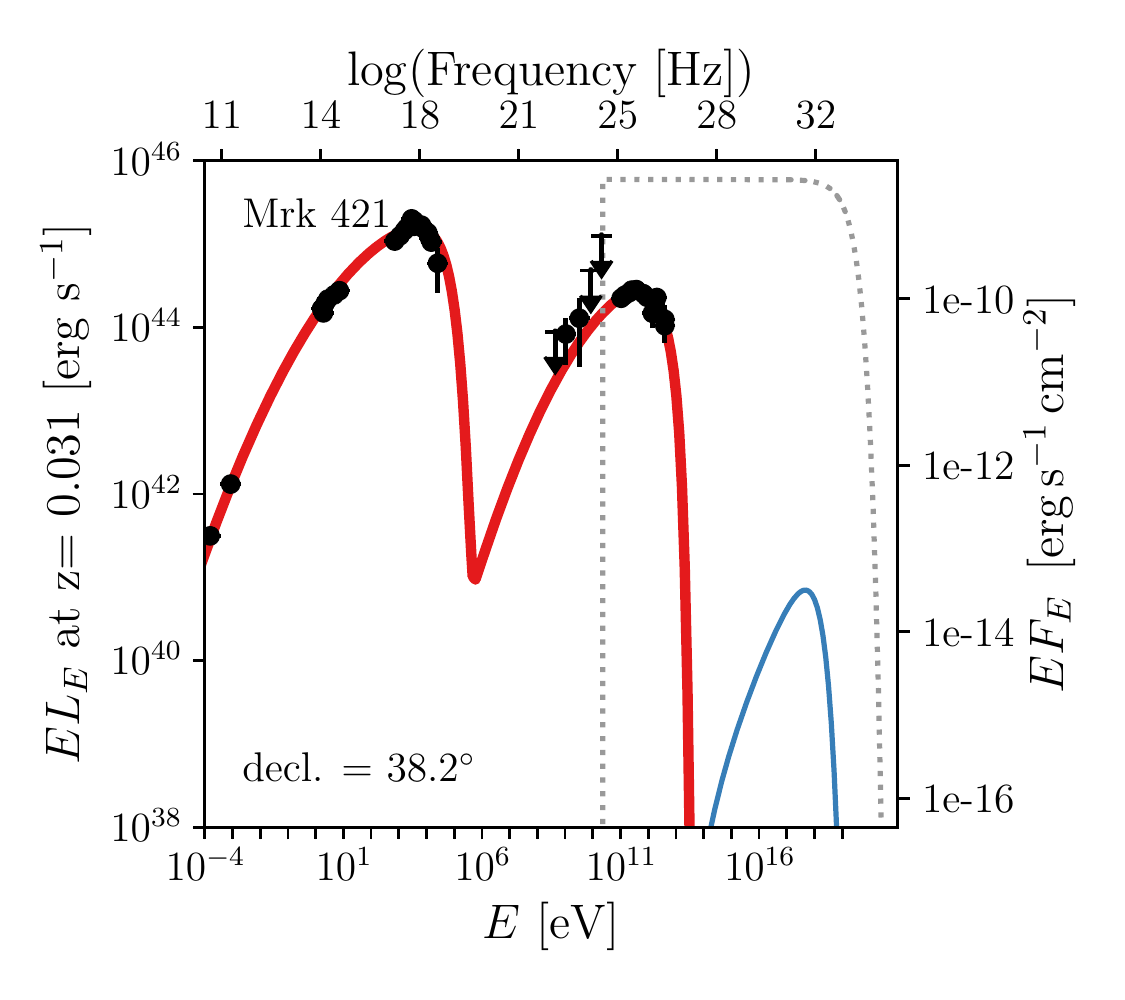}
\includegraphics[width=5.8cm,clip]{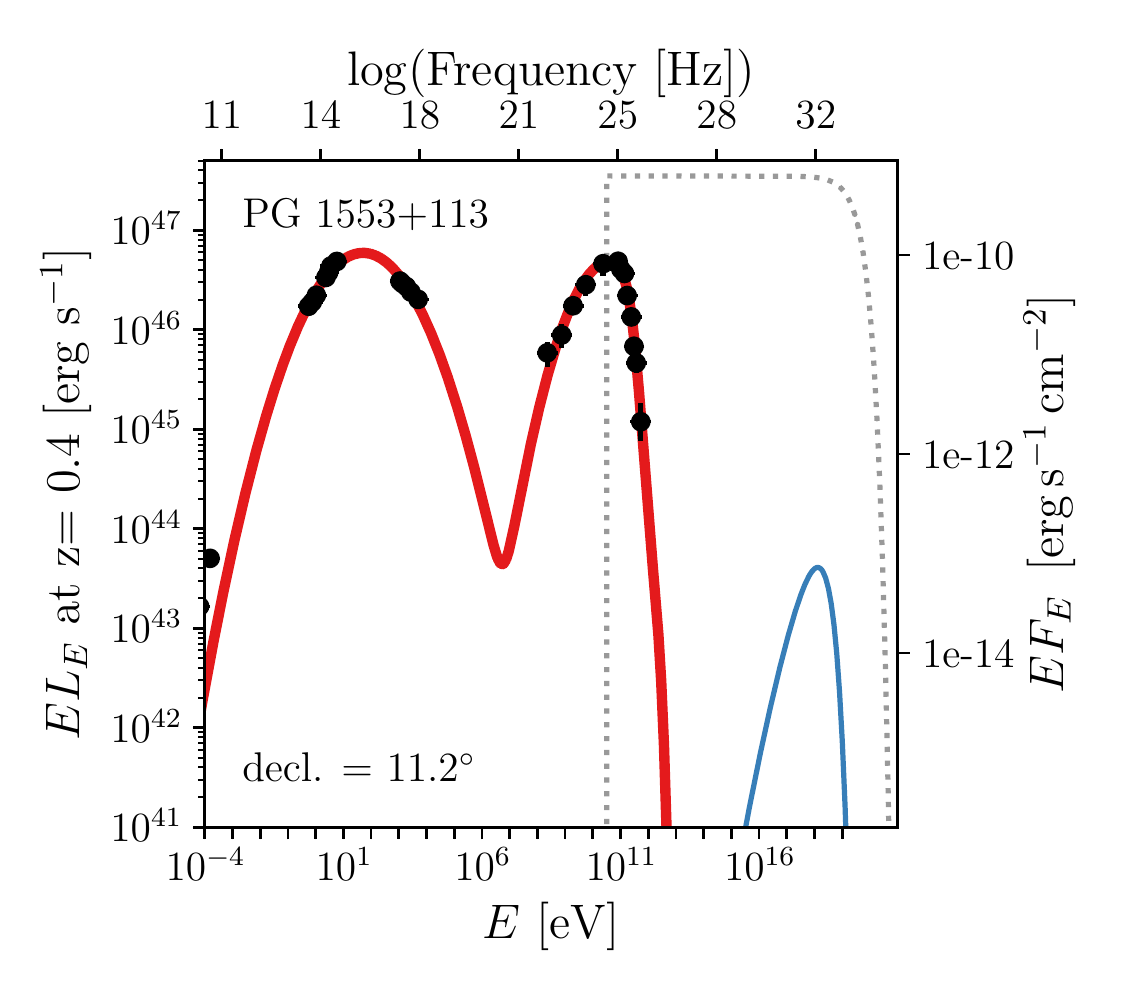}
\includegraphics[width=5.8cm,clip]{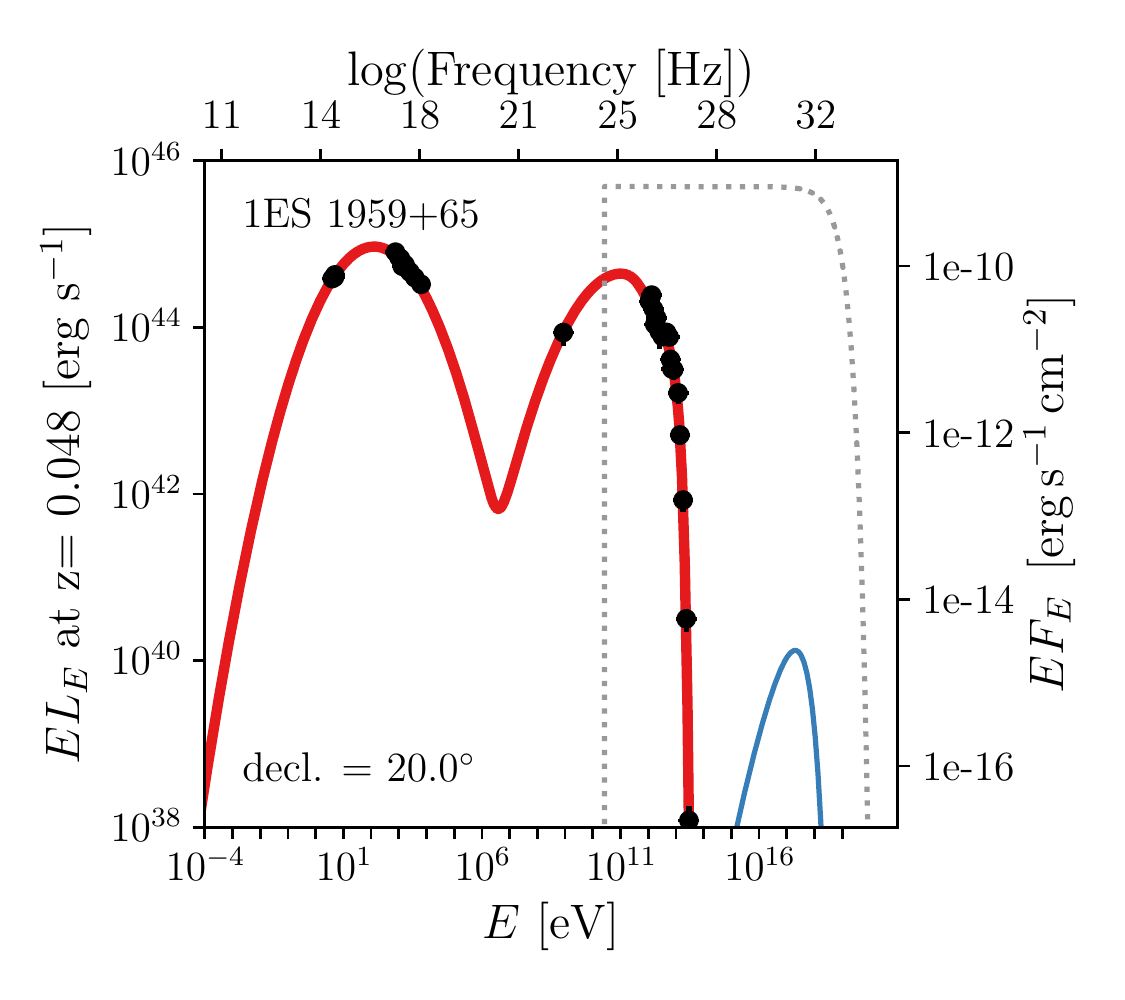}
\includegraphics[width=5.8cm,clip]{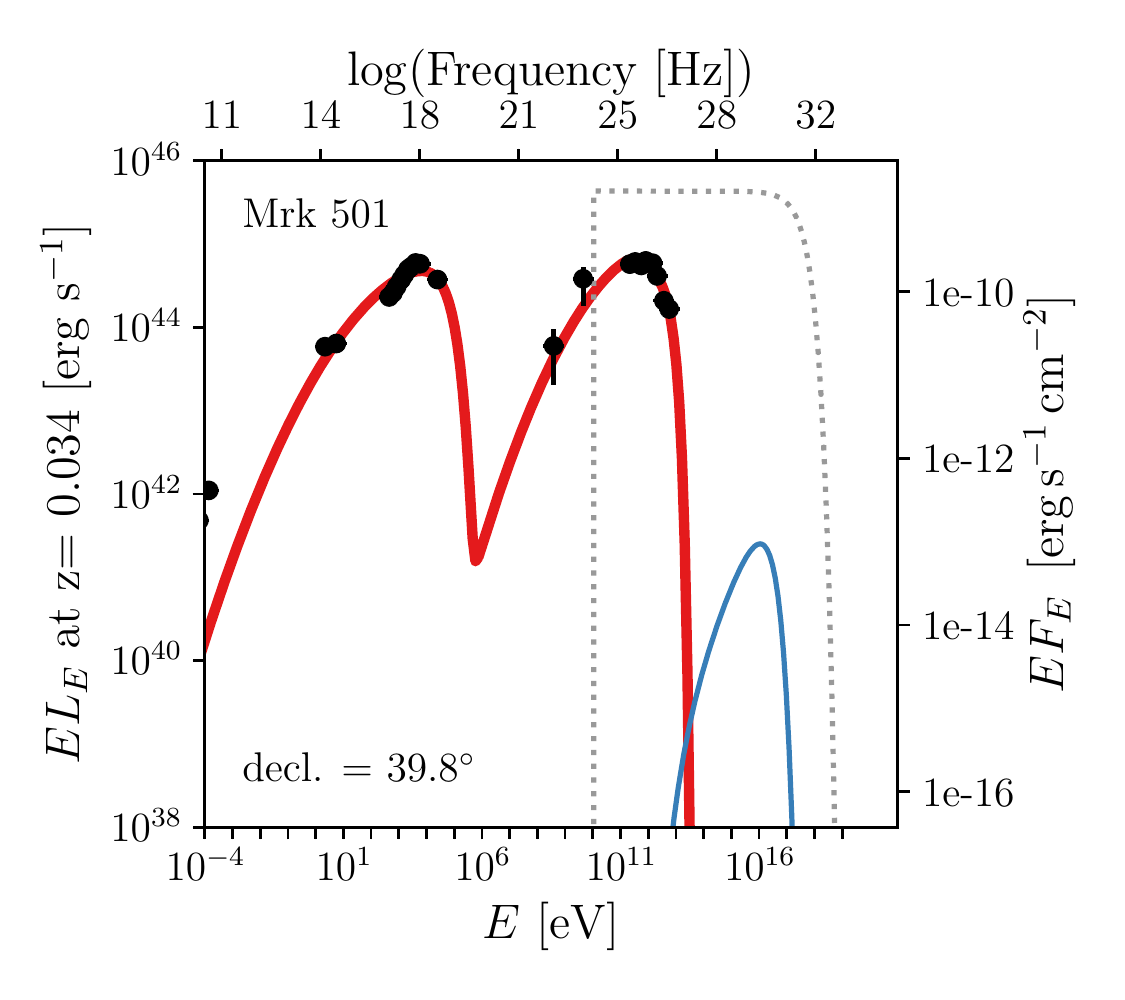}
\includegraphics[width=5.8cm,clip]{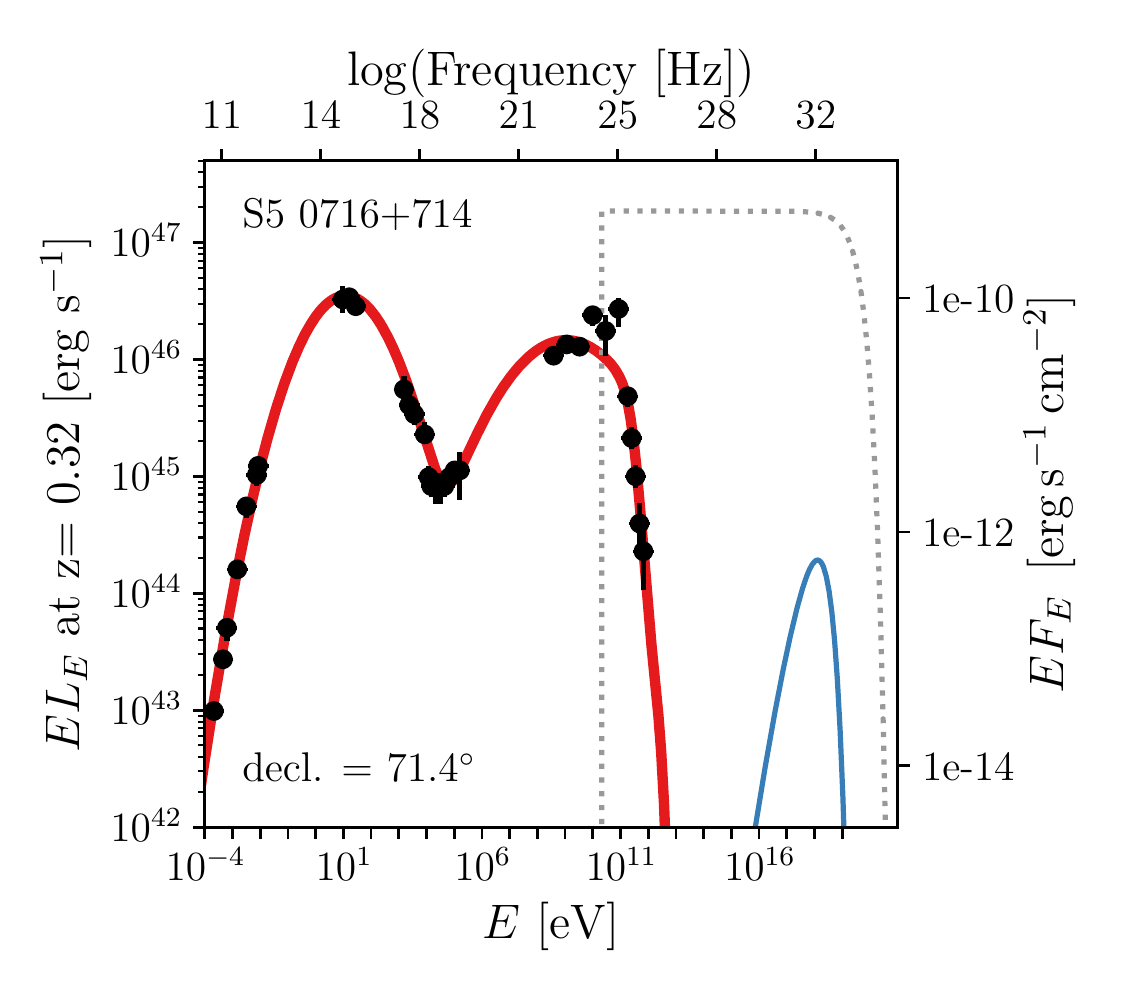}
\includegraphics[width=5.8cm,clip]{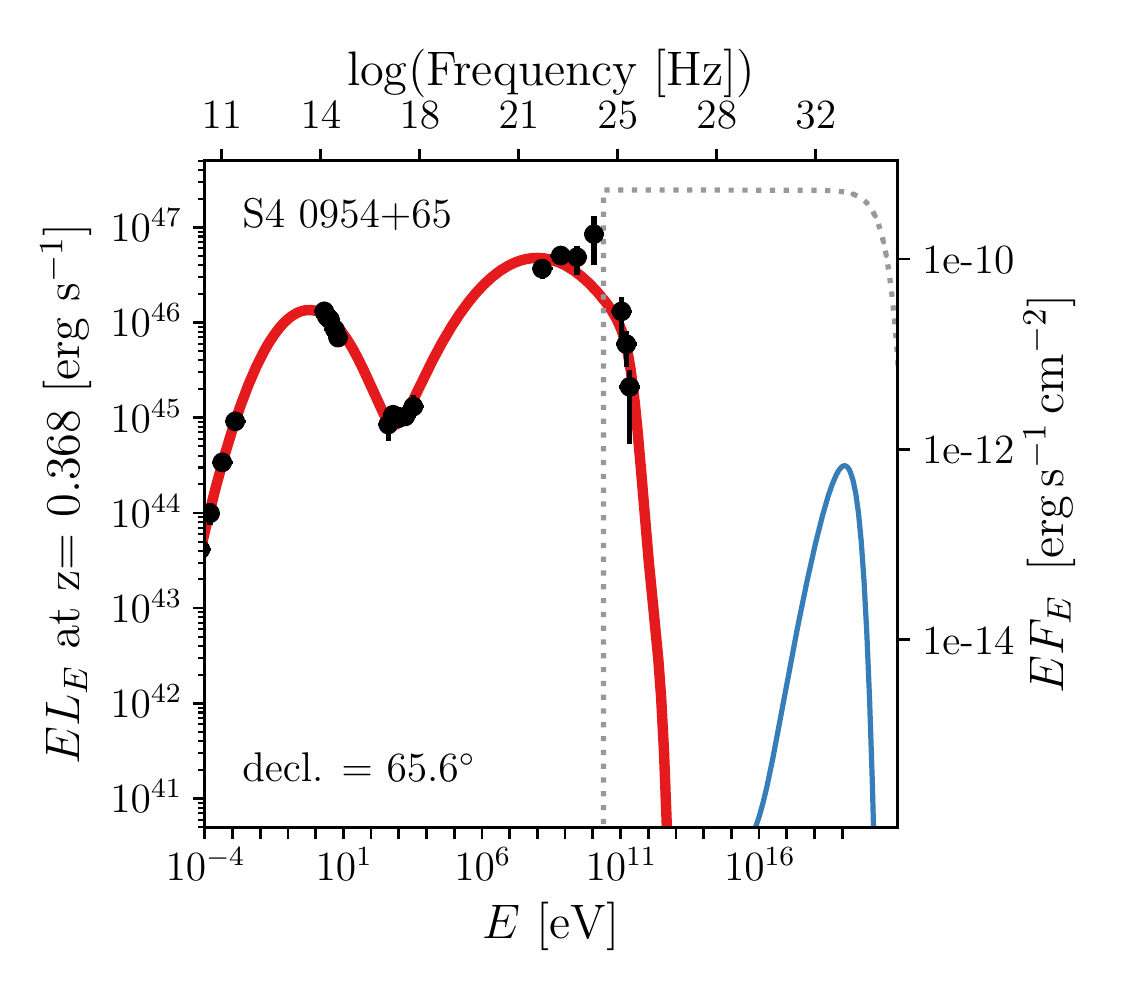}
\includegraphics[width=5.8cm,clip]{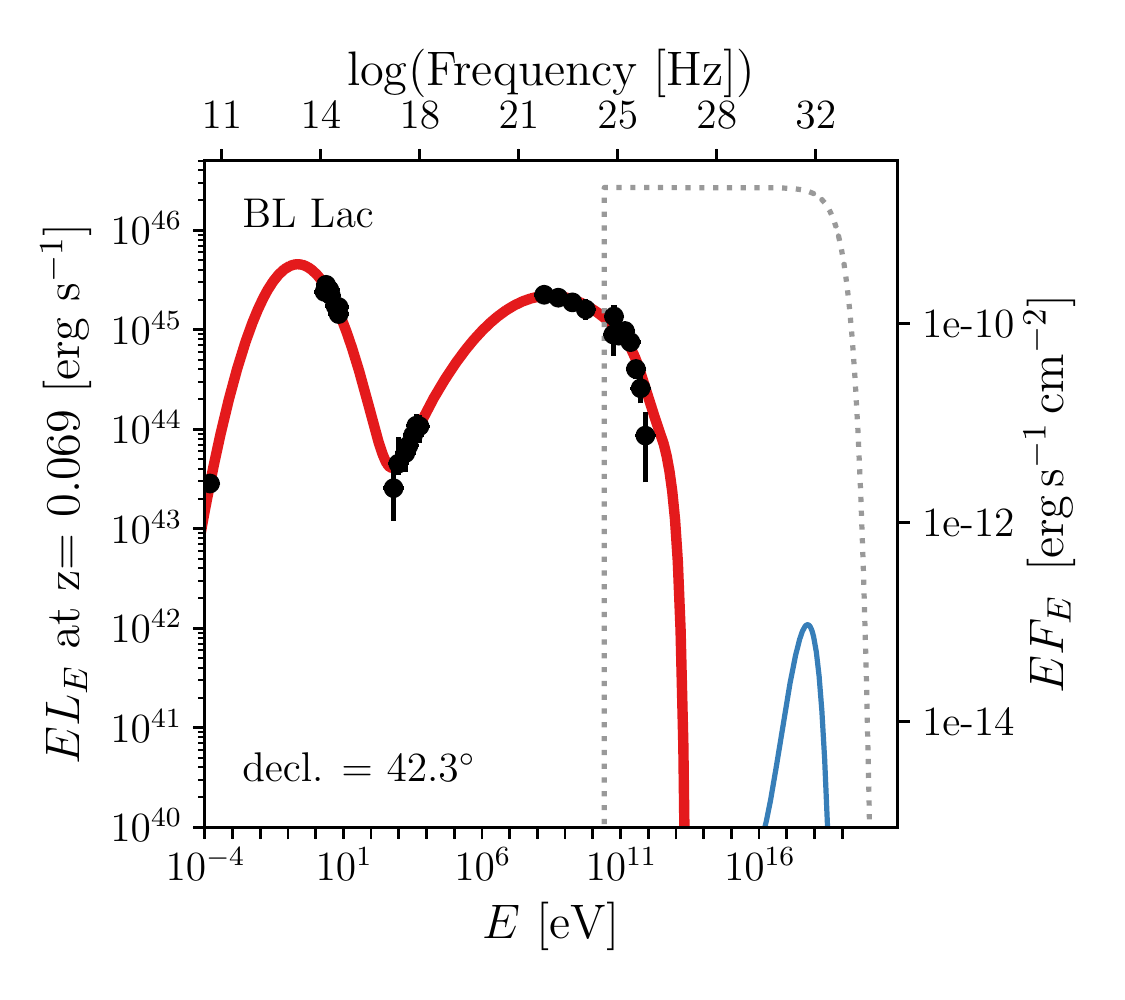}
\includegraphics[width=5.8cm,clip]{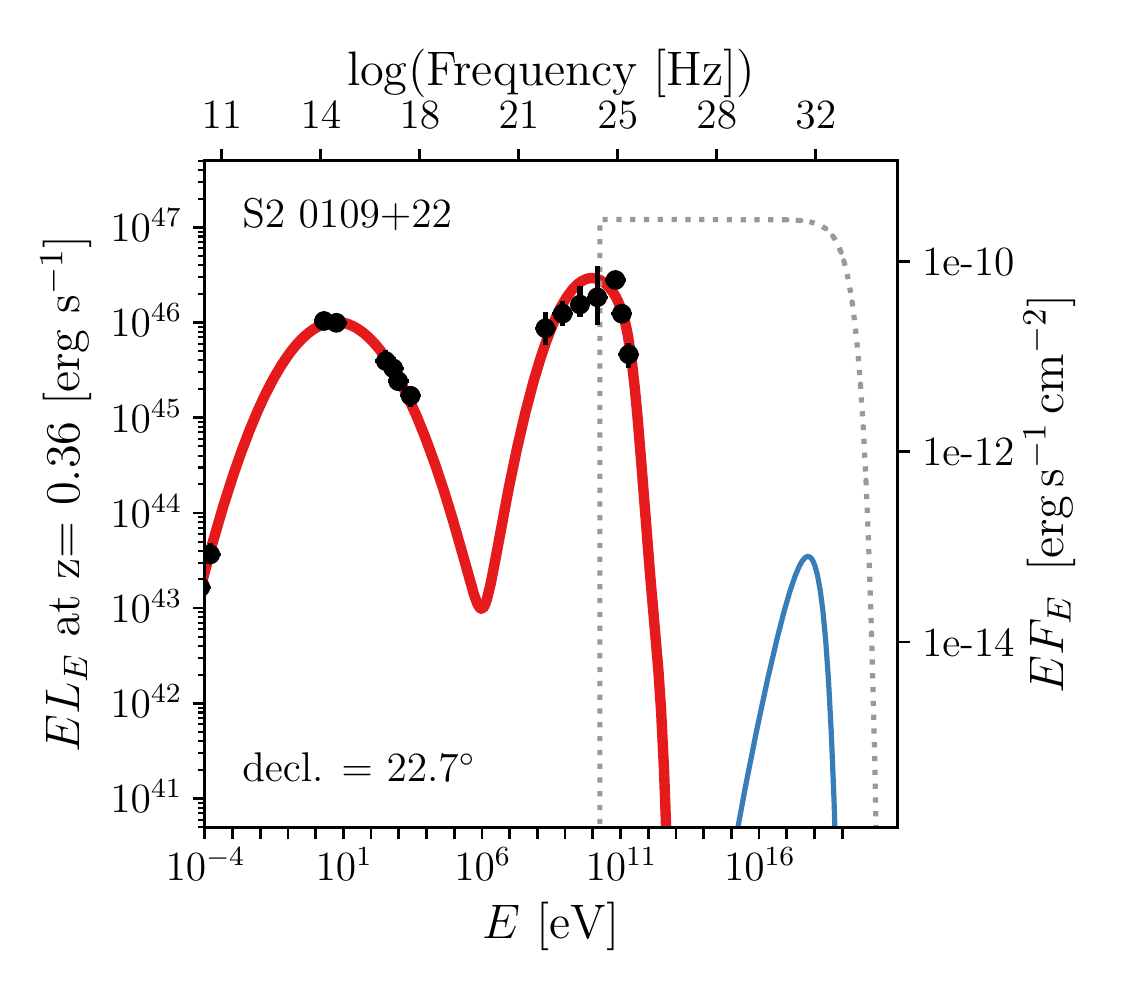}
\includegraphics[width=5.8cm,clip]{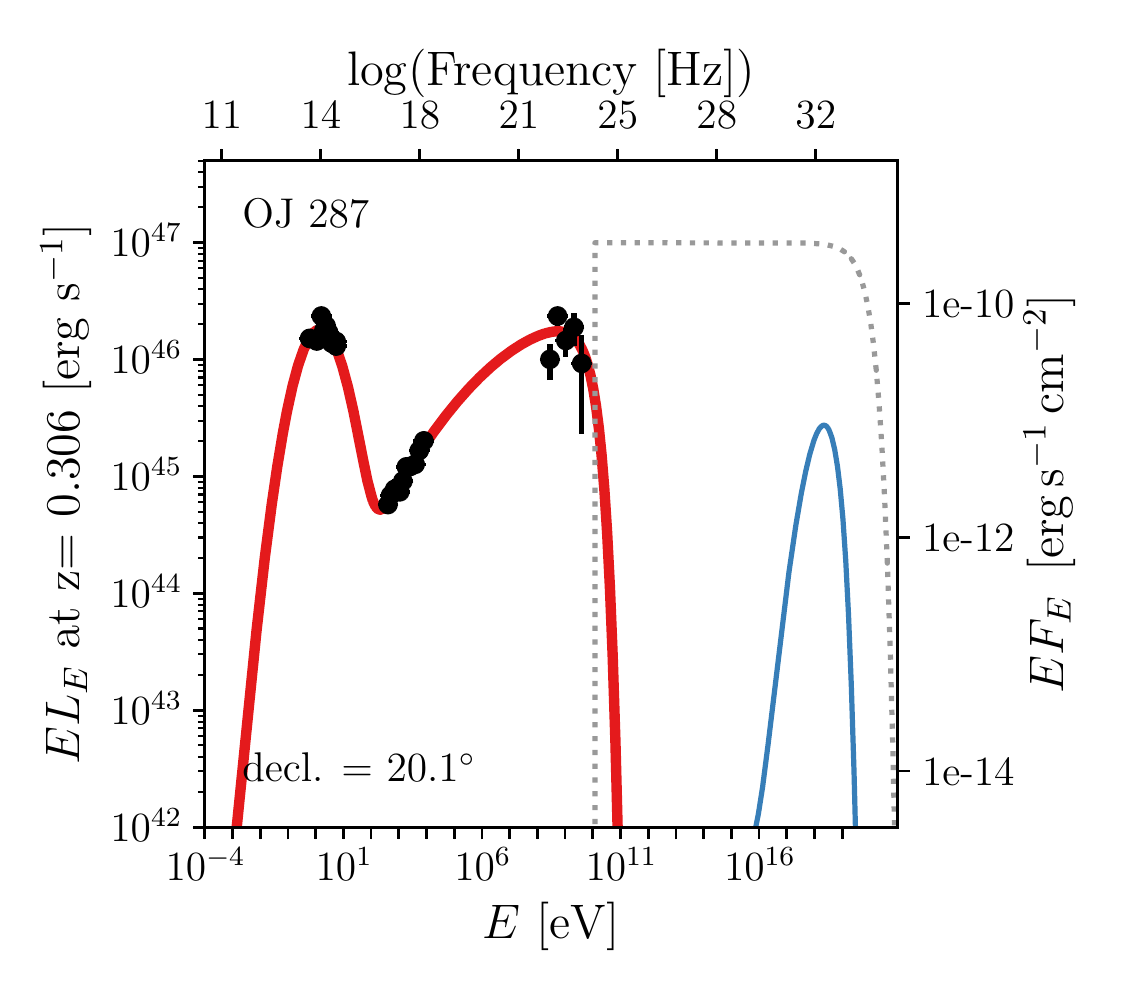}
\includegraphics[width=5.8cm,clip]{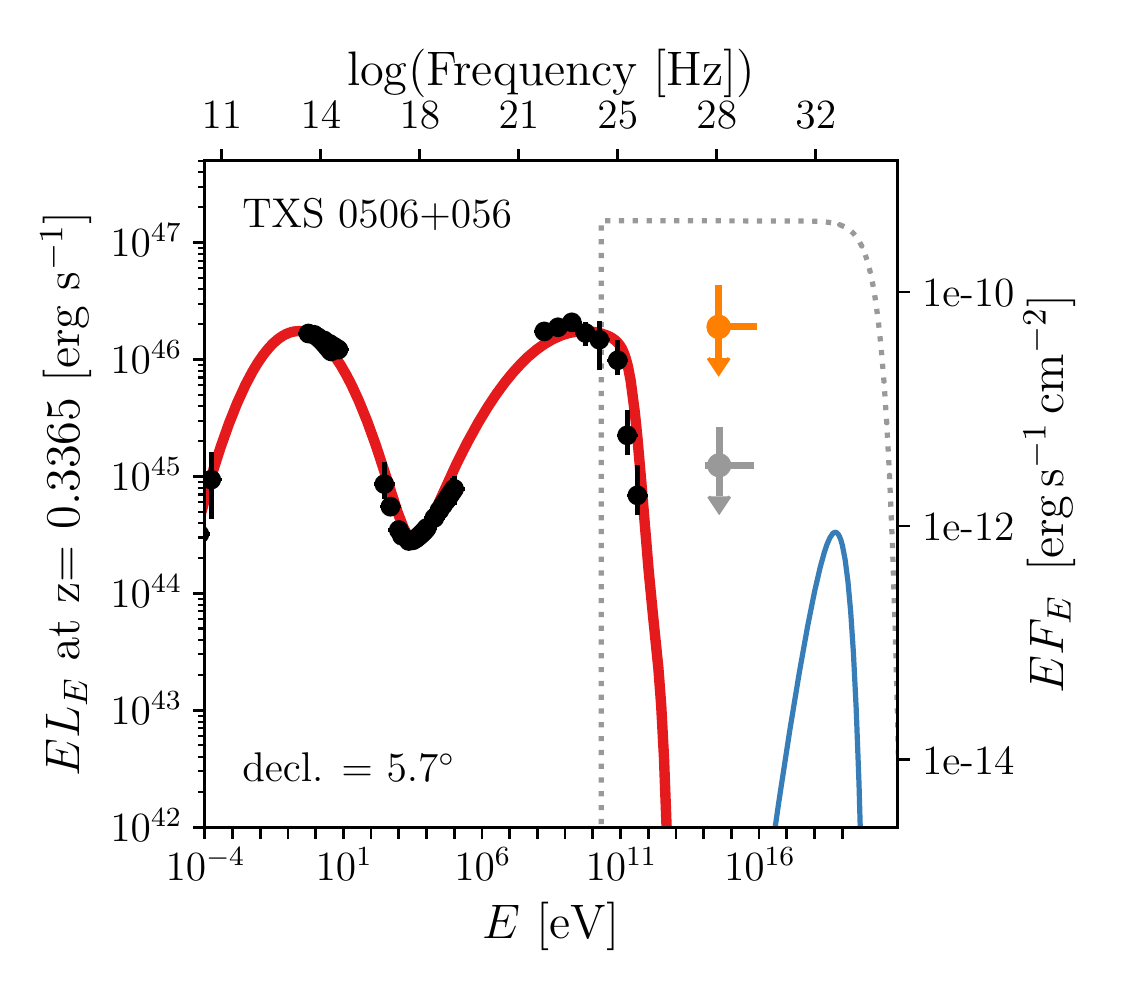}
\caption{Broadband SED (red), injected proton flux (black dotted
  lines), and expected instantaneous neutrino flux (blue) for each for
  the flares in our sample in Normalisation A in the observer frame. Only
  neutrinos produced in the interactions of protons with photons in
  the accelerating region, assumed to be a spherical blob are
  shown. The cosmic rest frame luminosity for an observer at redshift
  equal to the redshift of each source is given on the left vertical
  axis.\label{fig:FITSA} For \TXS, the data-points give the muon
  neutrino flux upper limits that would produce on average one
  detection like IceCube-170922A over a period of 0.5 (orange) and 7.5
  years (grey) at the most probable neutrino energy as calculated in
 ~\citet{IceCube:2018dnn}.  }
\end{figure*}

\begin{figure*}
\includegraphics[width=5.8cm,clip]{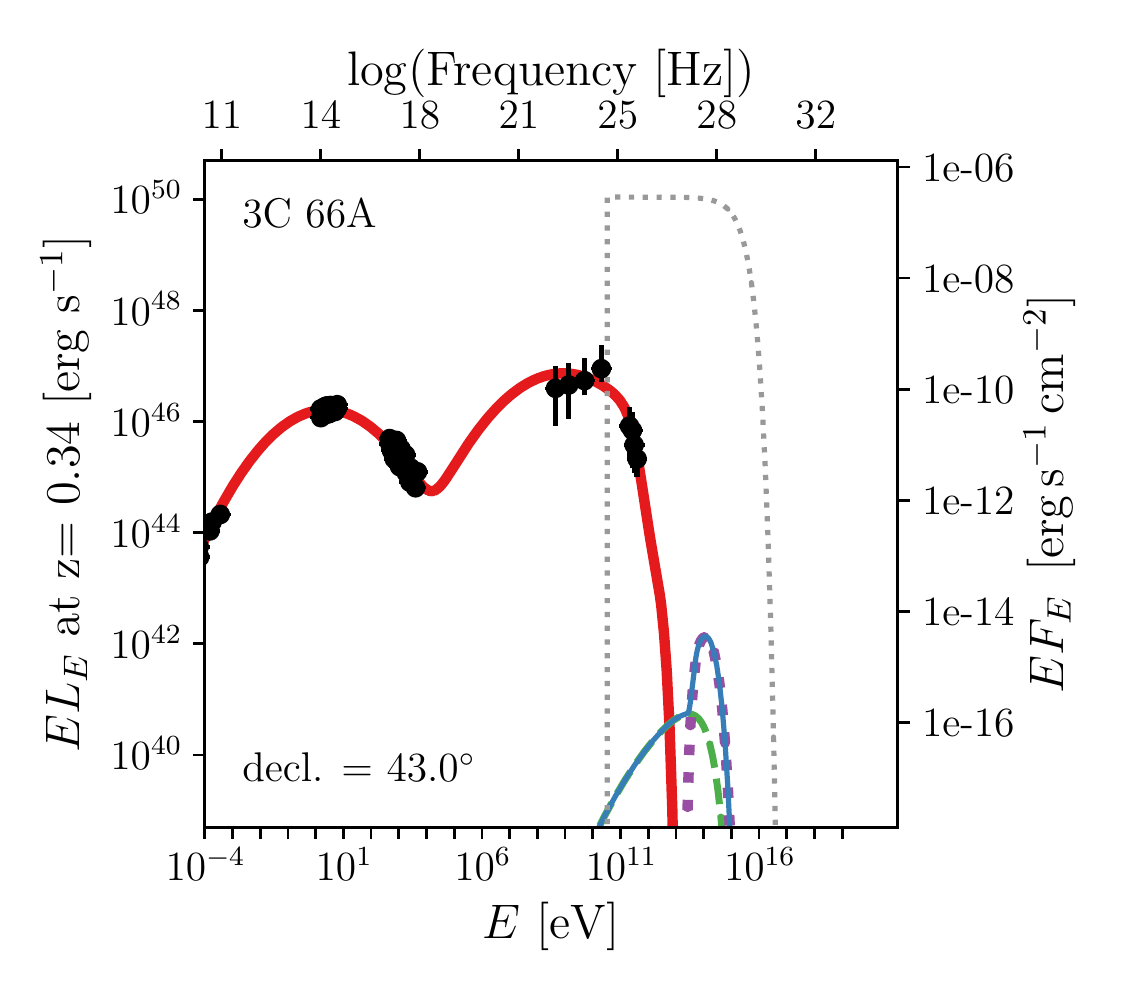}
\includegraphics[width=5.8cm,clip]{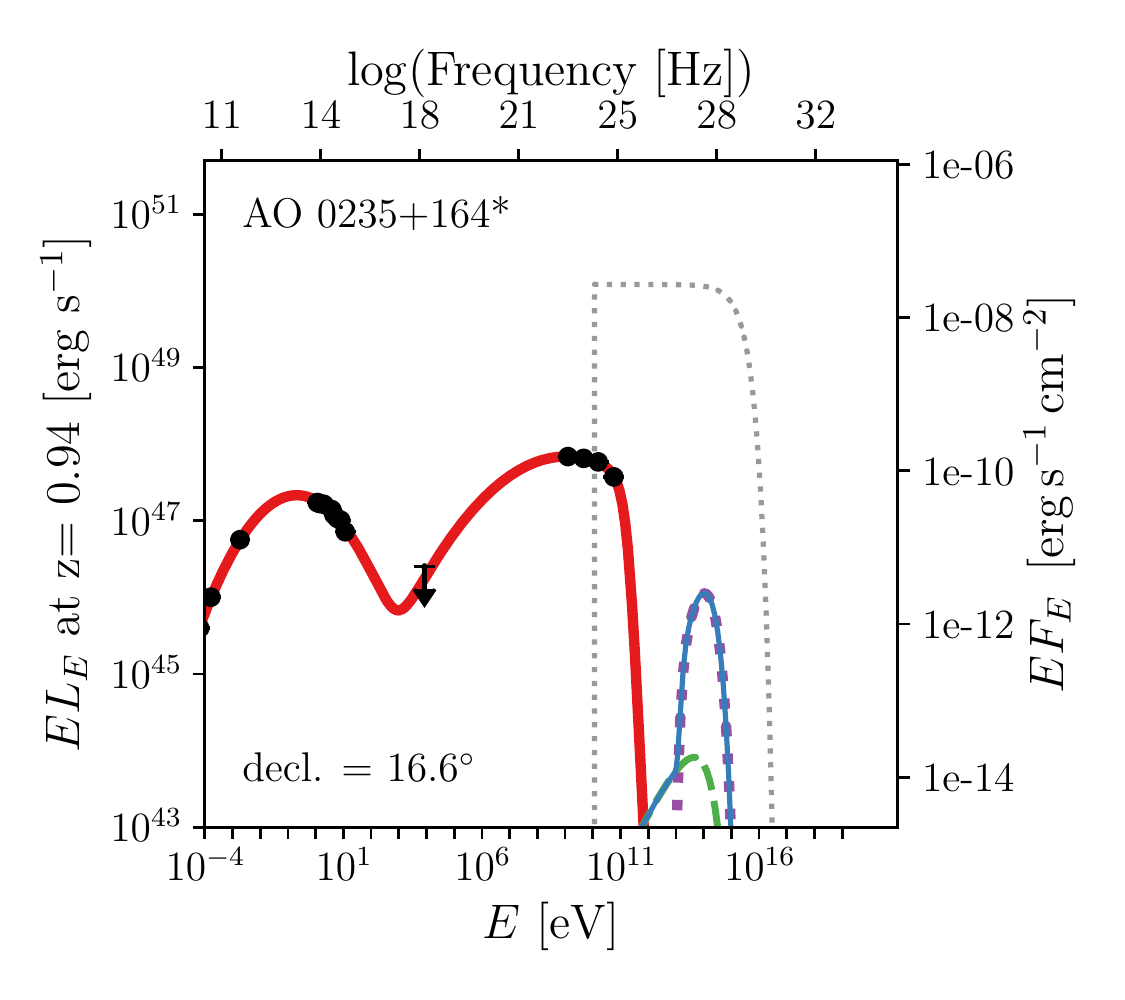}
\includegraphics[width=5.8cm]{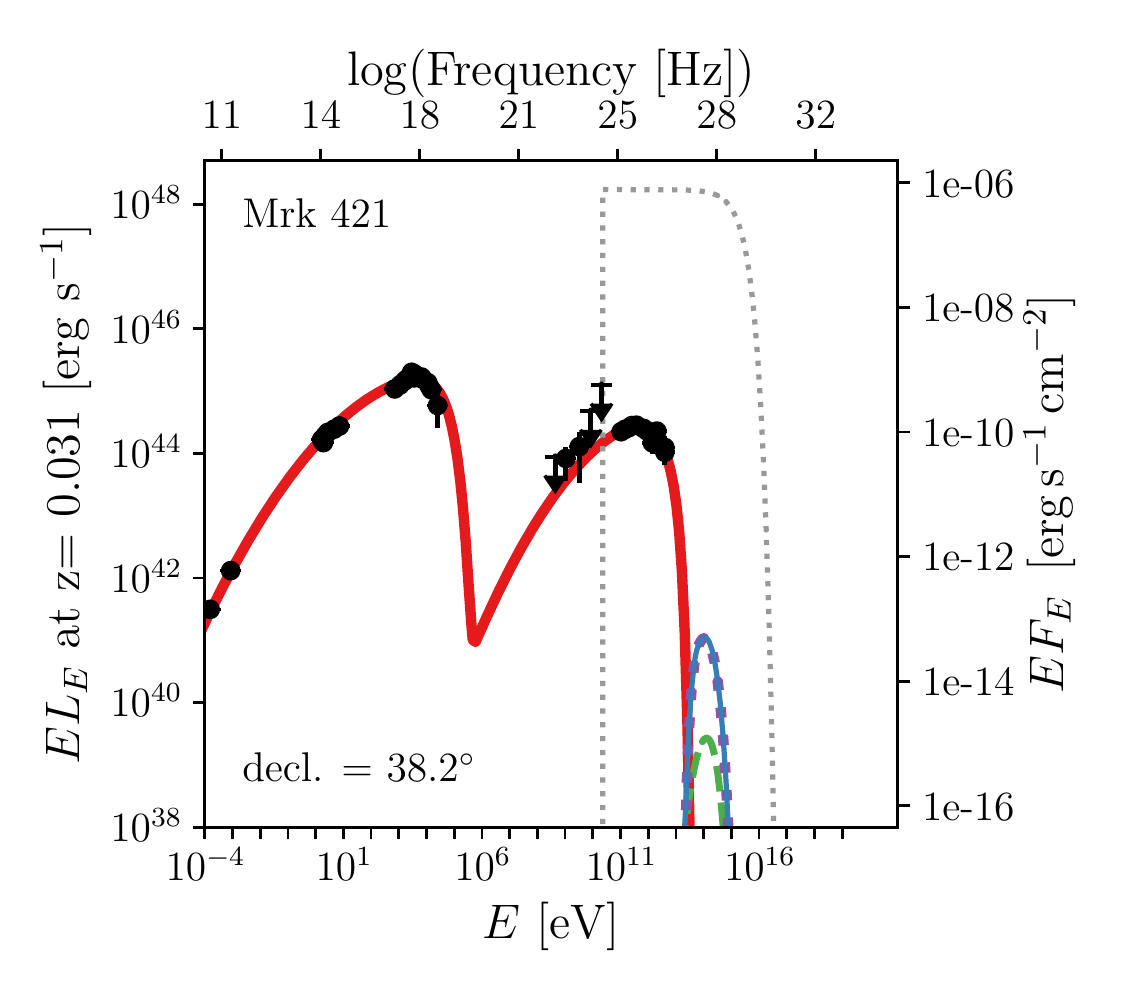}
\includegraphics[width=5.8cm,clip]{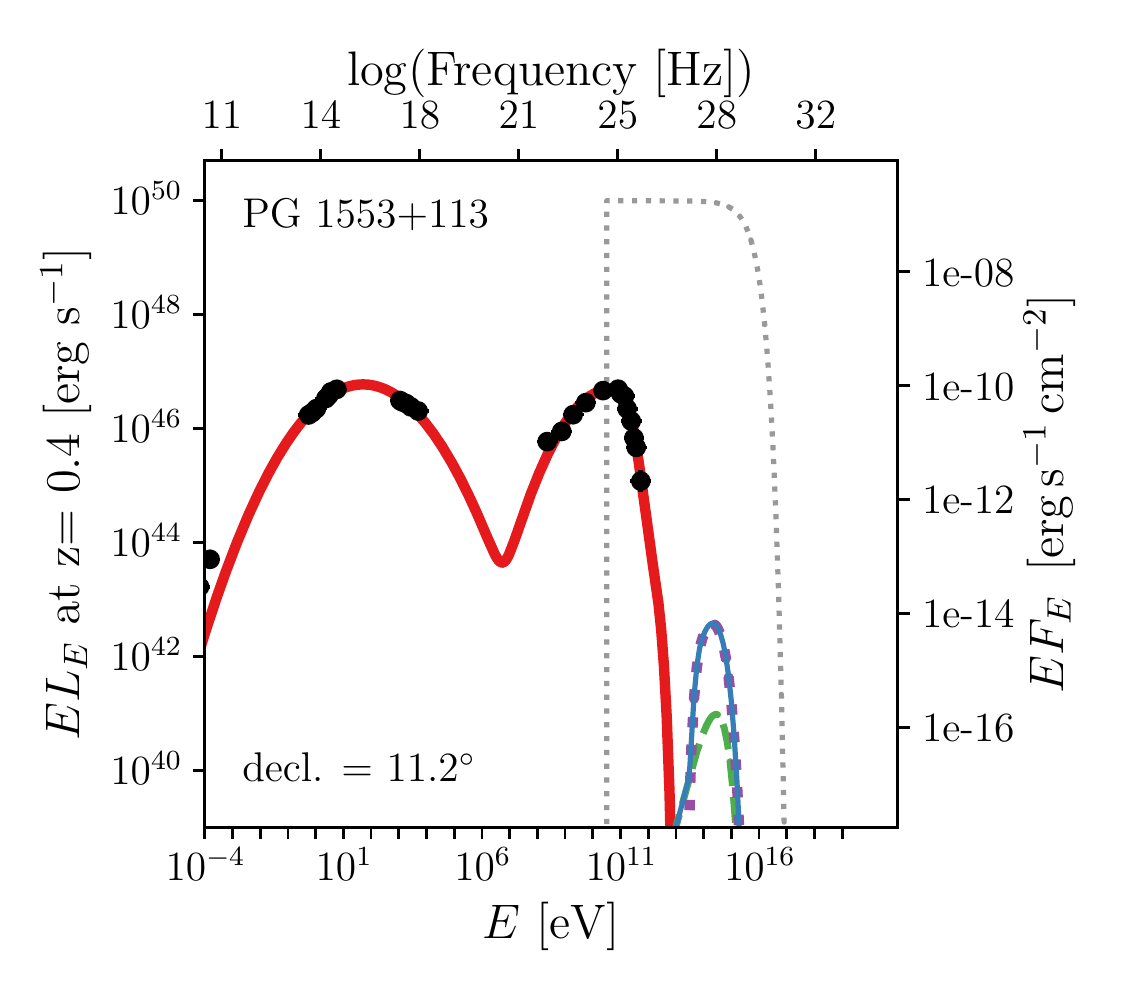}
\includegraphics[width=5.8cm,clip]{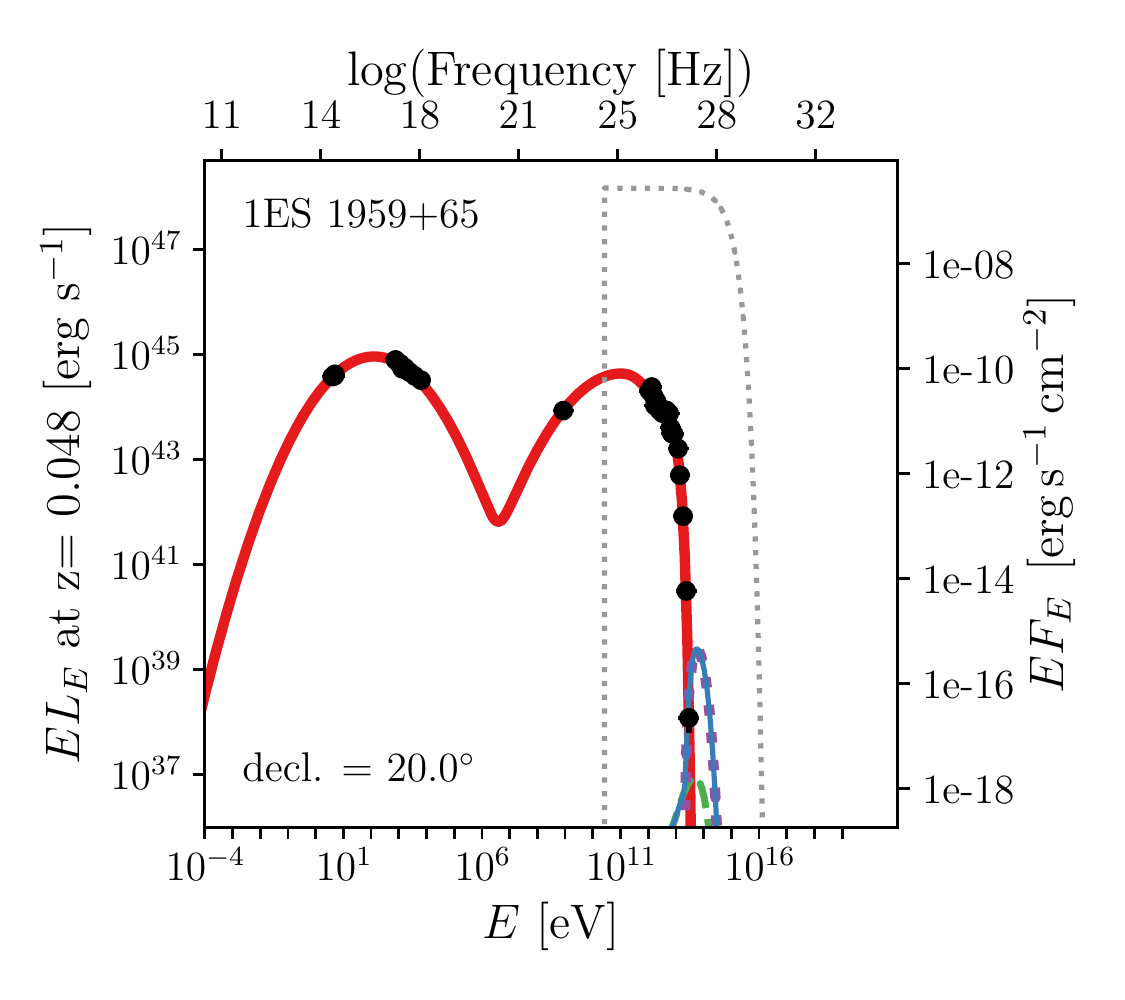}
\includegraphics[width=5.8cm,clip]{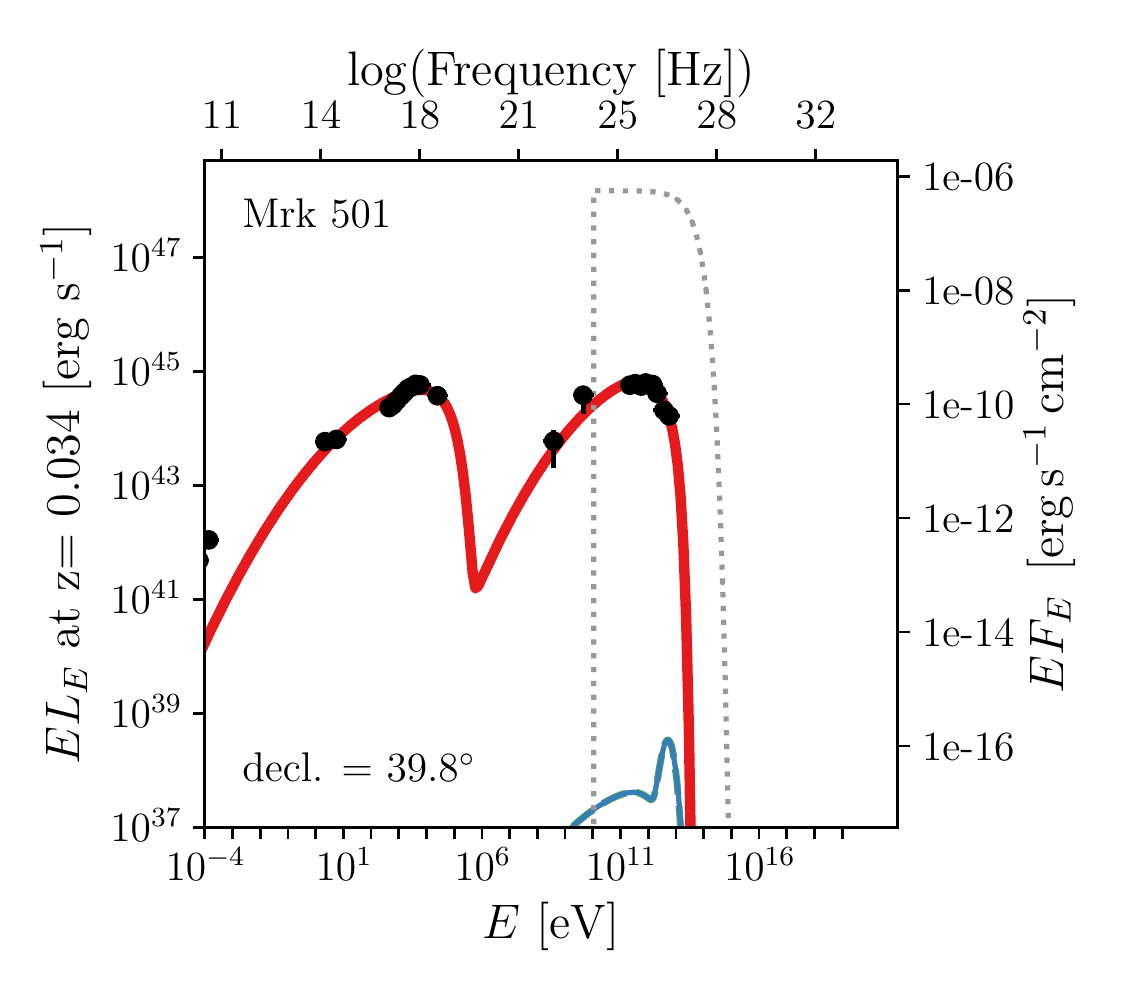}
\includegraphics[width=5.8cm,clip]{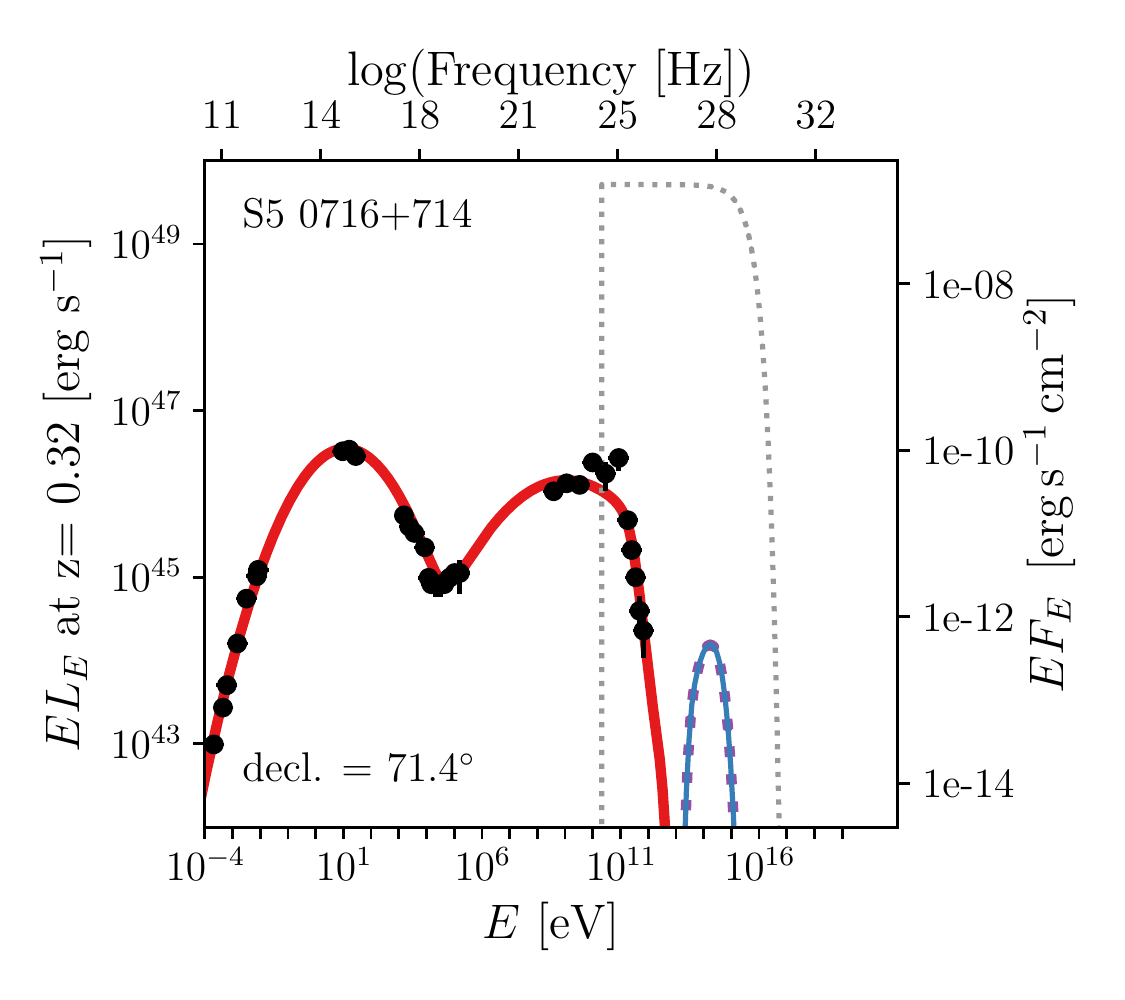}
\includegraphics[width=5.8cm,clip]{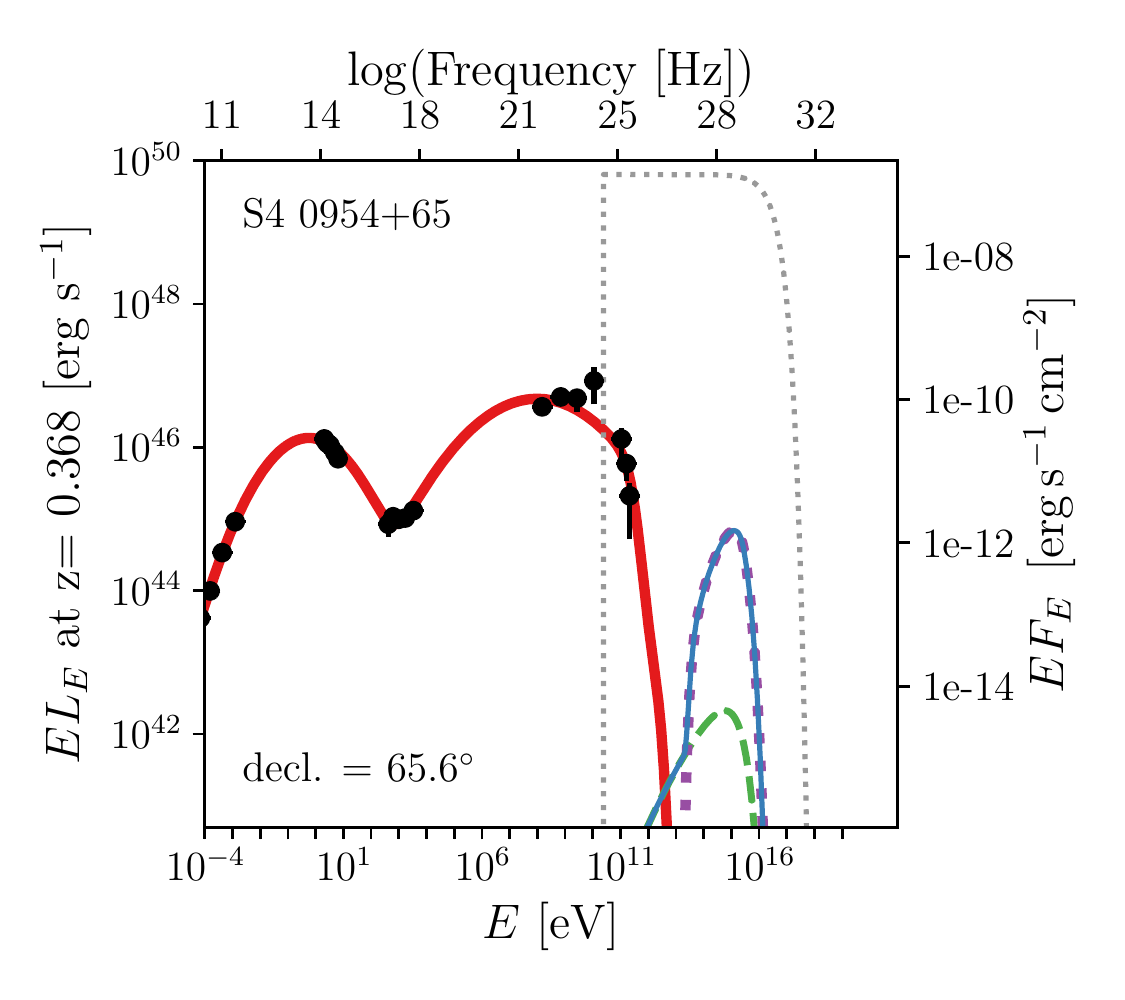}
\includegraphics[width=5.8cm,clip]{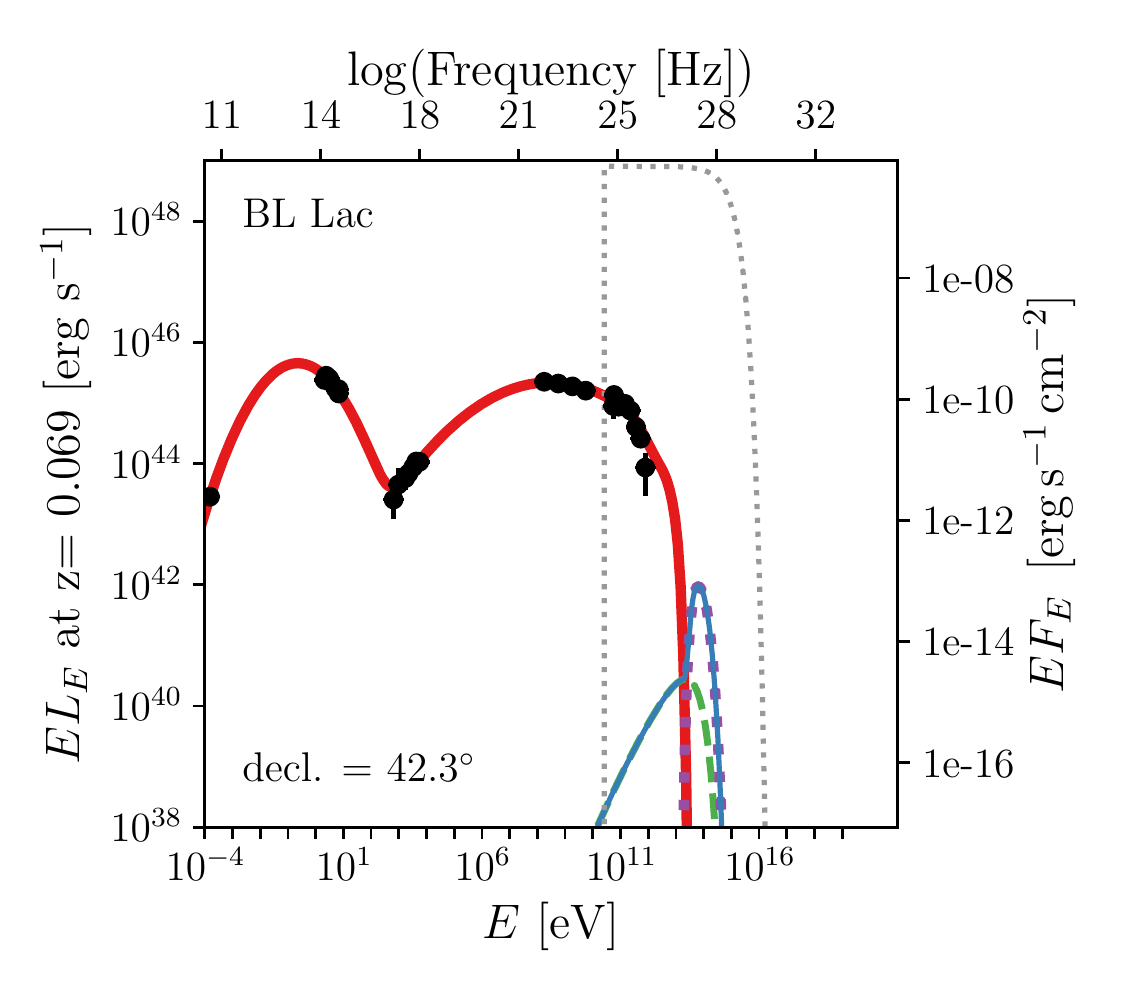}
\includegraphics[width=5.8cm,clip]{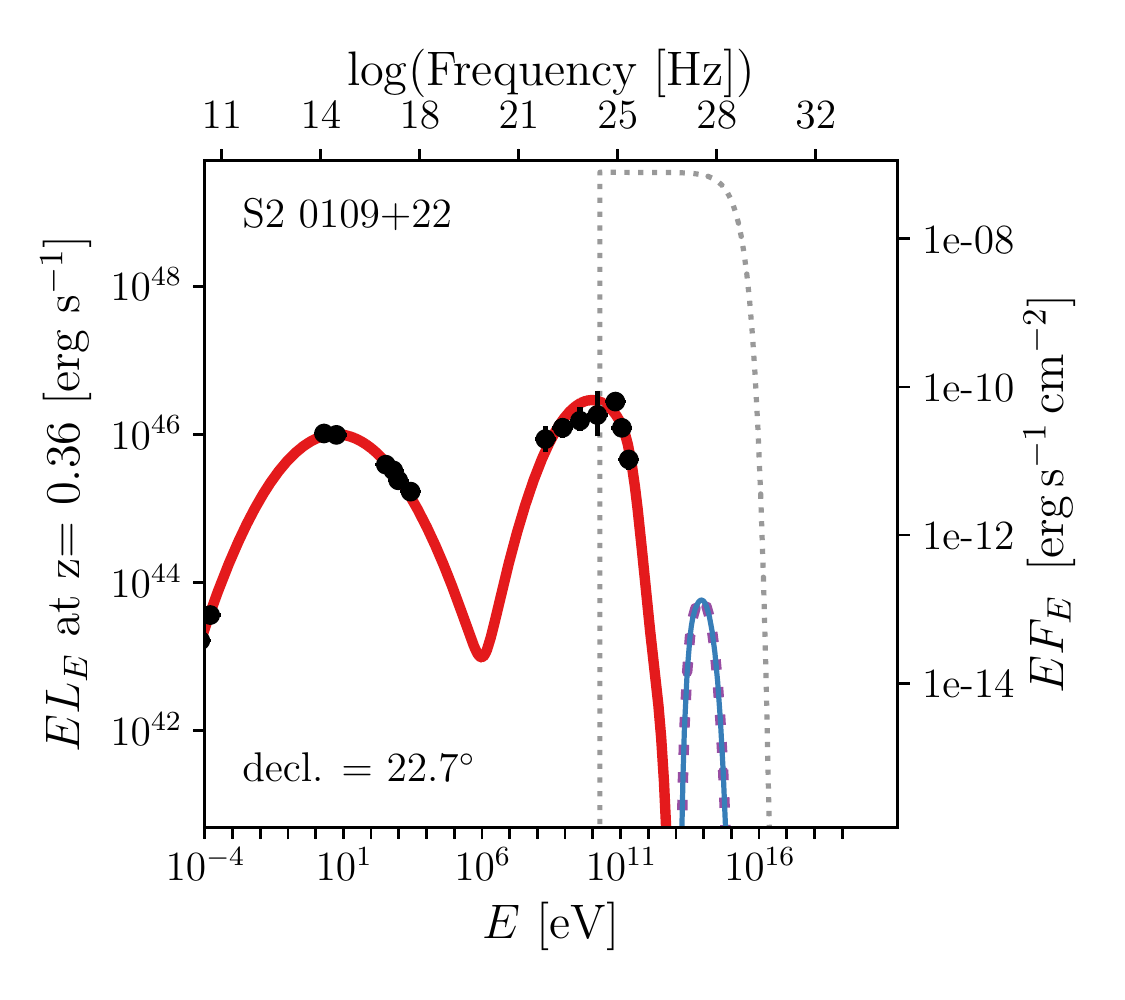}
\includegraphics[width=5.8cm,clip]{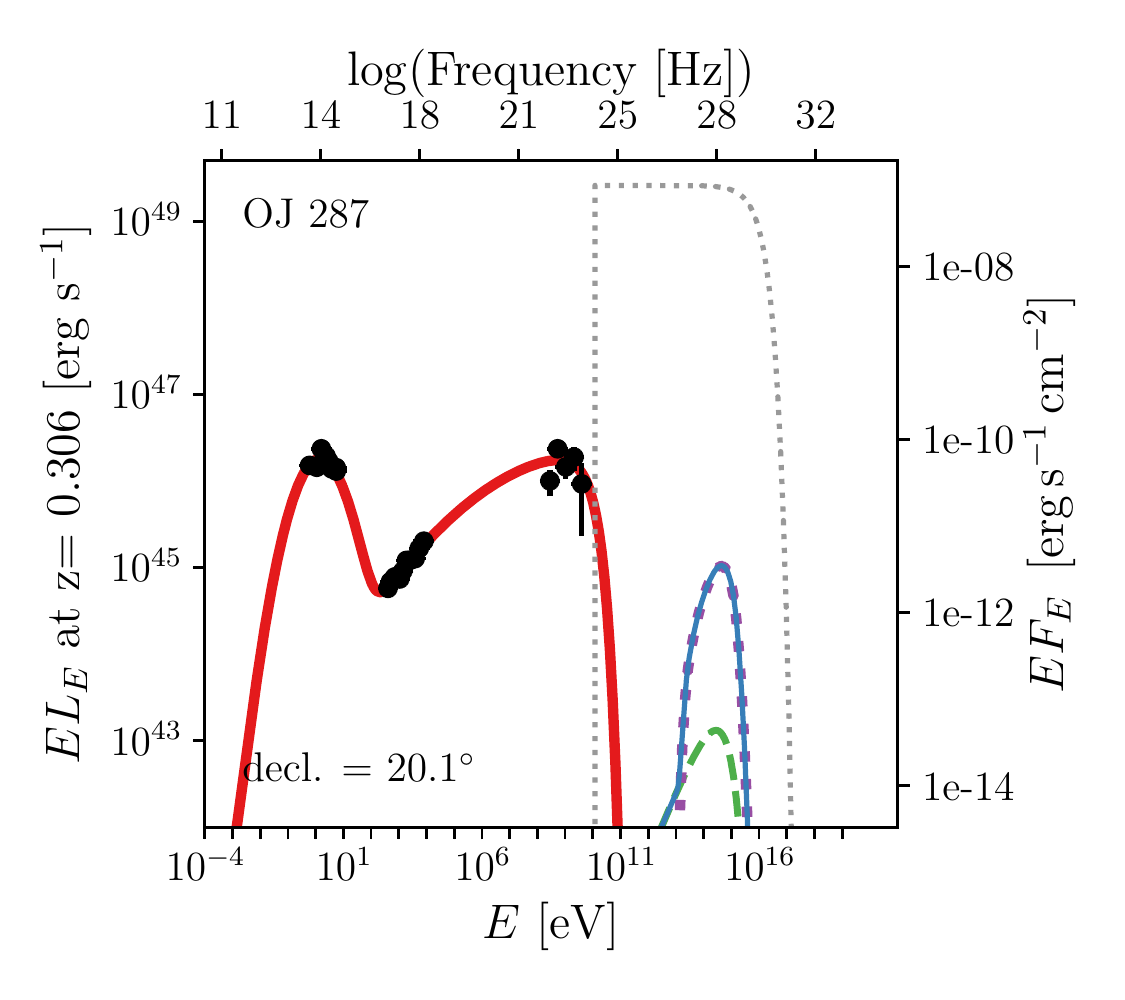}
\includegraphics[width=5.8cm]{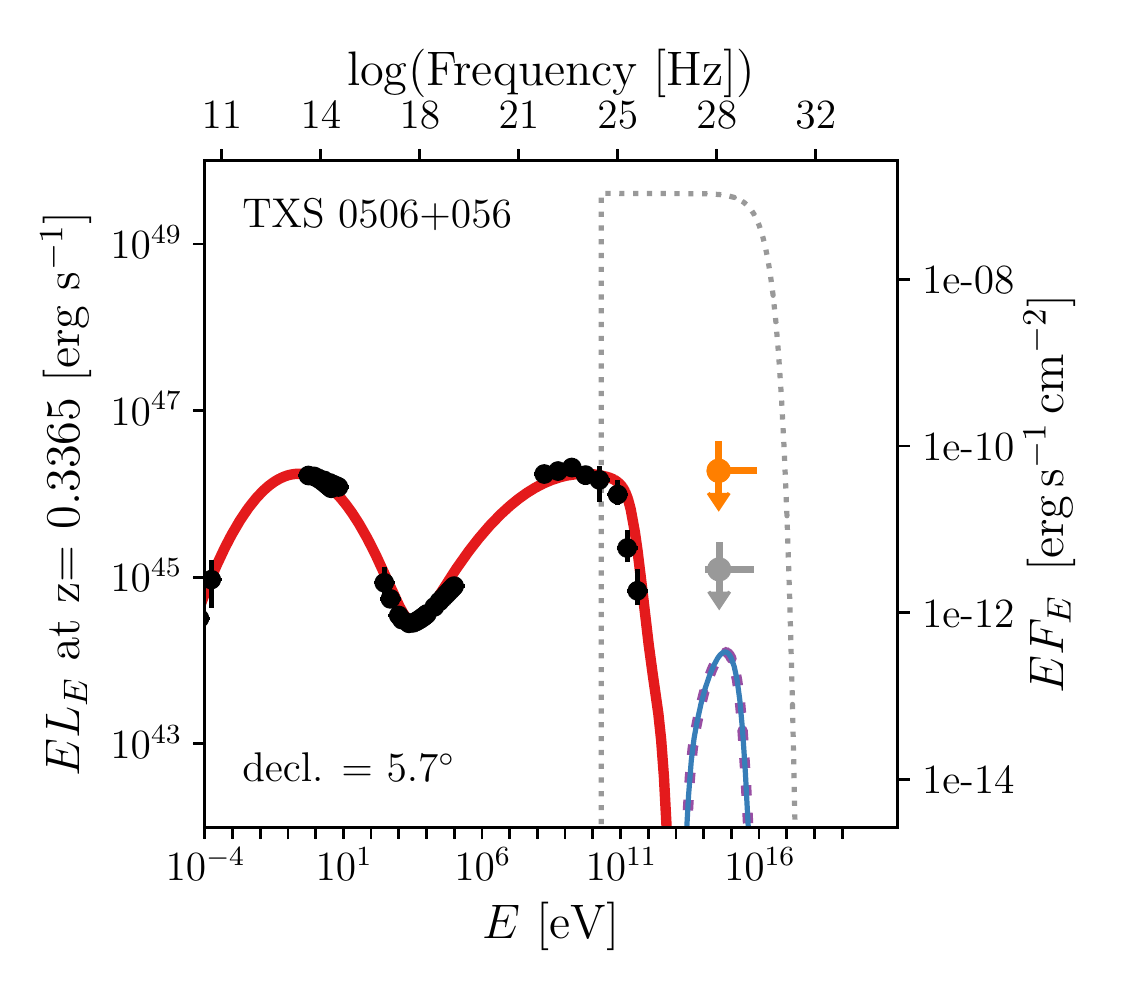}

\caption{Same as Figure~\ref{fig:FITSA} but for Normalisation B. Green dashed lines give the
  neutrino emission produced in interactions of protons with photons
  the blob. For some of the SEDs these are very suppressed and not seen in the plots. Purple dotted lines give the neutrino flux produced in
  interactions of protons with photons of the stationary external
  field. Blue solid lines give the total neutrino emission. $^\star$The star marks that for \AO we have used baryon loading ten times lower than all other sources, $\xi_{\rm cr} = 150$. \label{fig:FITSB}}
\end{figure*}

\begin{table*}
\caption{ For each source the table gives the
  time duration of the studied flare in days $\Delta T$, the
  declination, the assumed proton luminosity, $\mathcal{L}^{46}_{p}$, in units of
  $10^{46}$~erg s$^{-1}$ in the laboratory frame, the corresponding predicted laboratory frame muon-neutrino
  luminosity, $\mathcal{L}^{40}_{E_{\nu_{\mu}}}$, in units of $10^{40}$~erg s$^{-1}$, the
  predicted laboratory frame muon-neutrino energy output, $\mathcal{E}^{44,\mathrm{> 100\,TeV}}_{\nu_{\mu},\rm flare}$,
  in units of $10^{44}$~erg  for the flare studied, and
  the corresponding number of predicted muon-neutrino archival events
  in IceCube, $N^{\mathrm{IC},{\mathrm{> 100\,TeV}}}_{\nu_{\mu},\rm flare}$ with energy $\geq 100$~TeV for the model
  parameters of Normalisation A, which corresponds to baryon loading $\xi_{\rm cr} = 10$,
  high maximum proton energy, and SSC scenario with no external photon
  fields. In addition, the number of muon neutrinos per year of flaring activity $N^{\mathrm{IC},{\mathrm{> 100\,TeV}}}_{\nu_{\mu}, \rm/year~flare}$ is given.  
  The three rightmost columns give the number of muon neutrinos produced during all the flares identified through the FAVA ten-year-lightcurve 
  analysis, $N^{\mathrm{IC, > 100\,TeV}}_{\nu_{\mu}, \rm all~flares}$, the neutrino energy output of each source per average year, 
   $\mathcal{E}^{44,\mathrm{> 100\,TeV}}_{\nu_{\mu}, /\mathrm{yr}}$, and the total number of expected neutrinos 
    over the entire ten years of FAVA data
  available, $N^{\mathrm{IC, > 100\,TeV}}_{\nu_{\mu},\rm 10~yr}$, in Normalisation A, under the
  assumption that $N_{\nu_{\mu}} \propto F_{\rm HE}^2$ (see
  Section~\ref{subsec:NeutrinoCounts} for details).
\label{tab:modelA}}
\begin{tabular}{cccccccc|c|cc}

Source & $\Delta T$ & decl. & $\mathcal{L}^{46}_{p}$ & $\mathcal{L}^{40}_{{\nu_{\mu}}}$ &
 $\mathcal{E}^{44, \mathrm{> 100\,TeV}}_{\nu_{\mu},\rm /flare}$ & 
 $N^{\mathrm{IC},{\mathrm{> 100\,TeV}}}_{\nu_{\mu},\rm flare}$ & 
$N^{\mathrm{IC},{\mathrm{> 100\,TeV}}}_{\nu_{\mu}, \rm/year~flare}$ &
 $N^{\rm IC, \mathrm{> 100\,TeV}}_{\nu_{\mu}, \rm all~flares}$ &
 $\mathcal{E}^{44,\mathrm{> 100\,TeV}}_{\nu_{\mu}, /\mathrm{yr}}$ &
  $N^{\mathrm{IC, > 100\,TeV}}_{\nu_{\mu},\rm 10~yr}$ \\
\hline
3C 66A  &  14  &  43.0  &  700  &  400  &  1e+5  &  1e-8  &  2e-7  &  2e-7  &  1e+6  &  1e-6  \\
AO 0235+164  &  84  &  16.6  &  9e+3  &  8e+6  &  2e+10  &  5e-4  &  2e-3  &  2e-3  &  1e+10  &  3e-3  \\
Mrk 421  &  13  &  39.8  &  10  &  7 &  9e+2  &  3e-7  &  7e-6  &  6e-5  &  4e+4  &  1e-4  \\
PG 1553+113  &  30  &  11.2  &  700 &  4e+3  &  1e+6  &  6e-6  &  5e-7  &  1e-5  &  2e+7  &  1e-4  \\
1ES 1959+65  &  46  &  20.0  &  9  &  1  &  600  &  2e-7  &  1e-6  &  5e-7  &  100  &  8e-7  \\
Mrk 501  &  21  &  38.2  &  7  &  30  &  4e+3  &  4e-5  &  7e-4  &  1e-3  &  2e+4  &  3e-3  \\
S5 0716+714  &  14  &  71.4  &  400  &  2e+4  &  1e+6  &  2e-8  &  4e-7  &  6e-7  &  8e+6  &  1e-6  \\
S4 0954+65  &  28  &  65.6  &  600 &  3e+4  &  7e+6  &  1e-7  &  1e-6  &  1e-7  &  1e+6  &  2e-7  \\
BL Lac  &  7  &  42.3  &  50  &  70  &  1e+4  &  5e-7  &  2e-5  &  1e-5  &  7e+4  &  3e-5  \\
S2 0109+22  &  21  &  22.7  &  200  &  3e+3  &  9e+5  &  5e-7  &  1e-5  &  2e-5  &  6e+6  &  4e-5  \\
1ES 1959+65  &  84  &  20.0  &  4  &  3e-1  &  200  &  1e-7  &  5e-7  &  5e-7  &  100  &  8e-7  \\
OJ 287  &  7  &  20.1  &  200  &  2e+5  &  2e+8  &  2e-5  &  1e-3  &  9e-5  &  3e+8  &  3e-4  \\
TXS 0506+056  &  175  &  5.7  &  300  &  3e+4  &  2e+9  &  4e-5  &  7e-5  &  4e-5  &  4e+8  &  8e-5  \\
\hline
\end{tabular}
\end{table*}

\begin{table*}
\caption{Same as Table \ref{tab:modelA} but for Normalisation B, which
  includes the expected number of neutrinos in interactions with an
  external photon field. $^{\star}$star superscripts denote that for \AO we have used 
  baryon loading factor ten times lower than all other sources $\xi_{\rm cr} = 150$. 
\label{tab:modelB}}
\begin{tabular}{cccccccc|c|cc}
Source & $\Delta T$ & decl. & $\mathcal{L}^{46}_{p}$ & $\mathcal{L}^{40}_{\nu_{\mu}}$ & 
$\mathcal{E}^{44, \mathrm{> 100\,TeV}}_{\nu_{\mu},\rm /flare}$ & 
 $N^{\mathrm{IC},{\mathrm{> 100\,TeV}}}_{\nu_{\mu},\rm flare}$ & 
$N^{\mathrm{IC},{\mathrm{> 100\,TeV}}}_{\nu_{\mu}, \rm year~flare}$ &
 $N^{\rm IC, \mathrm{> 100\,TeV}}_{\nu_{\mu}, \rm all~flares}$ &
 $\mathcal{E}^{44,\mathrm{> 100\,TeV}}_{\nu_{\mu}, /\mathrm{yr}}$ &
  $N^{\mathrm{IC, > 100\,TeV}}_{\nu_{\mu},\rm 10~yr}$ \\
\hline
3C 66A  &  14  &  43.0  &  1e+5  &  70  &  1e+4  &  5e-6  &  1e-4  &  5e-5  &  1e+5  &  5e-4  \\
AO 0235+164  &  84  &  16.6  &  1e+5$^{\star}$   &  8e+5$^{\star}$   &  5e+8$^{\star}$   &  3e-2$^{\star}$  &  1e-1$^{\star}$  &  1e-1$^{\star}$  &  3e+8$^{\star}$  &  2e-1$^{\star}$  \\
Mrk 421  &  13  &  39.8  &  2e+3  &  5  &  6e+2  &  6e-5  &  1e-3  &  1e-2  &  3e+4  &  3e-2  \\
PG 1553+113  &  30  &  11.2  &  1e+5  &  200  &  6e+4  &  4e-5  &  5e-4  &  7e-4  &  8e+5  &  6e-3  \\
1ES 1959+65  &  46  &  20.0  &  1e+3  &  8e-2  &  3e+1  &  3e-7  &  2e-6  &  8e-7  &  10  &  1e-6  \\
Mrk 501  &  21  &  38.2  &  1e+3  &  2e-2  &  2  &  0  &  0  &  0  &  20  &  0  \\
S5 0716+714  &  14  &  71.4  &  6e+4  &  1e+4  &  2e+6  &  3e-4  &  8e-3  &  1e-2  &  1e+7  &  2e-2  \\
S4 0954+65  &  28  &  65.6  &  9e+4  &  6e+4  &  2e+7  &  9e-4  &  1e-2  &  1e-3  &  3e+6  &  1e-3  \\
BL Lac  &  7  &  42.3  &  8e+3  &  40  &  8e+3  &  5e-5  &  2e-3  &  1e-3  &  4e+4  &  2e-3  \\
S2 0109+22  &  21  &  22.7  &  4e+4  &  3e+3  &  7e+5  &  3e-4  &  6e-3  &  1e-2  &  5e+6  &  2e-2  \\
1ES 1959+65  &  84  &  20.0  &  700  &  2e-2  &  20  &  2e-7  &  7e-7  &  8e-7  &  10  &  1e-6  \\
OJ 287  &  7  &  20.1  &  3e+4  &  8e+4  &  4e+7  &  4e-2  &  2  &  2e-1  &  7e+7  &  7e-1  \\
TXS 0506+56  &  175  &  5.7  &  5e+4  &  2e+4  &  2e+7  &  2e-2  &  4e-2  &  2e-2  &  5e+6  &  4e-2  \\
\hline
\end{tabular}
\end{table*}

Figures \ref{fig:FITSA} and \ref{fig:FITSB} show the expected neutrino
flux from each flare in our sample for Normalisations A and B
respectively. The peak neutrino energy-flux is expected at $\gtrsim 10$~PeV
for Normalisation A and $\lesssim 1$~PeV for Normalisation B. 
The two figures show that the instantaneous
neutrino flux is not strictly proportional to the photon flux at a particular
wavelength since we have explicitly modelled the neutrino flux taking
into account the physical parameters in each source.

The two most important source characteristics that contribute to a
high expected instantaneous neutrino flux are a high bolometric
luminosity and a relatively low Doppler factor. These two
conditions ensure a higher number density of target photons and
$p\gamma$ interaction probability in the emitting region in the SSC
model which follows from Equations~\ref{eq:comoving_photon_density}
and~\ref{eq:pgammaRate}. In addition, a higher bolometric luminosity
and low Doppler factor imply a higher proton luminosity in the comoving frame. 
Within our model assumptions, these two conditions also lead to a
higher neutrino flux for Normalisation B, where external Compton interactions
occur, because we require that $u'_{\rm syn} > u'_{\rm ext}$ and
$L_{\rm ext} < L_{\rm syn}$. 

Tables \ref{tab:modelA} and \ref{tab:modelB} give the expected number
of neutrinos in each flare in our sample in the IceCube detector, for
Normalisation A and Normalisation B, respectively. We have calculated the expected
number of neutrinos by integrating the neutrino spectrum above 100 TeV where the atmospheric background is low. 
The highest neutrino counts under Normalisation A are expected for \AO, which is the
most luminous source in our sample, despite being the most distant, at
redshift $z = 0.94$. In the case of Normalisation B, \AO has been modelled with 
$\xi_{\rm cr}$ ten times lower than all other sources. 
The highest neutrino counts are expected in this case from \AO 
and from the flares of \OJ and \TXS which are some of the brightest sources in our
sample. 

Significantly lower neutrino counts are expected for the
nearby HSP sources (Mrk 421, Mrk 501, 1ES 1959+65,
\PG). These are generally less bright than the LSP/ISPs studied. 
\PG is is an exception among the HSPs we modelled because it
is one of the brightest sources considered, however, it has a high Doppler factor
and thus a low predicted neutrino luminosity.
 
In addition to sources having a high bolometric luminosity, high-neutrino counts can
be expected for the longest \gRay flares, for relatively nearby sources, 
and for sources located in a position in the sky favourable for detection within IceCube. 
Within the range of declinations spanned by the sources in our sample, in the $\sim 5^{\circ} - 72^{\circ}$ range,
the effective area of IceCube in the direction of different sources varies in fact by up to 
a factor of 20.
 
The predicted neutrino event counts for Normalisation A are in general
modest. Comparison with Table \ref{tab:modelB} reveals that the
expected neutrino counts increase by up to three orders of magnitude
for Normalisation B, due to the much higher baryon loading and neutrino production in
interactions of protons with an external photon field.

It is interesting to note that the relative neutrino signal between
the different flares is not the same under Normalisations A and B. For
example, \mrkfive produces the second strongest signal in our sample
for Normalisation A, but a very weak neutrino flare in Normalisation B consistent with zero above 100 TeV. This is
because \mrkfive, with detected variability of 4 hours, is assumed to
have a small emitting region radius, $r_{b}'$, in our model. In the
SSC model this increases the density of the relevant photon field for
neutrino interactions significantly, whereas in the external-Compton
model which generally dominates the SSC component as shown in Figure
\ref{fig:FITSB} the small $r^{\prime}_{b}$ implies a low proton
maximum energy, such that the neutrinos produced in interactions of protons 
with the external photon field are very 
suppressed with respect to cases with a higher proton maximum energy.

If the same physical mechanism is in operation in \TXS as in \AO and \OJ 
and the proton content in these sources is the same, one expects a stronger 
signal from these two past flares in the archival IceCube data than in the 2017 flare of \TXS.
However, the expected number of neutrinos is below the sensitivity of IceCube even in Normalisation B. Therefore, even in the absence of neutrino signal from these two flares, it is not possible to constrain this model. 

Comparison of time integrated neutrino flux column, $N^{\mathrm{IC, > 100\,TeV}}_{\nu_{\mu},\rm 10~yr}$, on Tables \ref{tab:modelA} 
and \ref{tab:modelB} with the flux produced by a single flare, $N^{\mathrm{IC},{\mathrm{> 100\,TeV}}}_{\nu_{\mu},\rm flare}$,
reveals that under our assumed relation between $L_{\nu}$ and $L_{\gamma}$ a short neutrino flare can produce 
a sizeable fraction of the time-integrated flux of a neutrino emitting blazar. 
In the case of \Sfour the 40-fold flux enhancement as seen in the FAVA data, 
 means that $\sim$half the expected neutrino flux
 from this source in the last ten years comes from the 28-day window
 around the 2015 flare. This is also true for the 2017 flare of 
 \TXS . 
  
Tables \ref{tab:modelA} and \ref{tab:modelB} also give the required cosmic-rest-frame proton luminosity
 $\mathcal{L}_p$ during the flares in our model. After converting it to the absolute 
 beaming-corrected proton luminosity, $\mathcal{L}_p/2 \Gamma^2$, this can be compared to the Eddington luminosity for a sanity check. 
Assuming a typical black-hole mass of $10^{8.5} M_{\sun}$~\citep{2011MNRAS.413..805P}, 
while with Normalisation A, all proton luminosities are safely below the Eddington limit, with Normalisation B 
we find that the beaming-corrected proton luminosity is in the range $\sim 0.2 - 30$ times 
the Eddington luminosity. This is not necessarily problematic, 
as during outbursts the Eddington limit can be temporarily exceeded~\citep[e.g.][]{10.1093/mnras/stv1802}. 
  
\begin{figure}
    \includegraphics[width=8cm,clip]{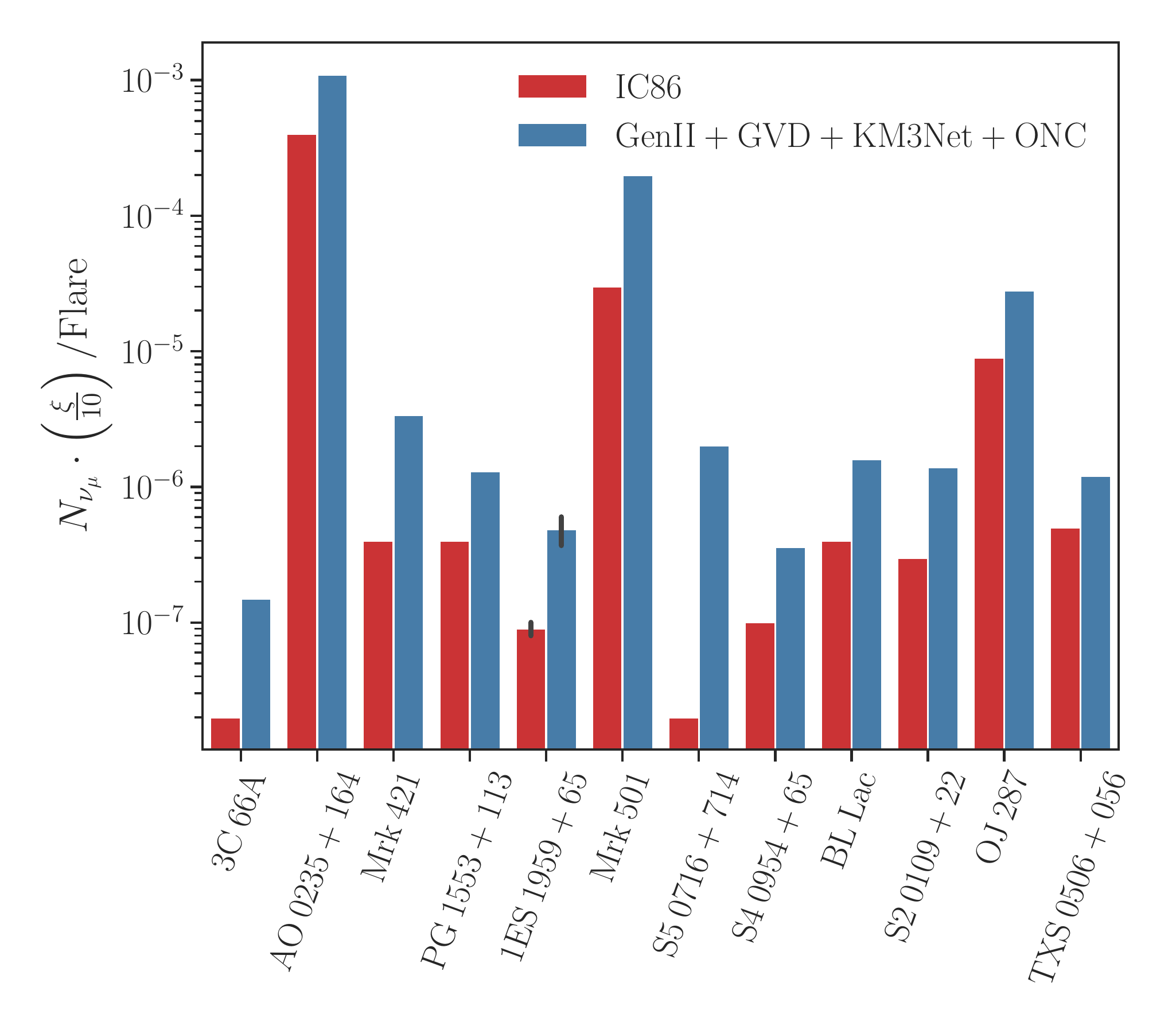}
    \includegraphics[width=8cm,clip]{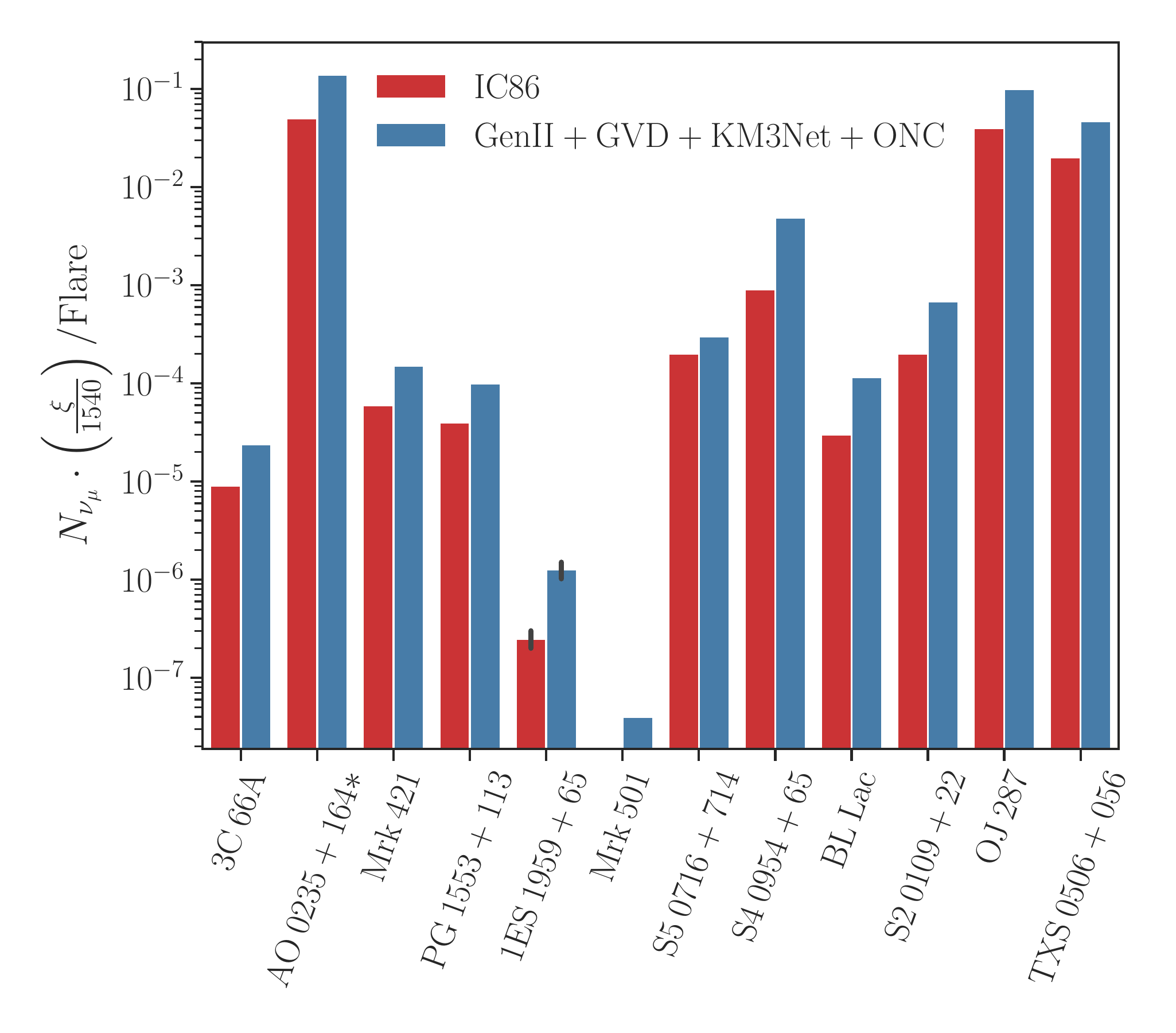}	
 \caption{Predicted neutrino counts for each of the flares in our
   sample under the Normalisation A (top) and Normalisation B parameters
   (bottom). The red bars give the expected number of muon neutrinos
   as seen in IceCube, in the IC86 configuration, and comparison to
   the expected number of neutrinos for identical flares if GenII,
   GVD, ONC, and KM3Met are in operation. For \IES the error bars are
   statistical and derive from the modelling of two separate
   flares. $^\star$The star marks that for \AO we have used baryon loading ten times 
   lower than all other sources, $\xi_{\rm cr} =150$. \label{fig:bar}}
\end{figure}

In Figure~\ref{fig:bar}, the expected number of muon neutrinos in
IceCube in full configuration (IC86) is contrasted with the number of
muon neutrinos that would have been detected for the studied flares in
planned future neutrino facilities. Our assumptions about the effective area of these 
future detectors are presented in Section~\ref{subsec:NeutrinoCounts}. 
The number of expected neutrinos
is given as a function of the baryon loading factor, which could be as
much as one order of magnitude higher assuming Normalisation A as discussed in 
Section~\ref{subsec:neutrino_production}.
For the sources with high positive declination future
neutrino detectors in the Northern hemisphere will strongly increase
the expected neutrino counts from individual flares as demonstrated by
the case of \Sfive.

\begin{figure*}
\includegraphics[width=7.9cm,clip]{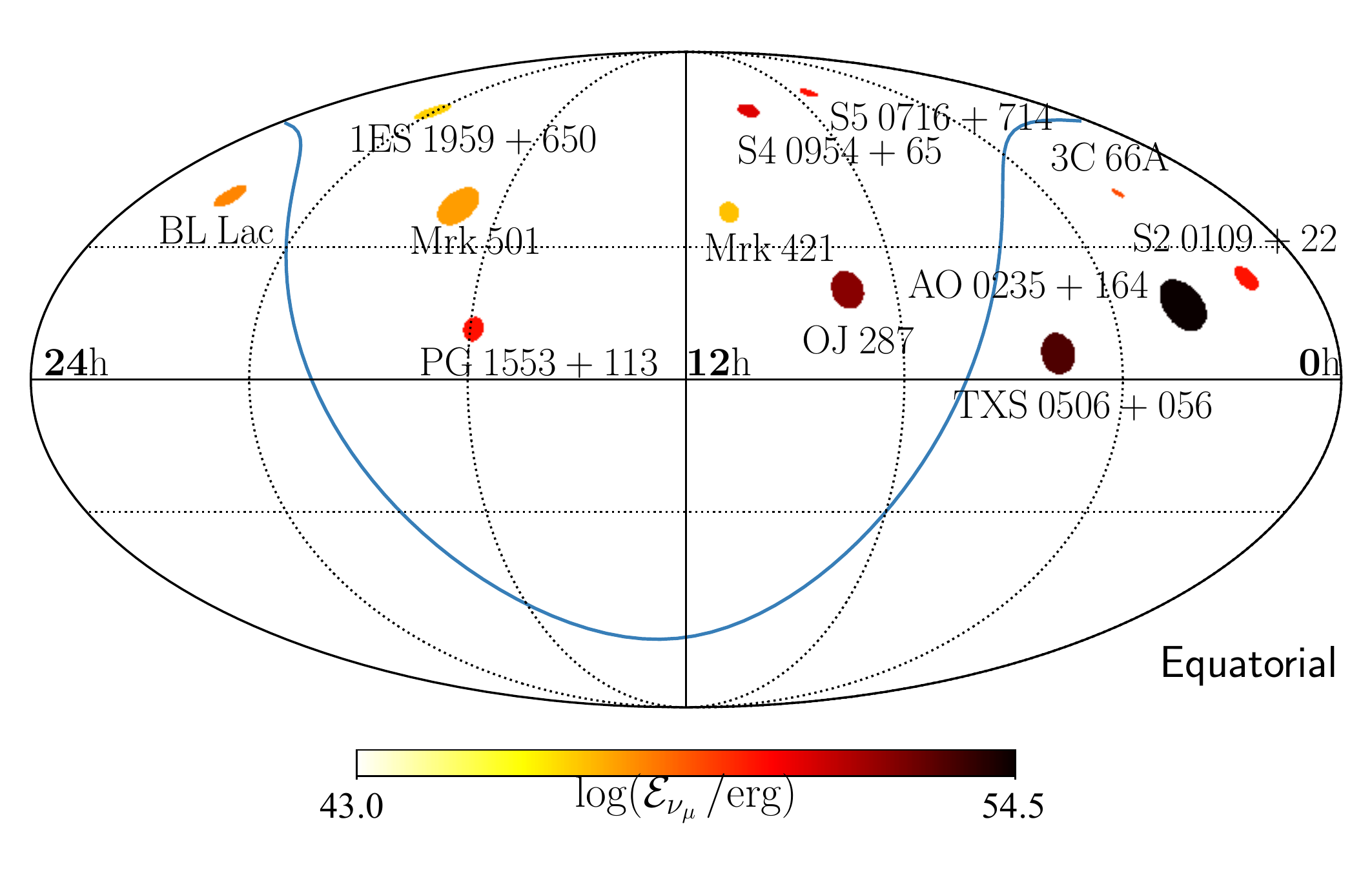}
\includegraphics[width=7.85cm,clip]{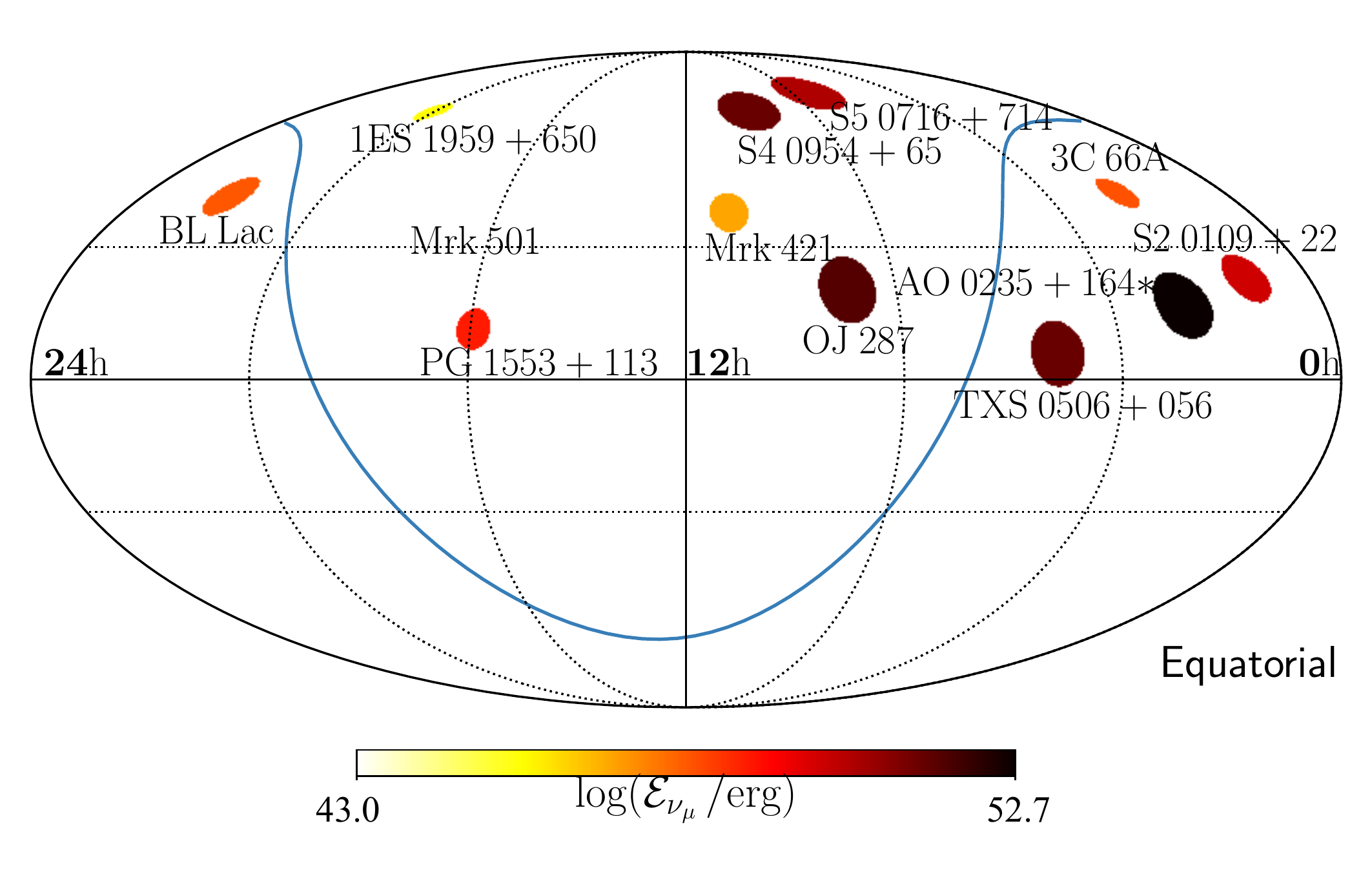}
\caption{Left: Predicted per-flavour neutrino energy output during the
  studied flare with Normalisation A. The size of the circles is proportional to
  the number of neutrinos expected in each flare. For flares that
  occurred prior to 2011, the appropriate incomplete IceCube
  configuration is taken into account. The blue solid line marks the
  Galactic plane. Right: Same as on the left but for Normalisation B. $^\star$The
   star marks that for \AO we have used baryon loading ten times lower than all other sources, $\xi_{\rm cr} = 150$. \label{fig:map}}
\end{figure*}    

Figure \ref{fig:map} shows a sky map of the energy released in muon
neutrinos, $\mathcal{E}^{44, \mathrm{> 100\,TeV}}_{\nu_{\mu},\rm /flare}$, during each of the flares in our
sample under the assumptions of Normalisations A and B, in equatorial
projection. The size of the circles is proportional to the number of
neutrinos expected in each flare. Comparison of the same source across two plots, reveals that in some
cases the expected energy flux is larger assuming Normalisation A while more
neutrinos are expected from Normalisation B. When this is the case, it is
because the $E_{\nu} F_{E_{\nu}}$ neutrino energy flux with Normalisation A continues to
increase up to higher energies as illustrated in Figure~\ref{fig:FITSA}, 
but the expected number of neutrinos depends on the
$F_{E_{\nu}}$ flux, which peaks at lower energies for some of the sources.

The bulk of the neutrino flux from blazar flares in our formalism is expected at energies 
larger than 100 TeV where the atmospheric neutrino background is low~\citep{Aartsen:2016xlq}. 
Constraining the search for blazar-flare neutrinos to energies greater than 100 TeV, where the search 
is approximately background free, a future neutrino detector can constrain the baryon-loading factor of the jet to be, 
\be
\xi_{\rm cr} \leq \frac{N_{\rm FC}^{90} \cdot \xi_0}{\sum_{\rm source~i}{N_{\nu_{\mu}}^{i}}},
\label{eq:xilim}
\ee
where $N_{\rm FC}^{90} = 2.44$ is the Feldman-Cousins upper limit at $90\%$
confidence, in the absence of neutrino signal, for a background free search, and $\xi_0$ the assumed baryon loading factor to obtain $N_{\nu_{\mu}}^i$, the number of expected muon neutrinos from flaring periods for the $i$th source.

In our conservative model (Normalisation A), the flare-only neutrino signal is too
low to be constrained by Gen2+GVD+KM3NeT+ONC or similar detectors, and
time-integrated searches for neutrino signal from BL Lacs are
preferable.

Assuming Normalisation B, for the two most powerful neutrino emitters in our sample, 
\AO and \OJ (scaled down in the case of \AO, with $\xi_{\rm cr} = 150$ to bring 
the cascade to a non-detectable level), a future neutrino network with effective volume ten times 
larger than IC86, (denoted $\rm IC \times 10$) with the exposure
of Gen2+GVD+KM3NeT+ONC would detect $N^{\rm IC \times 10, 10~{\rm yr}}_{\nu_{\mu}, \rm all~flares, > 100~TeV}\,\approx~3$ 
with our flare-period selection based on the HE FAVA data. In the absence of the muon-neutrino signal, 
assuming a flare pattern identical to that if the past decade, the baryon loading will be constrained
to $\xi_{\rm cr} < 1250$ if the energy density of the external photon target field can be probed. 
Aggressively reducing the energy threshold to 40 TeV, which would bring the background up 
to an acceptable level of $N_{\rm bg} \approx 0.2$, assuming the background model 
of~\citet{Aartsen:2016xlq}, we would expect $N^{\rm IC \times 10, 10~{\rm yr}}_{\nu_{\mu}, \rm all~flares, > 40~TeV} \approx 5 $ or limit $\xi_{\rm cr} \lesssim 835$. 

For a larger number of such powerful sources the constraint on $\xi_{\rm cr}$ 
scales down linearly with the number of expected neutrinos following Equation~\ref{eq:xilim}. 
To illustrate the possible reach of future instruments we use the 34 BL Lac objects 
extracted from the 1-Jy catalogue~\citep{1991ApJ...374..431S,1981A&AS...45..367K}. 
For these sources, which are predominantly LSPs, the ``bolometric'' luminosity of
the synchrotron peak, $L_{S, \rm bb}$, scales approximately as their radio luminosity.~\footnote{This is the case for
LSP objects because the peak of the synchrotron emission lies in the frequency range $10^{12}-10^{14}~{\rm Hz}$. 
Given this small range, one can extrapolate almost linearly between the radio flux and the flux at the synchrotron peak}. 
We make use of this approximate relation. Using the radio flux at 5~GHz, $F_{\mathrm{5\,GHz},i}$,
we find that for an assumed scaling of the neutrino luminosity as $L_{\nu} \propto L_{S, \rm bb}^{\alpha}$, each source, $i$, would produce of order, 
\be
N^{i}_{\nu_{\mu}} \sim N^{\rm \OJ}_{\nu_{\mu}} \left( \frac{F_{\mathrm{5\,GHz},i}}{F_{\mathrm{5\,GHz,\OJ}}} \right)^{\alpha}  \frac{A_{\mathrm{eff},i}}{A_{\rm eff,\OJ}}
\ee
muon neutrinos if it had an identical flare profile and neutrino spectrum to that of \OJ. 
In 10 years of observations with Gen2+ONC+GVD+KM3NeT, for $\alpha = 2$ which we have assumed 
throughout this work, the stacked number of expected neutrinos from the 1-Jy BL Lacs would be 
$N^{\rm IC \times 10, 10~{\rm yr}}_{\nu_{\mu}, \rm all~flares, > 100~TeV}\,\approx~22$ or otherwise 
constrain the baryon loading factor of these sources to 
$\xi_{\rm cr} < 170$. The limits on $\xi_{\rm cr}$ can be improved by increasing the number 
of sources and/or by stacking longer duration flares. 
Meanwhile, X-ray and proposed MeV \gRay telescopes will provide complementary 
measurements or constraints to the baryon loading factor of blazars by monitoring 
the part of the SED where the secondary cascade emission is expected. 

It is interesting to note that the FAVA lightcurve of \AO shows
another $\sim$year-long flare in 2015. The IceCube data for this time
are not publicly available and could not be examined. It has not escaped our attention that
the Very high-energy Gamma-ray follow-up (GFU) program of IceCube
events registered a neutrino triggered alert in April 2015 from the
direction of \AO~\citep{Aartsen:2016qbu}. The GFU alerts are produced only 
in response to an excess of neutrinos in the direction of the source and are 
by construction independent of any electromagnetic flaring activity. 
 The time of the alert with
respect to the FAVA lightcurve is illustrated in Figure~\ref{fig:Lightcurves}. 
\AO is one of the $\sim 100$ monitored sources within
the GFU program, and the alert was triggered by the observation of
eight neutrinos within 16.4 days. An alert with such significance is
expected at a rate $\sim 0.01$/source/year, and is thus consistent with
background expectations, even after accounting for the independent
observation that it was flaring in the FAVA analysis. 

We have modelled the FAVA flare as a double gaussian 
with duration $\approx444$~days. Assuming that the SED of \AO was in the 
same state on MJD 57427, which is when the source reached peak flux in the
FAVA data, 
 as it had been on MJD 54736, which is our fitted peak of the 2008 flare, 
 we can obtain an estimate of the neutrino 
 signal from this flare, following the procedure outlined in Section~\ref{subsec:NeutrinoCounts},
 which turns out to be $N_{\nu_{\mu},>100~\rm TeV} \approx 4.\cdot 10^{-4} (\xi_{\rm cr}/10)$ 
 with Normalisation A, and $N_{\nu_{\mu},>100~\rm TeV} \approx 0.05 (\xi_{\rm cr}/150)$ with Normalisation B. 

\subsection{Systematic uncertainties}
\label{subsec:systematics}

The expected neutrino flux from a blazar flare is extremely sensitive
to the Doppler factor of the motion of the emitting region of the
blazar jet. As demonstrated in Section \ref{sec:method}, in the blob
formalism the observed neutrino luminosity is amplified by a factor of
$\delta^4$ with respect to the emitted neutrino luminosity.  The
Doppler factor of the motion of the emitting region can be constrained
by blazar observables such as the time variability, and the
requirement that the source is not opaque to emitted \gRays, but the
most detailed measurements of jet kinematics come from radio galaxy
observations~\citep{2017ApJ...846...98J,Lister:2019ttx}. In the SSC
scenario, additionally, the magnetic field can be uniquely specified
for a given value of $\delta$ in the absence of measurement
uncertainties. The value of $B$ relates to the maximum energy of
protons, and determines the balance between cooling timescales in the
emitting region and thus also the shape of the emerging neutrino
spectrum.

At present, there exist large uncertainties in the determination of
the Doppler factor, Lorentz factor, and viewing angle of individual
blazar jets, though future jet observations may narrow down the
allowed parameter region. Broadly speaking the bulk of BL Lac jets are
expected to have Doppler factors in the range $10 \leq \delta \leq 50$
\citep{Hovatta09,2015MNRAS.454.1767L}. Typically, for an individual
blazar SED an uncertainty range can only be quoted within an assumed
model for the emission from the source in question. The addition of
free parameters, including more than one emitting region and external
photon fields, makes it even more challenging to specify $\delta$ and
$B$.

A final parameter of importance for the expected number of neutrinos,
once $\delta$ and $B$ are specified, is the baryon loading factor
$\xi_{\rm cr}$. This can be constrained with neutrino or X-ray observations once
$\delta$ and $B$ are measured as detailed in the previous section. In what follows we explore the effect
of systematic uncertainties on $\delta$ and $B$ on the expected number
of neutrinos from individual blazar flares for fixed $\xi_{\rm cr} = 10$, and
other model parameters fixed as per Normalisation A.

\begin{figure}
 \includegraphics[width=9cm,clip,rviewport=0 0 1 1]{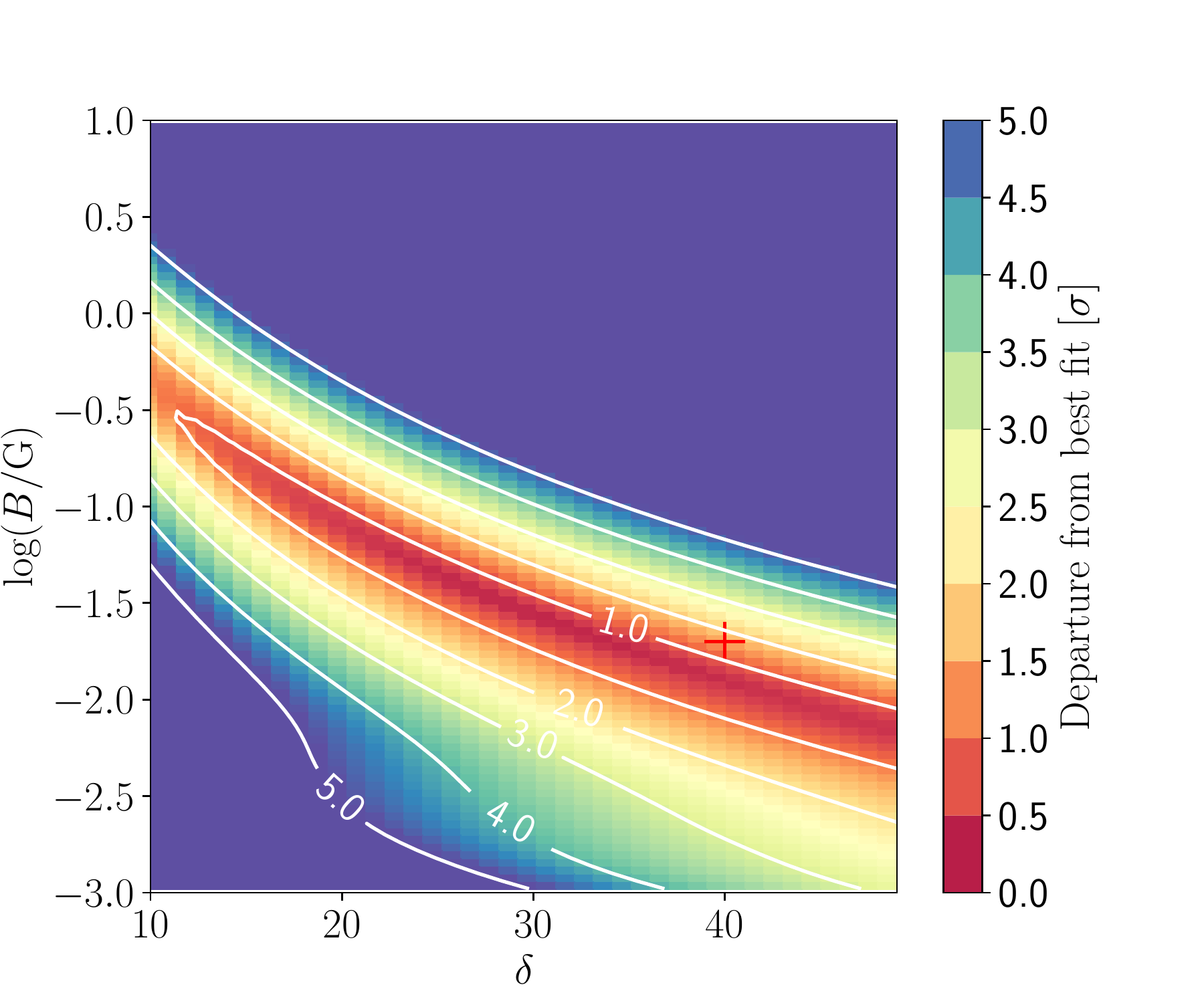}	
  \includegraphics[width=9cm,clip,rviewport=0 0 1 1]{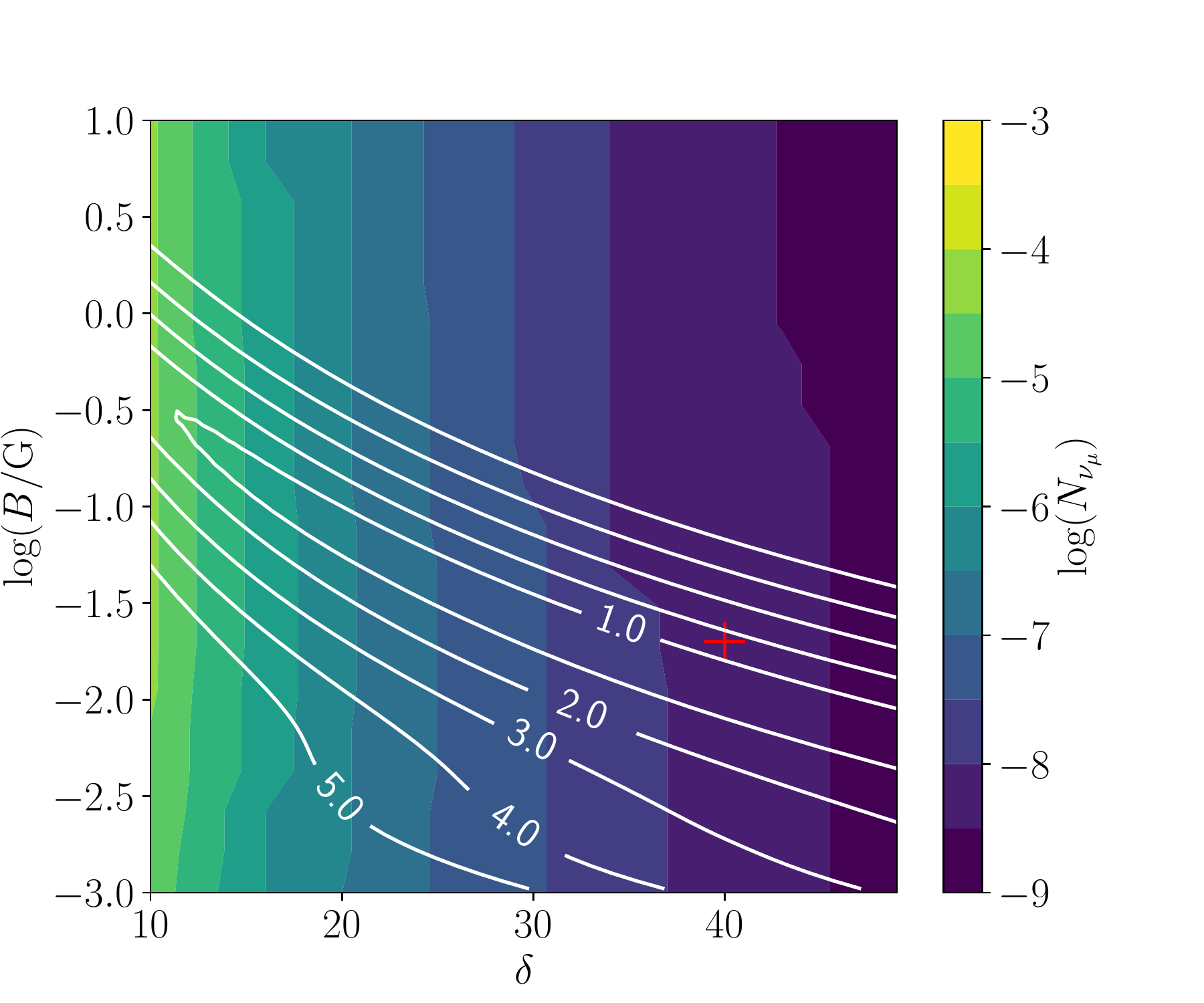}	
 \caption{Top: Systematic uncertainty on the determination of $\delta$
   and $B$ (in Gauss) during the 14-day, October 2008 flare of \ThreeC. The colormap
   gives the departure from the best fit in units of $\sigma$ (see main text for details),
   i.e. realisations with departure $\leq 2 \sigma$ encompass the $95\%$
   allowed region assuming a Gaussian distribution. Bottom: Number of
   muon neutrinos expected to be seen in
   IceCube (IC86) for each combination of $\delta$ and $B$. The red cross (in top and bottom panels) marks the values of $\delta$ and $B$ assumed in
   the present study.
 \label{fig:systematics}}
 \end{figure}

The upper panel of Figure~\ref{fig:systematics} illustrates the
parameter space allowed for $\delta$ and $B$ from the SED of \ThreeC
during its 2008 flare.  For each investigated combination of $\delta$
and $B$ we fit the SED of \ThreeC using Equations~\ref{eq:DopplerFactor} 
and~\ref{eq:B_SSC}, for fixed values of
$\delta$, $B$ and $t_{\rm var,d}$. We perform a numerical scan over $\delta$ and
$B$, calculating the $\chi^2$ of every fit.The SED fits
are done with the log-parabolic model presented in Section~\ref{subsec:SEDFitting}. 
The colormap on the upper panel of
Figure~\ref{fig:systematics} gives the deviation of the fitted SED
with respect to the best-fit SED in units of,
\begin{equation}
n \sigma = S \sqrt{\chi^2 - \chi^2_\text{min}}
\label{eq:significance}
\end{equation} 
with $\chi_{\rm min}^2$ the $\chi^2$ of the best-fit realisation. 
The scale factor $S~=~1/\sqrt{\chi^2_\text{min}/\text{ndf}}$, 
where ndf the number of free parameters of the fit, is an approximate correction~\citep{Rosenfeld:1975fy} 
to enlarge the uncertainty because of a poor minimum $\chi^2$ that might either signify
 underestimated experimental uncertainties or simplified model assumptions.
 Equation~\ref{eq:significance}, defines the confidence regions illustrated in
Figure~\ref{fig:systematics} in units of standard deviations. The
precise meaning of the plotted confidence regions is that the
$1\sigma$ contour gives the $68.3\%$ containment region of $\delta$ at
a fixed value of $B$ and vice versa, assuming that the two variables
are normally distributed~\citep{James:2006zz}. The contour lines in
Figure \ref{fig:systematics} give the number of standard-deviations
with respect to the best-fit SED. The red cross gives the literature value of $\delta$ and $B$ assumed
throughout the rest of this work, as summarised in Table
\ref{tab:deltaB}.

The lower panel of Figure~\ref{fig:systematics} gives the expected
number of muon neutrinos from the flare of \ThreeC, with Normalisation A, while
scanning over $\delta$ and $B$, integrated over neutrino
energies above 100~TeV. We fix the SED here to the best-fit SED, and vary only
$\delta$ and $B$ i.e. we do not simultaneously investigate the effect
of the variation of the shape of the target photon field, which we
expect to be subdominant.

\begin{figure}
 \includegraphics[width=9cm,clip,rviewport=0 0 1 1]{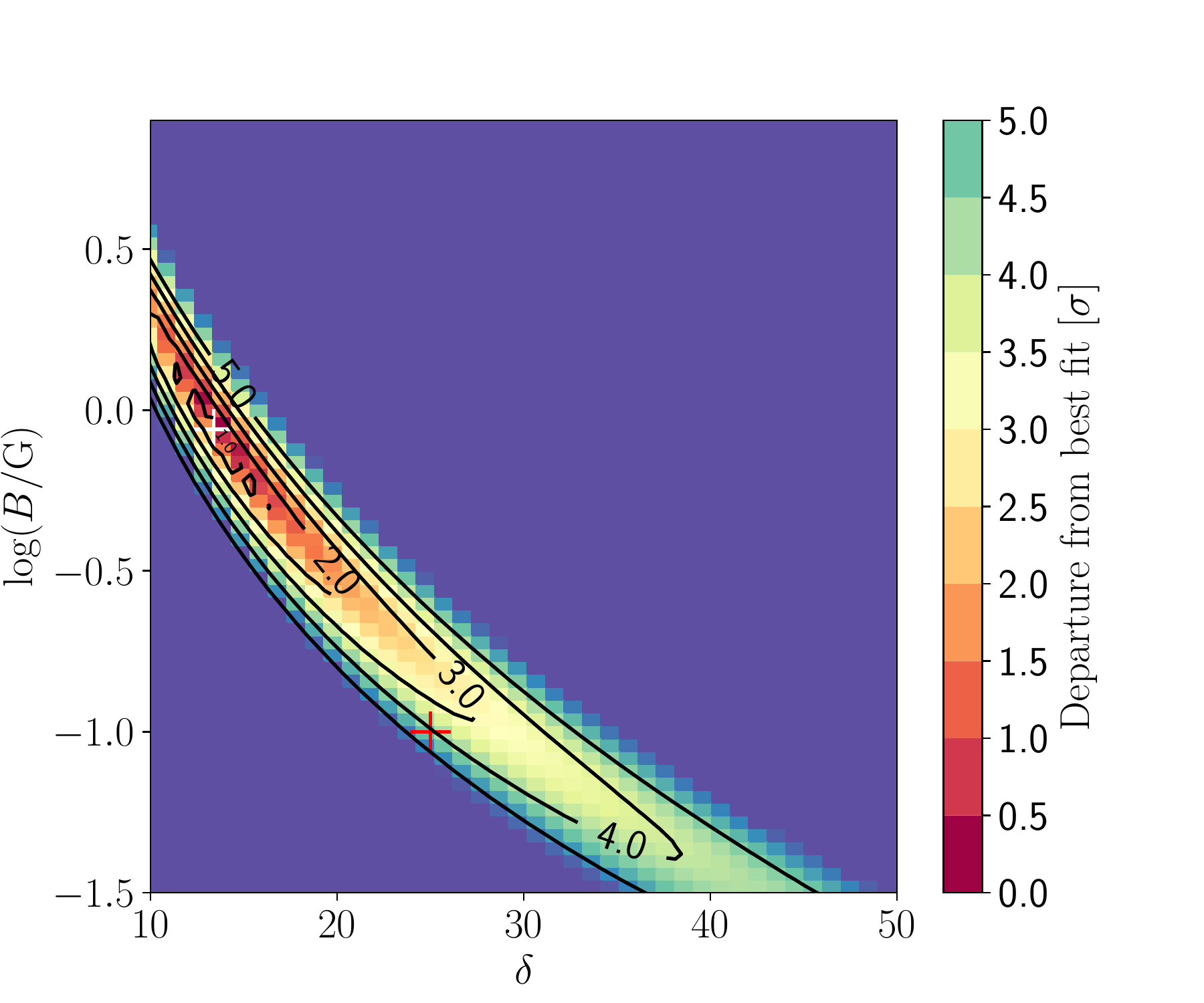}	
  \includegraphics[width=9cm,clip,rviewport=0 0 1 1]{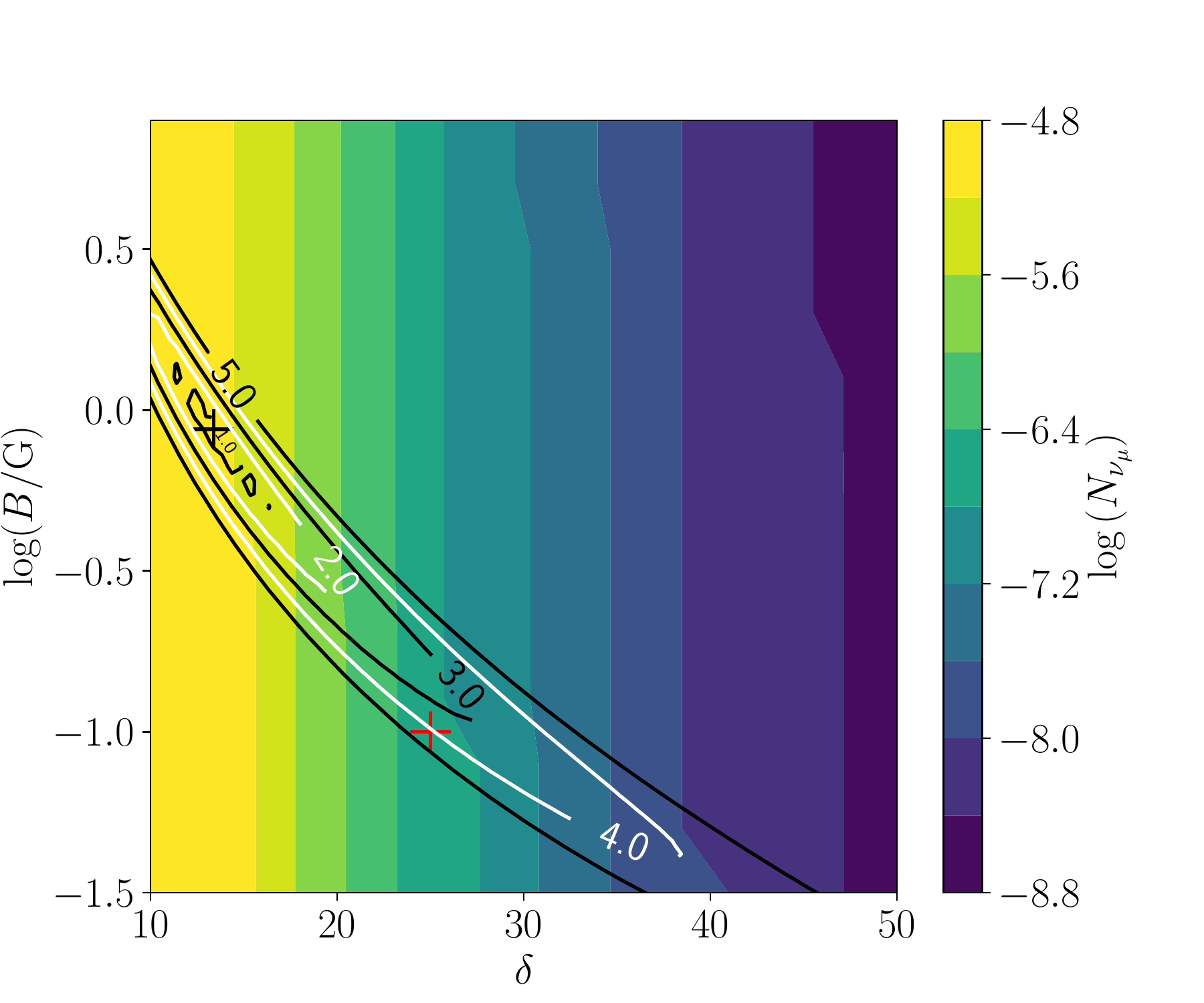}	
 \caption{Same as Figure \ref{fig:systematics} but for the 14-day January 2015 flare of \Sfive. The white (black) cross marks the values of $\delta$ and $B$ that give the best-fit $\chi^2$ in the top (bottom) panel.
  \label{fig:systematicsS5}}
 \end{figure}

\begin{figure}
 \includegraphics[width=9cm,clip,rviewport=0 0 1 1]{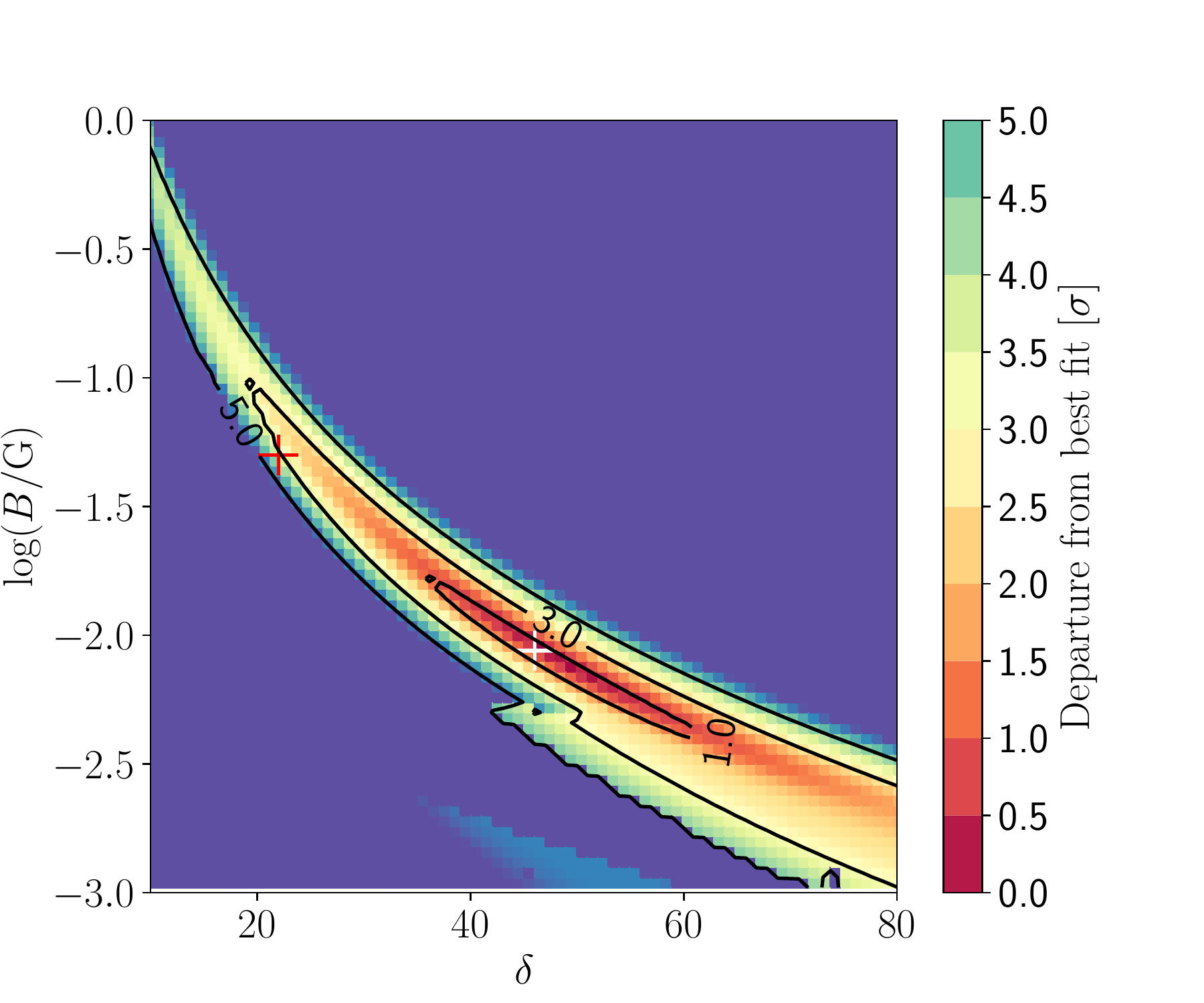}	
  \includegraphics[width=9cm,clip,rviewport=0 0 1 1]{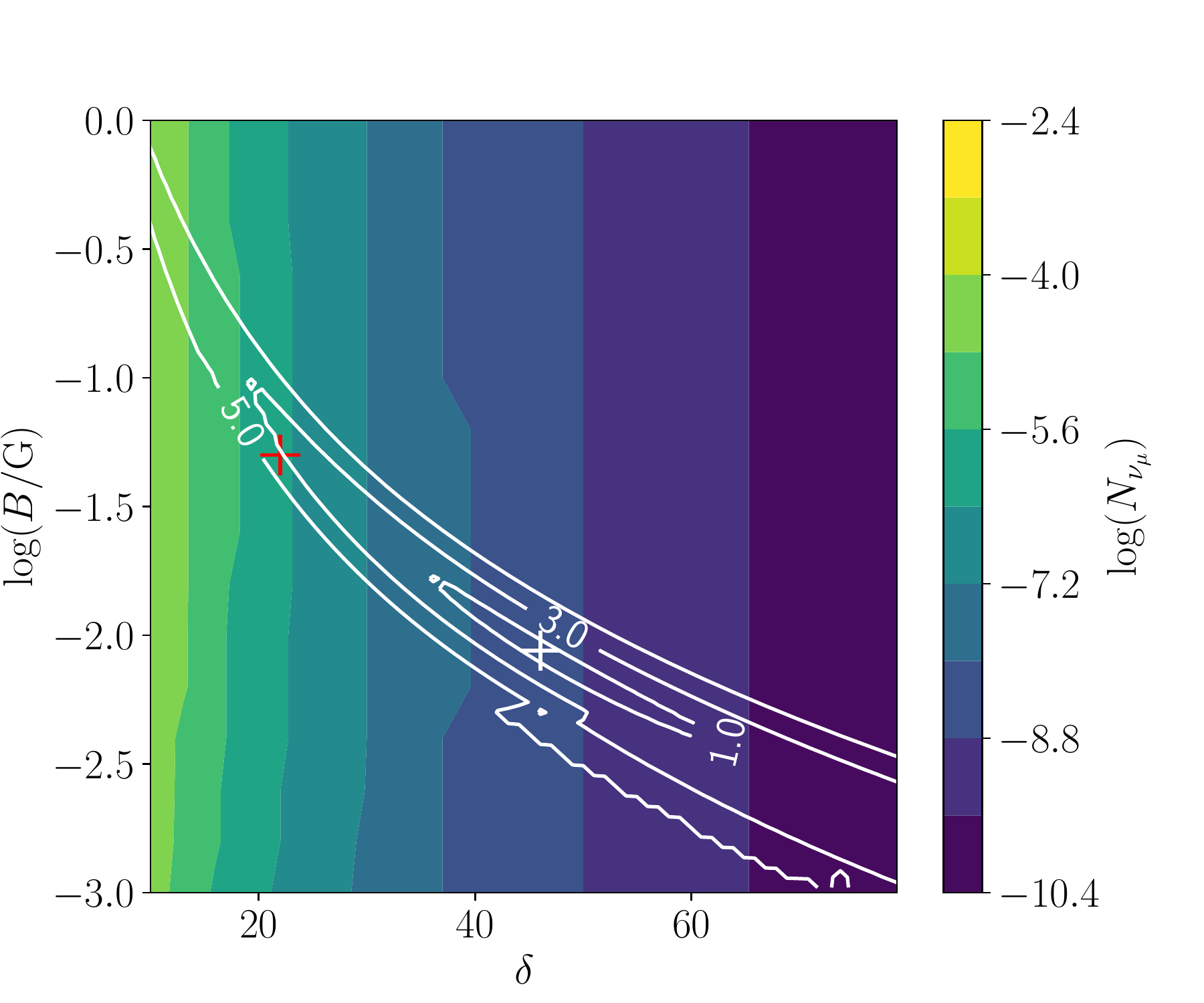}	
 \caption{Same as Figure \ref{fig:systematics} but for the 21-day July 2015 flare of \Stwo. The white cross marks the values of $\delta$ and $B$ that give the best-fit $\chi^2$.
  \label{fig:systematicsS2}}
 \end{figure}

Note that unlike in all earlier sections we have not calculated the
number of expected neutrinos using Equation \ref{eq:NnuT} but instead
use the box approximation, \ie,
\begin{equation}
N_{\nu} = \frac{\mathrm{d}N_{\nu, t_0}}{\mathrm{d} t} \cdot \Delta T, 
\end{equation}
with $\mathrm{d}N_{\nu, t_0}/\mathrm{d} t$ given by Equation \ref{eq:Nnu}, and $\Delta T$ the time duration of the \Fermi flare
given in Table~\ref{tab:timescales}. Figures \ref{fig:systematicsS5}, and \ref{fig:systematicsS2} give the
same results but for the flares of \Sfive and \Stwo.

Within the $1\sigma$ error-region of \ThreeC which has a long and shallow minimum
the neutrino expectation varies by more than four orders of magnitude. For this source the best $\chi^2$ 
in the log-parabolic model is outside the figure bounds at $\delta = 93$ and $B = 0.001$~G. 
For \Sfive the literature value estimate is about $\sim4\sigma$ away from
the best-fit estimate obtained with the log-parabolic model, and in
this case the expected number of neutrinos in these two models varies by
$\sim 1.5$ orders of magnitude. For \Stwo the literature value falls
$3\sigma$ away from the best-fit log-parabolic parameters. In this
case also the expected number of neutrinos in the two models varies by
$\sim 1.5$ orders of magnitude. It is important to note that
additional constraints beyond the shape of the SED investigated here
can be imposed to exclude certain regions of the parameter space shown
in these plots; for example, rapid variability may impose a limit on
the Doppler factor of the source on a case by case basis. Here, our
aim is simply to illustrate the strong dependence of the neutrino flux
expectations on the physical conditions inside the source and the
model dependence of the derived values of these parameters.

\section{Discussion and Conclusions}
\label{sec:discussion}

We have studied the expected number of neutrinos from individual
blazar flares, focusing on BL Lac objects. We presented the expected 
number of neutrinos from past flares from twelve sources that were detectable by IceCube. 
We employed a model based on the SSC formalism to keep the number of free parameters minimal.  
We considered a broad range of the allowed parameter space 
for the the efficiency of proton acceleration in BL Lac jets, 
the jet baryon loading, and the presence of external fields. To model 
the expected neutrino fluence of each of the flares from the twelve sources 
studied we used simultaneous multi-wavelength observations of the SEDs of the sources. 
We used the FAVA lightcurves to homogeneously define the duration of the flares. 

We calculated the expected number of high-energy neutrinos from each of the flares, 
under conservative and optimistic assumptions, 
with the aim to address the question of what are the characteristics that make
a flare particularly strong in terms of expected neutrino counts, and
whether the origin of neutrinos in \gRay bright blazar flares,
which is suggested by the 2017 observations of \TXS, is further
testable with future neutrino observations. 

We found that, in general, the largest neutrino signal is expected from the flares
of the most luminous sources, as long as they have moderate Doppler factors. 
In this sense, the neutrino flux is dominated by the most powerful
sources/flares, and thus given the rarity of BL~Lac objects the expected
neutrino signal in stacked searches does not increase significantly by including faint sources/flares.  
 
In the context of the SSC model, under conservative assumptions about the baryon loading
of BL Lac jets, $\xi_{\rm cr} \sim$10, which corresponds to the average jet baryon loading 
expected if the diffuse UHECR flux is powered by blazars, 
and assuming the only available target photon field for neutrino interactions 
is the photon-field comoving with the emitting region, 
the expected number of muon neutrinos in IceCube per flare is low, 
$N_{\nu_{\mu},\rm > 100 TeV} \lesssim 5 \cdot 10^{-4}$. 
The stacked signal over ten years of flare-periods for the most powerful sources in our sample
 is also low $ N^{\mathrm{IC,10 yr}}_{\nu_{\mu},{\mathrm{100 TeV}}} \lesssim 10^{-3}$. 
 If such a mechanism is exclusively in operation, we do not expect a strong neutrino signal in stacked 
 searches for neutrinos from blazar \gRay flares even with an order of magnitude larger neutrino detector. 

Under optimistic assumptions about the jet proton luminosity and in the presence of external photon fields, 
inspired by the modelling of the 2017 flare of \TXS, we found that the two most powerful sources
 in our sample, \AO and \OJ would produce, in total, $N^{\mathrm{IC \times 10,10 yr}}_{\nu_{\mu}, \rm all~flares, > 100~TeV} \approx 3$ muon neutrinos, 
 during \Fermi flaring periods, in future neutrino detectors with total instrumented volume $\sim$ten times larger than IceCube,
 or otherwise lead to a constraint on the proton luminosity of blazar jets, $\xi_{\rm cr} \lesssim 1250$. 
 Stronger constraints on the baryon loading can be obtained with the addition of more powerful sources 
 or by lowering the energy threshold of the neutrino search at the cost of additional atmospheric neutrino background.
We illustrated the expected reach of constraints on the jet proton luminosity with a stacking approach using the 1-Jy BL Lac sample. 
In all instances, for such a constraint, the energy density of the external field, Doppler factor, and magnetic field strength in the emitting 
region need to be independently determined. 

We demonstrated that the main uncertainties in predicting the neutrino 
output from individual flares are the Doppler factor of the motion of the jet emitting region, 
the magnetic field strength, the presence of external target photon fields, and the baryon content of the jet. 
We studied the systematic uncertainty in the expected neutrino flux from individual blazar flares introduced by the uncertainty of $\delta$ and $B$. 
We found that $\delta$ and $B$ can vary a significant amount, even in the relatively well constrained SSC model, and within this uncertainty range 
the expected number of neutrinos also varied by up to $\sim 2$ orders of magnitude for the flares that we studied. 
The Doppler factor, magnetic field strength, and presence of external photon fields, can be 
constrained by radio, $\gamma$-ray and other astronomical observations. With these parameters specified, 
future neutrino detectors will be able to constrain or measure the baryon loading of blazar jets. Thus, astronomical observations are of paramount importance for understanding neutrino production in blazars.  

Further to the two benchmark models we studied, and the systematic uncertainty on $\delta$ and B, there are two ways to increase the 
neutrino production in the studied flares: 
\begin{enumerate} 
\item increase the baryon loading
\item increase the energy density of the stationary field. 
\end{enumerate} 

In general, the baryon loading cannot be increased arbitrarily because of the associated electromagnetic cascade
 emission expected from the interactions of the protons, which would alter the shape of the observed SED. 
 For some of the SEDs we studied there is room to increase the baryon loading with respect to our optimistic benchmark 
 Normalisation B, but blazars with such strong baryon loading are not expected to be typical. 
 Firstly, the radiative feature that would be produced in the $\sim$keV-MeV energy range of the SED 
 (as in e.g. Figure 1 of~\citealp{petro_2015}) is not broadly observed. The strong upper limits on the baryon loading factor on this source-by-source basis, were demonstrated extensively during the follow-up campaign of the 2017 flare of \TXS.
Additionally, if all blazar jets had such strong baryon loading, they would overproduce the observed cosmic ray spectrum and the diffuse \gRay background measured by \Fermi, since very-high and ultra-high-energy cosmic rays produce \gRay cascades during their intergalactic propagation~\citep[e.g.][]{Murase:2012df}. 
  
With regard to item (ii) we have only investigated scenarios where $u'_{\rm syn} > u'_{\rm ext}$. This is not necessarily the case. 
However, for most of our sources, which emit TeV $\gamma$-rays, the external radiation field energy density cannot be much higher
than we investigated here in a one-zone setup, or the opacity to TeV \gRays would be too large to allow the observed \gRays to escape.
An additional constraint on the energy density of any external radiation 
field in BL Lac objects comes from the relative strength and peak frequency of the two bumps 
of the SED~\citep{Tavecchio:2019nvg}. In the case of HSP objects the upper limit on $u'_{\rm ext}$ is often stronger than the constraint imposed by requiring transparency to TeV photons, and generally $u'_{\rm ext} \leq 10 \cdot u'_{\rm syn}$ can be considered a generous upper limit. 
In summary, neutrino production cannot be systematically larger than we have investigated in this work in a one-zone model. 

We did not study FSRQ objects explicitly, though our sample contains sources which are intrinsically FSRQs.  
These will be the subject of subsequent work. Being more powerful, with stronger external fields, 
FSRQs likely can produce more neutrinos per flare for similar values of the baryon loading factor.
On the other hand, since the SSC mechanism is likely not in operation in FSRQs, the number of free parameters
needed to model neutrino production is larger, and the model parameters less well constrained. It is also possible that
there are systematic differences in the proton content of BL Lac and FSRQ jets as shown in e.g.~\citet{Croston:2018ssc}. 

A key ingredient for future neutrino analyses to be able to constrain, or in the case of neutrino detection in coincidence with blazar flares, 
determine the properties of the sources, such as the baryon loading of the the jet, is the availability of a
 simultaneous multi-wavelength observations of the spectral energy distribution at the time of the flare, 
 and determination of the Doppler factor and magnetic field of the jet from precise astronomical observations. 

\section*{Acknowledgements}
We thank Jaime Alvarez-Mu\~niz, Rolf Buehler, Matteo Cerruti, Yu-Ling Chang, Theo
Glauch, Paolo Giommi, Shigeo Kimura, Ioannis Myserlis, Maria
Petropoulou, Andrew Taylor and Michael Unger for useful discussions. This work is 
supported by the Deutsche Forschungsgemeinschaft through grant 
SFB\,1258 ``Neutrinos and Dark Matter in Astro- and Particle Physics'' (F.O., P.P., E.R.),
the Alfred P. Sloan Foundation and NSF grant No. PHY-1620777 (K.M.) 
and the Eberly Foundation (P.M.)

\bibliography{OMPPR19}

\appendix
\section{Flare Definition} 
\label{sec:FAVA}

\begin{figure*}

 \includegraphics[width=8.6cm,clip,rviewport=0 0 1 1]{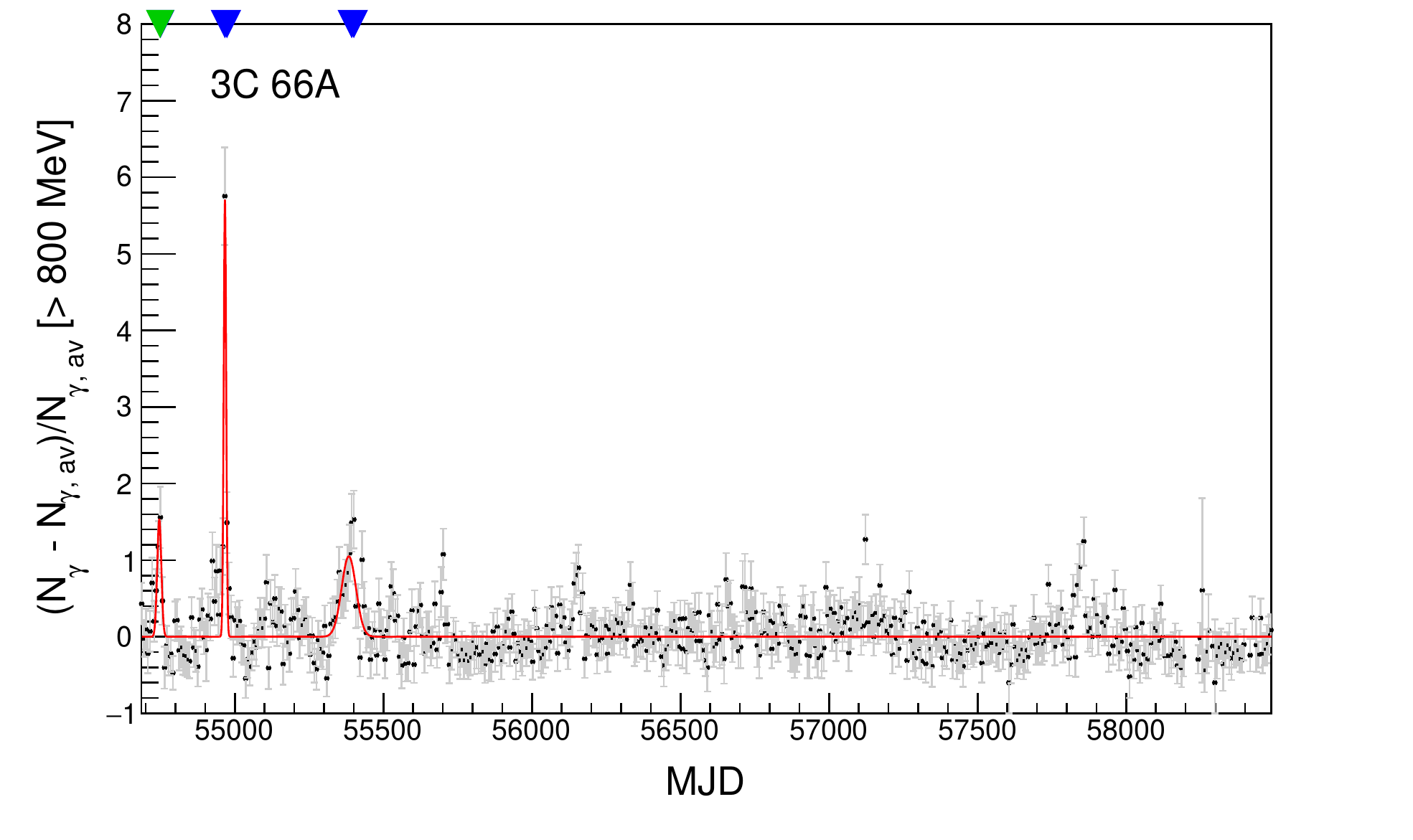}	
  \includegraphics[width=8.6cm,clip,rviewport=0 0 1 1]{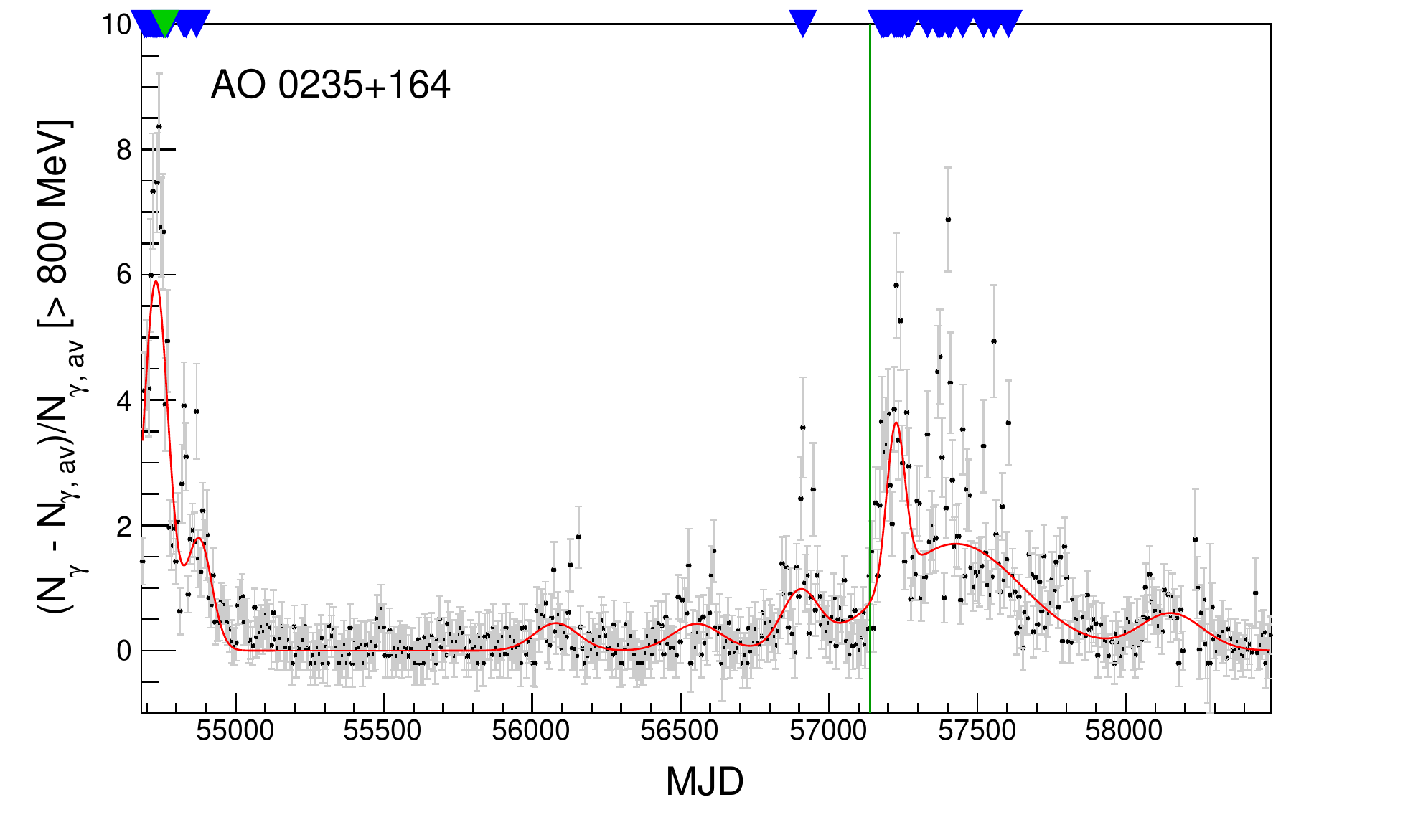}		 
 \includegraphics[width=8.6cm,clip,rviewport=0 0 1 1]{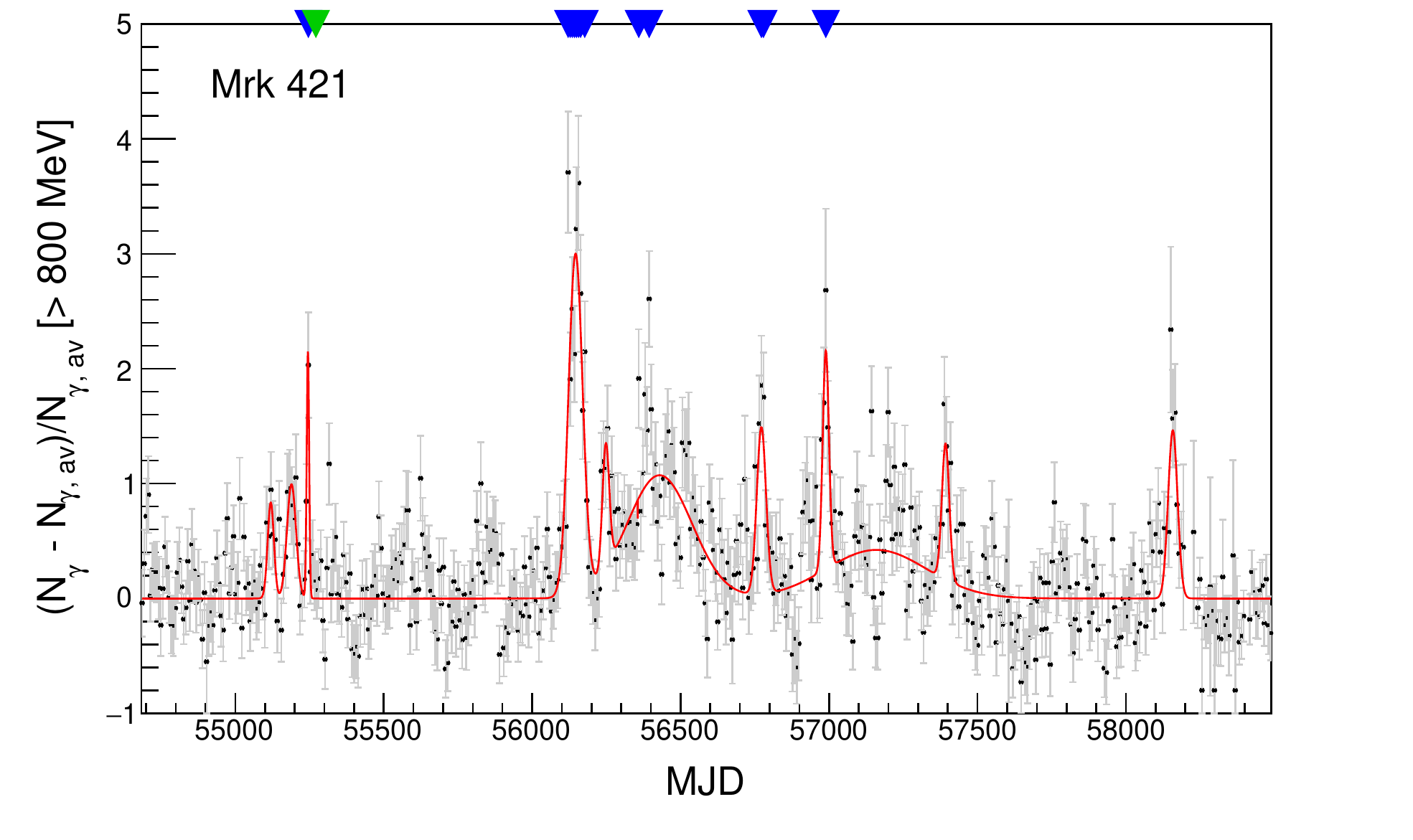}
  \includegraphics[width=8.6cm,clip,rviewport=0 0 1 1]{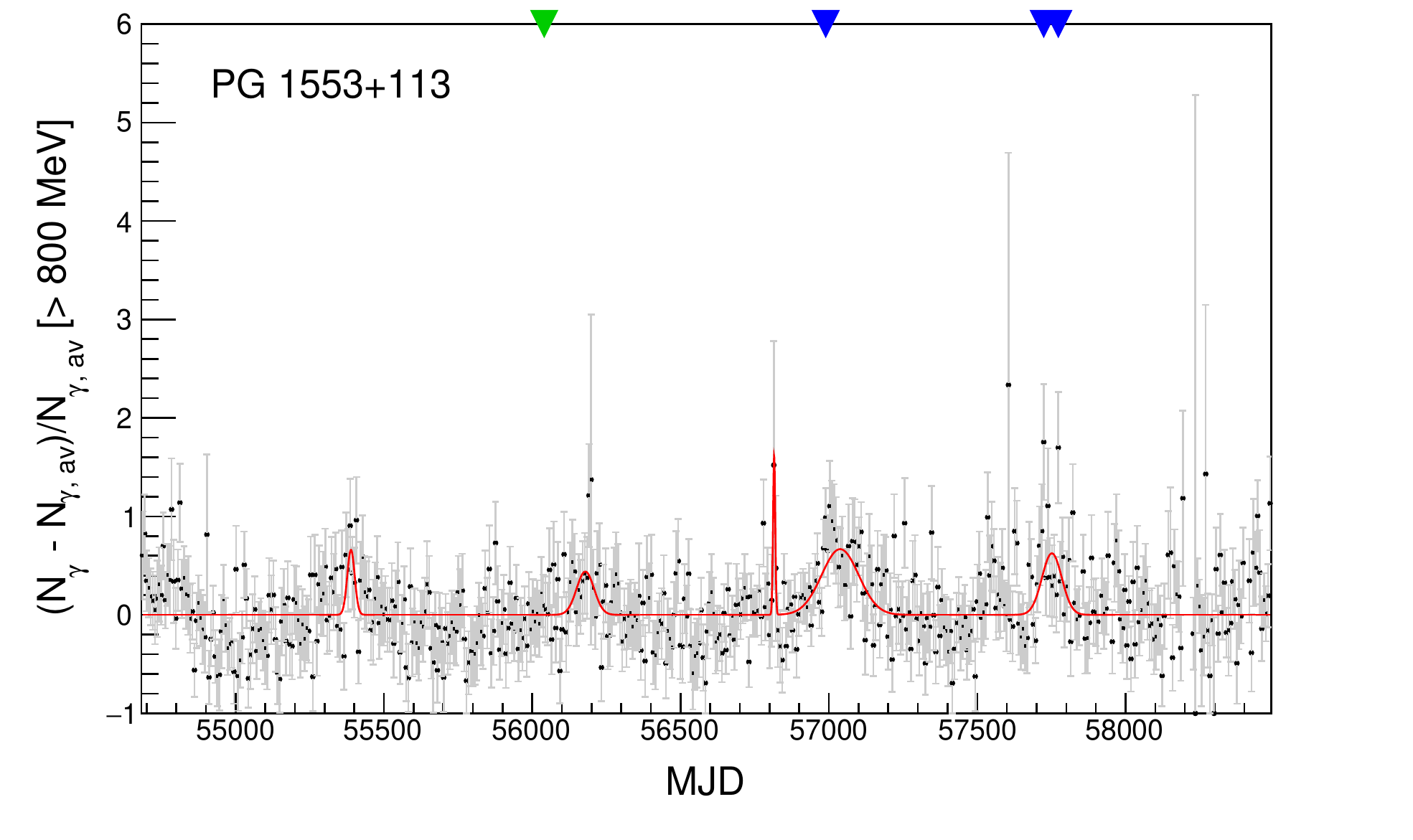}		
 \includegraphics[width=8.6cm,clip,rviewport=0 0 1 1]{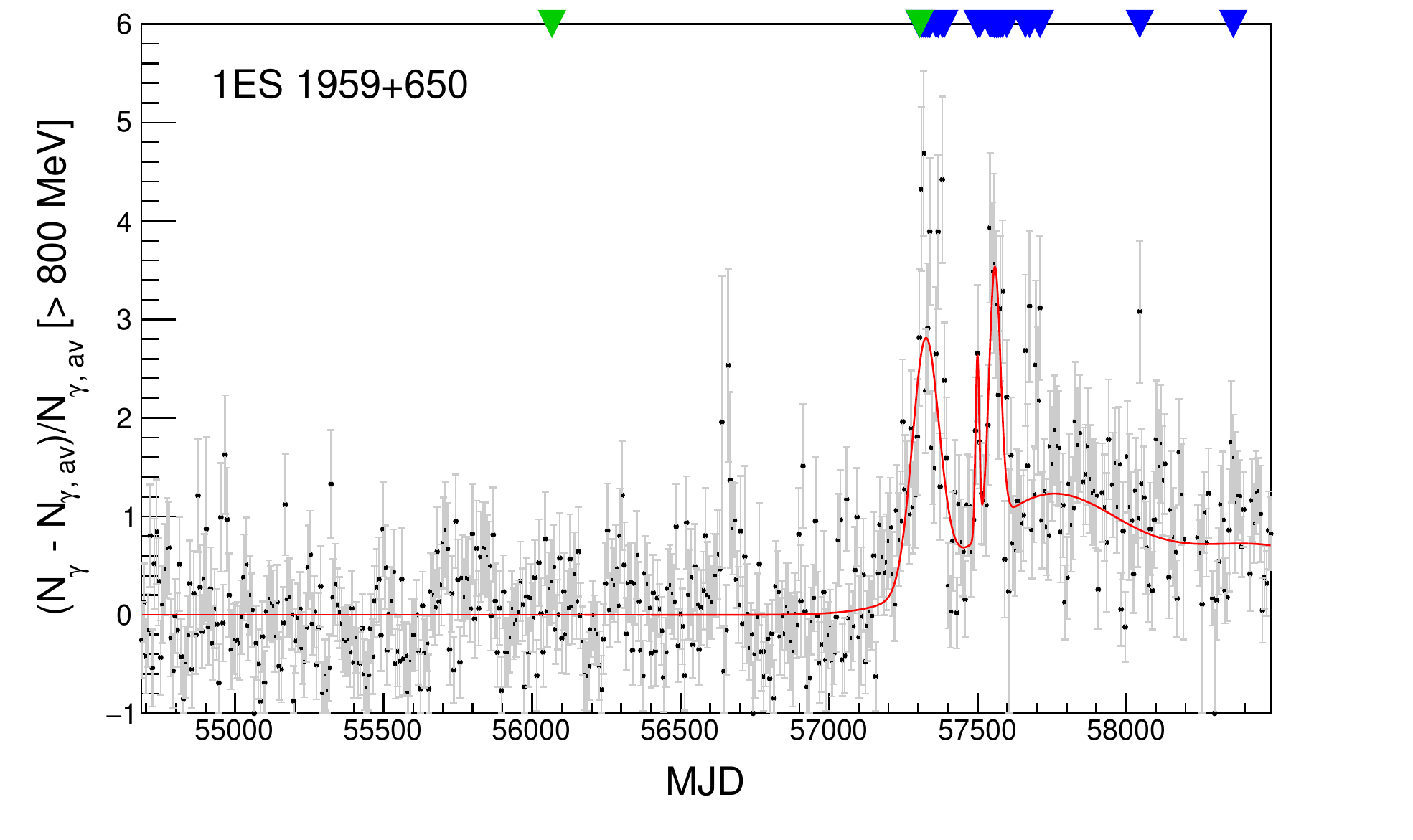}
 \includegraphics[width=8.6cm,clip,rviewport=0 0 1 1]{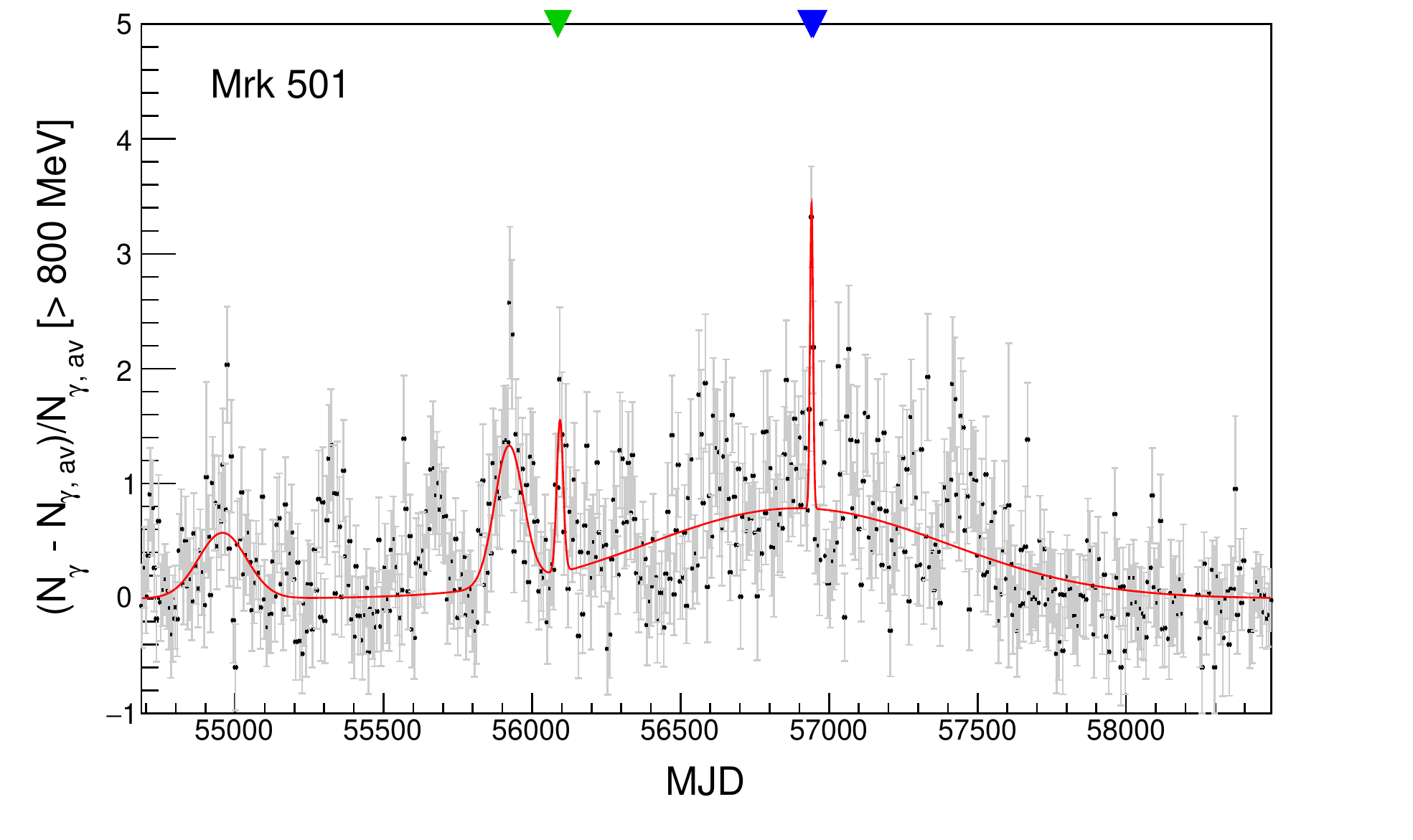} 

 \caption{FAVA lightcurves and fitted flares. Blue triangles mark the times when the
  FAVA lightcurves exhibit a $\geq 5\sigma$ flare in the high-energy FAVA bin. In the case of \PG, which did not exhibit any 
  $\geq 5\sigma$ flares, the $\geq 4\sigma$ flares are shown instead. 
Green triangles mark the time of the flare(s) modelled in the present work.   
For \AO the solid green vertical line gives the time of the GFU neutrino
   alert.~\label{fig:Lightcurves}}
 \end{figure*}

 \begin{figure*}

  \includegraphics[width=8.6cm,clip,rviewport=0 0 1 1]{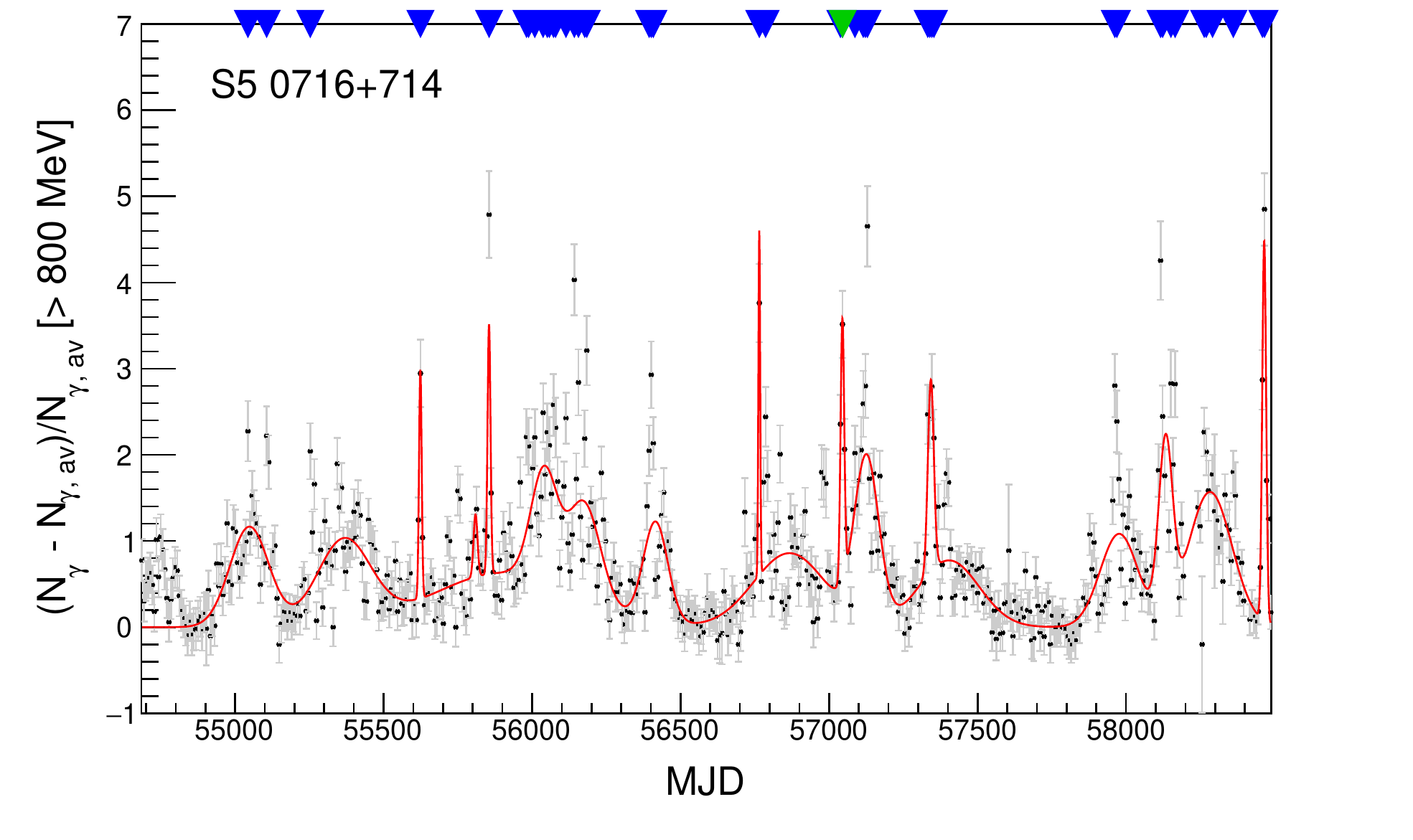} 
 \includegraphics[width=8.6cm,clip,rviewport=0 0 1 1]{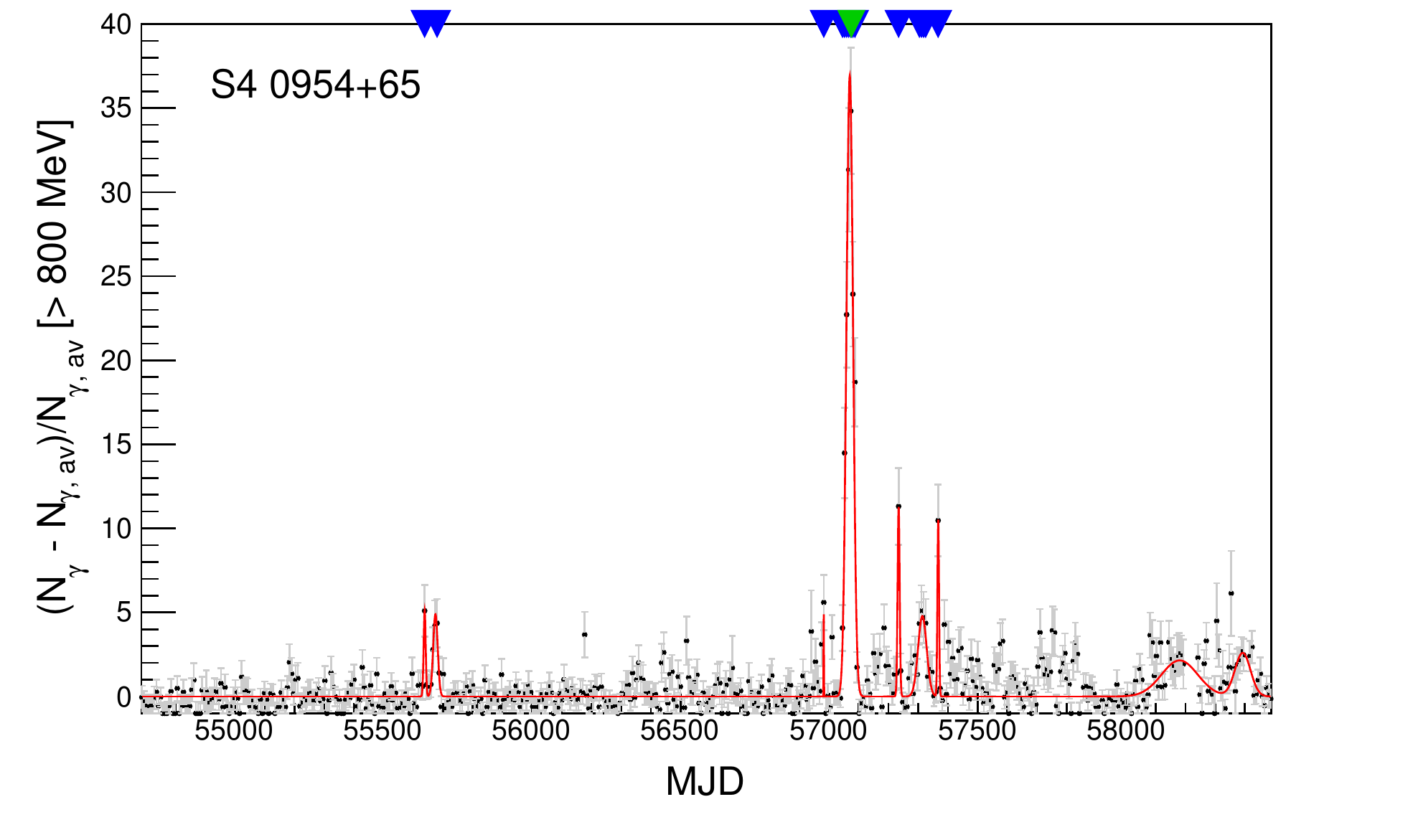} 
 \includegraphics[width=8.6cm,clip,rviewport=0 0 1 1]{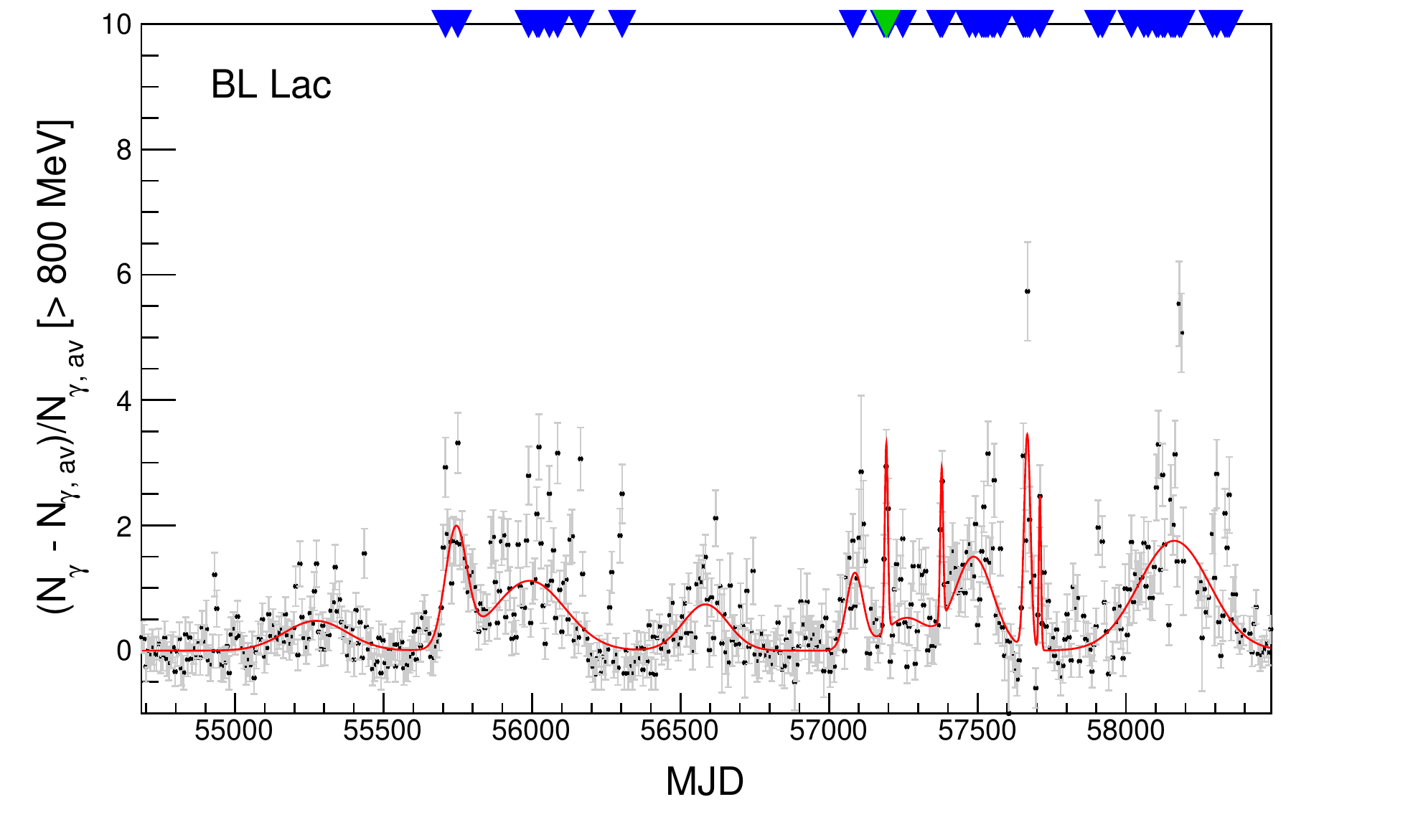}
 \includegraphics[width=8.6cm,clip,rviewport=0 0 1 1]{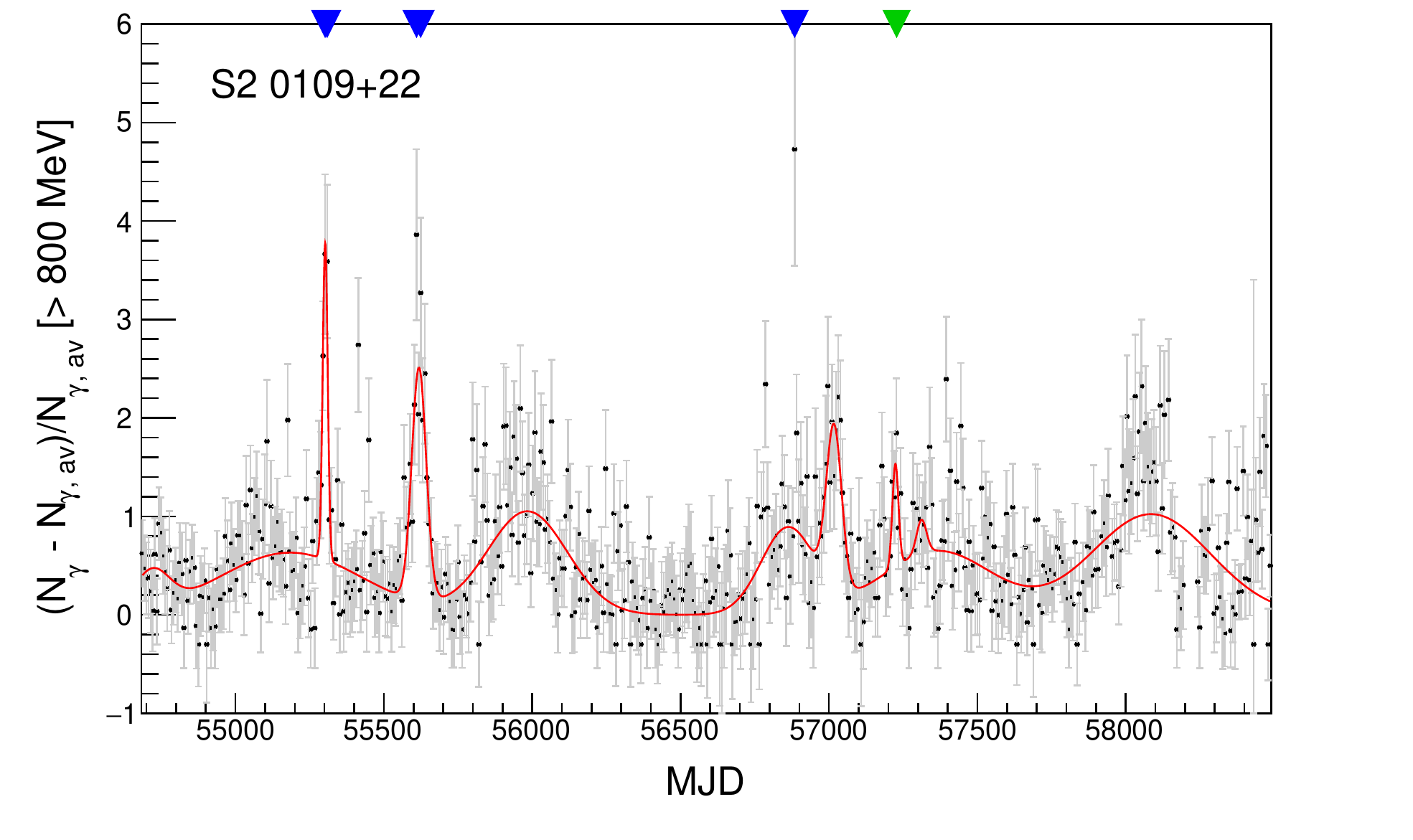} 
  \includegraphics[width=8.6cm,clip,rviewport=0 0 1 1]{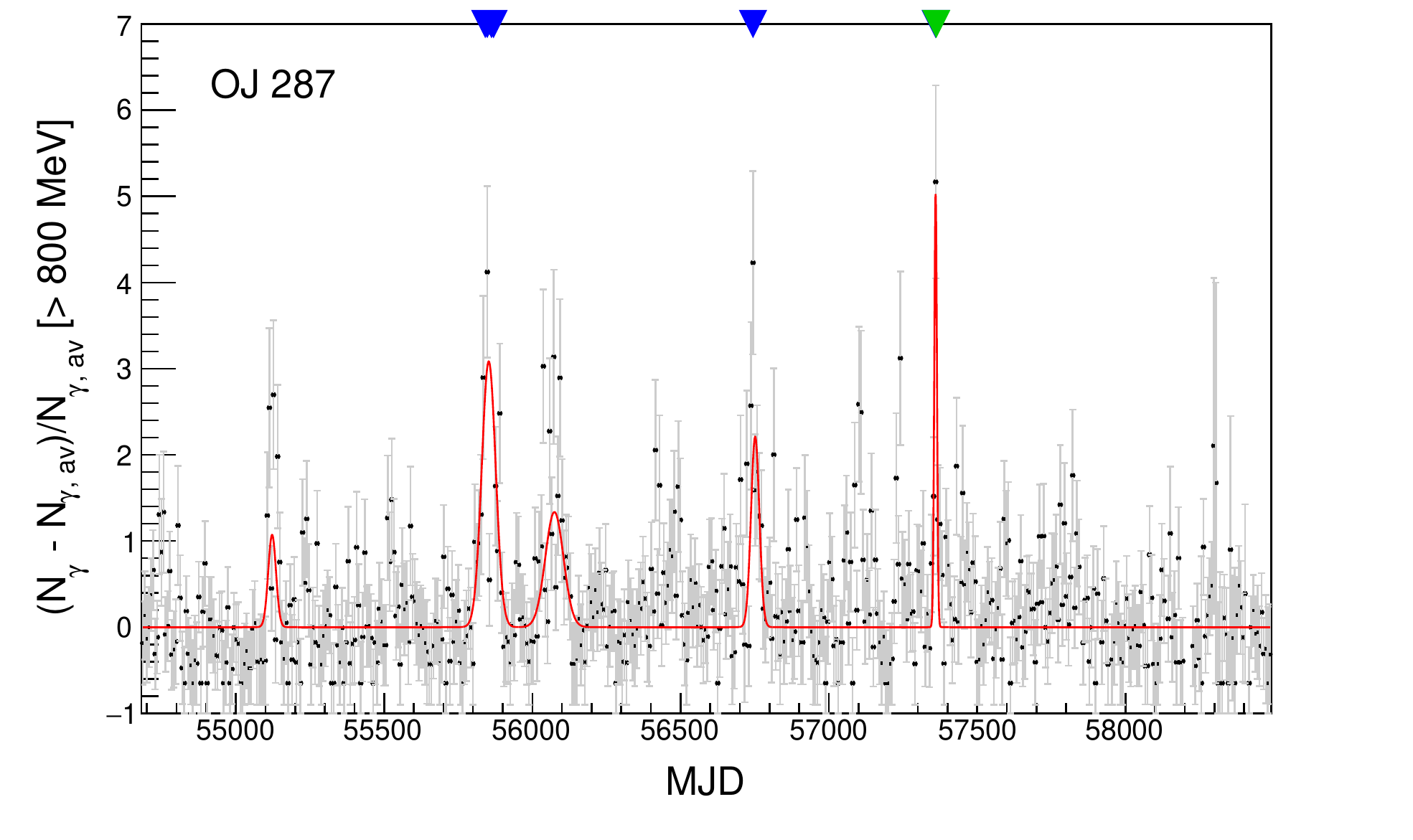}		
  \includegraphics[width=8.6cm,clip,rviewport=0 0 1 1]{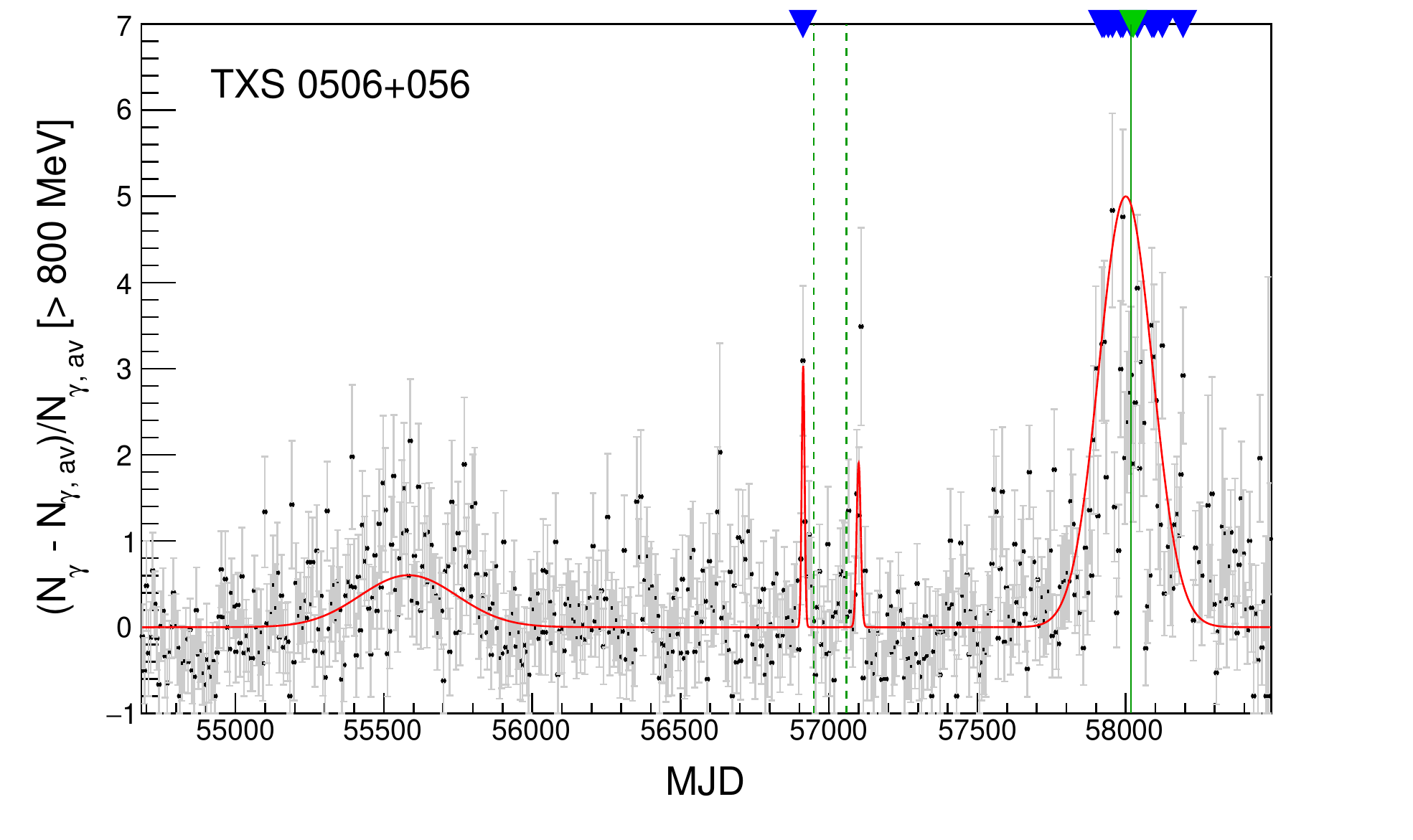}
 
  \contcaption{FAVA lightcurves and fitted flares. For \TXS the green dashed lines give the duration of the 2014-2015 neutrino flare. The green solid line gives the time of detection of IC170922A.}
\end{figure*}

There is no unambiguous definition of blazar flaring states. In this work, in
order to homogeneously define the flare periods we used the FAVA
data~\citep{2017ApJ...846...34A}. We consider only the high-energy
FAVA energy bin (> 800 MeV) to limit the uncertainties introduced to
our results by the larger point spread function of the \Fermi-LAT at
low energies. We look for flaring activity in the FAVA data, during
the flares as defined in Table~\ref{tab:1}. As a reminder, the flares
were selected in several different wavebands. In FAVA, we define the
flare period, as the two-tailed, gaussian, one-sigma region around the
mean. In addition to obtaining the flare period for each of the flares
studied in the earlier sections, we identify further periods of strong
flux enhancement in the FAVA light-curves of the sources in our
sample. We consider all periods during which flaring activity was
observed, at the $\geq 5~\sigma$ level, according to the standard FAVA
definition. The FAVA significance definition, which examines the
Poisson deviation of each bin with respect to the expected background
counts, is most sensitive to sharply peaked variations of the flux. We
therefore, also identify, longer but less sharply peaked flux
enhancements visually, and fit them with a gaussian function to obtain
their respective durations. Each of these selected periods typically
contain one or more $\geq 4\sigma$ upward fluctuations.  The FAVA
lightcurves, and fitted flares are shown in Figure~\ref{fig:Lightcurves}. The
data are binned in photon counts detected per week, $N_{\rm \gamma}$, and
are plotted as relative fluxes, $(N_{\rm \gamma}-N_{\rm \gamma,av})/N_{\rm \gamma,av}$,
with, $N_{\rm \gamma,av}$, the expected number of background photons per
week. 
For some sources (\TXS, \Sfive, \OJ, \Stwo, \mrkfive, \mrkfour, \bllac, \AO, \ThreeC) the baseline 
$N_{\rm \gamma,av}$ given by FAVA needed slight adjustment
to obtain relative fluxes of zero in non-flaring time periods. Changes of baseline in the FAVA 
lightcurves, where observed, are generally related to a change in the background model caused by, for example, changes in the 
\Fermi-LAT observing mode over time. 

Blue triangles denote weeks during which there was flaring
activity at the $\geq 5 \sigma$ level according to the FAVA
definition. Green triangles denote the mean of our canonical flare
periods studied in the earlier sections. We present combined fits to
the entire light-curves, which sometimes identify periods of low
plateaus of enhanced activity. For \TXS the green dashed lines give 
the duration of the 2014-2015 neutrino flare. The green solid line gives the time of detection of IC170922A.
For \AO the solid green vertical line gives the time of the GFU neutrino
alert.

Table \ref{tab:timescales} gives the estimated flare durations based
on the analysis of the FAVA lightcurves, or in the absence of
significant flux enhancement in the FAVA data, the flare timescale
determined in a different waveband as stated. The table also gives the
timescale of smallest detected time-variability in days, $t_{\rm
  var,Obs,d}$, which we use to determine the size of the emitting
region and waveband at which it was detected. Since the FAVA data are
by default weekly binned we used the shortest variability available at
any wavelength to determine $t_{\rm var,Obs,d}$. Source by source details are given below. 

\begin{itemize} 
\item For \ThreeC the studied flare was seen with $>5\sigma$
  significance in the HE FAVA data over a period of two-weeks. A
  coincident flare with similar duration was seen in the optical
  data. A shorter flare was seen in the VHE \gRay data. 
    
  \item For \AO the FAVA data show an $\sim 84$-day long high state in
  August 2008, soon after the launch of \Fermi, consistent with the
  \gRay analysis of~\citet{2012ApJ...751..159A}, followed by a lower-intensity second flare. We have additionally identified a long, $\gamma$-ray flare in 2015 as
  potentially interesting from the point of view of neutrino
  production as discussed in Section~\ref{sec:results}.
  
\item For \mrkfour the FAVA data show no significant flare, though as
  reported in earlier analyses
 ~\citep{Aleksic:2014rca,Petropoulou:2016ujj} the X-ray and VHE
  activity were remarkable. We therefore assume the VHE flare
  timescale for the present analysis.
  
\item  In the case of PG 1553+113, no 5$\sigma$ flares were recorded by the
  FAVA analysis. We nevertheless, flag the $\geq 4\sigma$ flares for
  this source.  The April
  2012 flare does not appear as a period of significant enhancement in
  the FAVA data. We thus assume a 30-day duration, as measured with
  the \Swift-XRT, noting that the 6-day VHE flare observed with MAGIC
  preceded the flare in the X-ray data.

\item For \IES the May 2012 flare does not coincide with any flaring
  activity in FAVA. In the VHE \gRay data there was a short-lived
  flare on May 20th 2012 (MJD 57067), and the X-ray data showed an
  enhancement approximately one week later. The optical data showed a
  historically high-state between MJD 57034 and MJD 57080. The optical, and UV flux did not change much during
  this entire period. We thus use the optical/UV duty cycle in the
  present analysis. For the October 2015 outburst of \IES, an 84-day
  long, high-significance flare is seen in the FAVA data, consistent
  with the analyis of~\citet{0004-637X-846-2-158}. Several flares
  follow and the source remains in a higher than average state until
  the end of the available FAVA data.

  \item  For \mrkfive the VHE flare was
  preceded by a flare at X-ray wavelengths by a few days, but the
  optical and low-energy data show no significant flux enhancement. A
  low significance $\sim21$-day flare ($\sim 2.5\sigma$) level can be
  identified in the HE FAVA data, which defines the timescale used for
  the present analysis.

\item For \Sfive
  a $\sim 2$~week long, $>5\sigma$ flare is seen in the FAVA
  data. Coincident flux enhancements were seen at other wavelengths
  including X-ray and optical and VHE \gRay observations. 
  
\item The flares of \Sfour are short and very intense with respect to
  the flares of the other sources in our sample, in general. During
  the February 2015 flare, the relative flux increased by as much as a
  factor of $\sim 40$ in the HE FAVA data.
  
  \item Multiple
  short flares can be seen in the FAVA data of \bllac during June
  2015. The VHE flare seen by MAGIC coincides with a $\sim 7$-day
  strong FAVA flare, which we assume as the relevant timescale in our
  analysis.

\item For \Stwo, the 2015 studied flare, does not coincide with as
  strong flare in the HE FAVA data. There is an enhancement at the
  $2\sigma$ level, over a period of $\sim 21$ days, consistent with
  the report of the \Fermi analysis of~\citet{2018MNRAS.480..879M},
  who confirm a flux doubling with respect to the baseline 3FGL flux
  during this period. We use this estimate for the flare duration.
  
\item The December 2015 flare of \OJ coincides with a week long
  enhancement in the HE FAVA data at the $5\sigma$ level. In other
  wavelengths, several flares were seen until the source returned to
  its pre-outburst state in May 2016. In addition, the source reached
  a historic maximum in the X-ray band in February 2017. It was
  followed up and observed in several wavelengths, and it was possible
  to detect it in the VHE band for the first time with VERITAS
 ~\citep{OBrien:2017noa}. This second enhancement was not visible in the
  FAVA data and has not been modelled here. 

\item The 2017 flare of \TXS in the FAVA data can be well described by a 
gaussian peaked at MJD58000 and duration 175 days. Our estimate is in 
agreement with the results of~\cite{IceCube:2018dnn}. 

\end{itemize}

\section{Table of acronyms}
Table~\ref{tab:acronyms} gives the definition of abbreviations and acronyms used in this article.

\begin{table}
\caption{Abbreviations and acronyms used in this article.\label{tab:acronyms}}
\begin{center}
\begin{tabular}{c c} 
\hline
GVD & Baikal Gigaton Volume Detector \\
\vspace{0.1cm}
Gen2 & \begin{tabular}{c} IceCube-Gen2: \\ The proposed extension of the IceCube detector \end{tabular}\\
\vspace{0.1cm}
IC40 & \begin{tabular}{c} Partial IceCube configuration  \\
  April 5 2008 - May 20 2009 \end{tabular} \\
  \vspace{0.1cm}
 IC59 & \begin{tabular}{c} Partial IceCube configuration \\ May 20 2009 - May 31 2010 \end{tabular} \\
 \vspace{0.1cm}
 IC79 & \begin{tabular}{c} Partial IceCube configuration \\ May 13 2010 - May 16 2011 \end{tabular} \\
\vspace{0.1cm}
IC86 & \begin{tabular}{c} Partial IceCube configuration \\ May 16 2011 - to date \end{tabular} \\
\vspace{0.1cm}
KM3NeT & Cubic Kilometre Neutrino Telescope \\	
\vspace{0.1cm}
ONC & Ocean Networks Canada \\
\hline
\end{tabular}
\end{center}
\end{table}

\bibliographystyle{mnras}
\label{lastpage}
\end{document}